\newcommand{\modbluetext}[1]{\ifdef{\showcomments}{\textcolor{purple}{#1}}{{#1}}}
\documentclass[acmsmall,screen]{acmart}
\usepackage{color}
\usepackage{xcolor}
\usepackage[justification=centering]{caption}
\usepackage{amsmath,amsfonts}
\usepackage{algorithmic}
\usepackage{graphicx}
\usepackage{textcomp}
\usepackage[T1]{fontenc}
\usepackage{tikz}
\usepackage{tabularx}
\usepackage{multicol}
\usepackage{multirow}
\usepackage{url}
\usepackage{booktabs} 
\usepackage[utf8]{inputenc}
\usepackage{hyperref}
\usepackage{listings}
\usepackage{soulutf8}
\usepackage[edges]{forest}
\usepackage[T1]{fontenc}
\usepackage{microtype}
\usepackage{mathpazo}
\usepackage{pifont}
\usepackage{stackengine}
\usepackage{resizegather}
\usepackage{mhchem}
\usepackage{fontawesome5}
\usepackage{subcaption}
\usepackage{environ}
\usepackage{tabularray}
\usepackage{makecell}

\definecolor{mattered}{RGB}{214, 26, 60} 
\definecolor{mattegreen}{RGB}{54, 159, 57}
\newcommand{\redcross}{{\color{mattered}\fontsize{12pt}{12pt}\selectfont\faTimes}}
\newcommand{\greencheck}{{\color{mattegreen}\fontsize{12pt}{12pt}\selectfont\faCheck}}
\usepackage{wrapfig}
\usepackage[colorinlistoftodos]{todonotes}

\newcommand{\nameoffline}{Offline Pre-Deployment Model Design Techniques}
\newcommand{\nameonline}{Online Runtime Inference Optimizations}
\newcommand{\nameapplication}{On-Device LLM-Based Applications}

\AtBeginDocument{%
  }

\setcopyright{cc}
\setcctype{by}
\acmJournal{CSUR}
\acmYear{2025} \acmVolume{1} \acmNumber{1} \acmArticle{1} \acmMonth{1} \acmPrice{}\acmDOI{10.1145/3719664}

\begin{document}

\title{A Review on Edge Large Language Models: Design, Execution, and Applications}

\author{Yue Zheng}
\affiliation{%
 \institution{Zhejiang University of Technology}
 \city{Hangzhou}
 \country{China}}
\email{zhengyue@zjut.edu.cn}

\author{Yuhao Chen}
\affiliation{%
  \institution{Zhejiang University}
  \city{Hangzhou}
  \country{China}}
\email{csechenyh@zju.edu.cn}

\author{Bin Qian}
\affiliation{%
  \institution{Zhejiang University}
  \city{Hangzhou}
  \country{China}}
\email{bin.qian0718@gmail.com}

\author{Xiufang Shi}
\affiliation{%
  \institution{Zhejiang University of Technology}
  \city{Hangzhou}
  \country{China}}
\email{xiufangshi@zjut.edu.cn}

\author{Yuanchao Shu}
\authornote{Corresponding authors.}
\affiliation{%
  \institution{Zhejiang University}
  \city{Hangzhou}
  \country{China}}
\email{ycshu@zju.edu.cn}

\author{Jiming Chen}
\authornotemark[1]
\affiliation{%
  \institution{Zhejiang University}
  \city{Hangzhou}
  \country{China}}
\email{cjm@zju.edu.cn}

\renewcommand{\shortauthors}{Zheng et al.}

\begin{abstract}
Large language models (LLMs) have revolutionized natural language processing with their exceptional understanding, synthesizing, and reasoning capabilities. However, deploying LLMs on resource-constrained edge devices presents significant challenges due to computational limitations, memory constraints, and edge hardware heterogeneity. This survey provides a comprehensive overview of recent advancements in edge LLMs, covering the entire lifecycle — from resource-efficient model design and pre-deployment strategies to runtime inference optimizations. It also explores on-device applications across various domains. By synthesizing state-of-the-art techniques and identifying future research directions, this survey bridges the gap between the immense potential of LLMs and the constraints of edge computing.
\end{abstract}




\maketitle

\section{Introduction}
Transformer-based large language models (LLMs) have made significant strides in recent years, reshaping the landscape of natural language processing (NLP). 
This rapid evolution has led to the emergence of several open-source LLMs, including Meta's LLaMA family~\cite{touvron2023llama,touvron2023llama2, dubey2024llama3herd}, Google's Gemma~\cite{team2024gemma,GoogleGemma2}, and more recently, DeepSeek AI's DeepSeek series~\cite{guo2025deepseekr1,liu2024deepseekv3}. 
The success of LLMs stems from their exceptional capabilities in natural language understanding, synthesizing, reasoning, and generation~\cite{dong2019naturalLanguageUnderstandingGeneration,dong2022surveyNaturalLanguageGeneration}, driving breakthroughs in applications like document summarization, question answering, and text reformulating~\cite{koh2022empiricalDocumentSummarization,liu2023preNaturalLanguageProcessing,lai2024surveyfigurative}. These advancements have had profound implications across both academic and industrial domains, enabling the development of widely adopted tools like ChatGPT~\cite{IntroducingChatGPTOpenAI}, Copilot~\cite{copilot} and Gemini~\cite{team2023gemini}. The continued progress in LLMs underscores their transformative impact on artificial intelligence~\cite{wenAutoDroidLLMpoweredTask2024,kang2024quantitativeAutoFL,chen2024drivingWithLLMs}, human-computer interaction~\cite{kim2024languageRCI,gurrealWebAgent,huang2023innermonologue}, and beyond.

While cloud-based deployment has traditionally supported LLMs' computational demands, there is a growing need to bring these models to resource-constrained edge devices, including personal agents~\cite{rawassizadehODSearchFastResource2022,wenAutoDroidLLMpoweredTask2024}, office assistants~\cite{gurrealWebAgent,team2023gemini} and industrial Internet of Things (IoT) systems~\cite{tian2024drivevlm,jeong2023winclip}. Edge-based LLMs — executed directly on devices — offer key advantages:
Firstly, local inference ensures faster responses and functionality without internet connectivity~\cite{caoMobiVQAEfficientOnDevice2022}, critical for applications in robotics and autonomous systems~\cite{chen2024drivingWithLLMs,xiang2024languageEmbodied,dai25hotmobile}. 
Secondly, processing sensitive data on-device eliminates risks associated with cloud transmission~\cite{fanTaskFusionEfficientTransfer2023,team2023gemini}.
Lastly, on-device learning enables models to adapt to user-specific preferences and contexts~\cite{qi2024interactive,bhardwaj2022ekya,khani2023recl,padmanabhan2023gemel}.

However, deploying LLMs on resource-constrained edge devices presents significant challenges.
Firstly, ~\textbf{the computational and memory constraints} impose substantial limitations on LLM loading and inference. LLMs often consist of billions of parameters, resulting in massive memory footprints that exceed the RAM capacities of most edge devices~\cite{chen2023confidant}. For example, a LLaMA-2~\cite{touvron2023llama2} model with 7B parameters requires over 8GB of memory even in FP16 precision. Without compression techniques, edge devices risk latency spikes and memory overflow during model loading~\cite{linAWQActivationawareWeight2024}. Moreover, the quadratic complexity of the self-attention mechanism with respect to sequence length exacerbates computational demands,  creating severe throughput bottlenecks on edge Central Processing Units (CPUs), Graphics Processing Units (GPUs) or Neural Processing Units (NPUs)~\cite{shuvo2022efficientaccelerationOnEdgeDevice}.

~\modbluetext{Secondly, \textbf{the heterogeneous nature of edge computing devices} complicates runtime inference optimizations. Edge devices range from smartphones with ARM CPUs and limited memory to IoT devices equipped with low-power chips. On mobile devices, frameworks like llama.cpp~\cite{llamacpp} and MLC LLM~\cite{mlc-llm} optimize computational operators, while edge GPUs adopt approaches like vLLM~\cite{kwonEfficientMemoryManagement2023} to alleviate memory bandwidth limitations and enhance throughput. Effective hardware-software co-design is critical to align workloads with hardware-specific capabilities. Additionally, the choice of hardware (e.g., CPUs, GPUs, or NPUs) and its integration with software frameworks directly affect inference efficiency, necessitating adaptable solutions tailored to diverse edge environments~\cite{zhang2024vulcan}}

Lastly, \textbf{developing practical edge applications} remains challenging, particularly in bridging centralized LLM processing with distributed edge scenarios. In personal and enterprise applications, frameworks such as AutoDroid~\cite{wenAutoDroidLLMpoweredTask2024} and WebAgent~\cite{gurrealWebAgent} demonstrate the complexities of maintaining responsiveness and accuracy for task automation. For industrial systems like autonomous vehicles~\cite{tian2024drivevlm,chen2024drivingWithLLMs}, precise task prioritization and dynamic resource allocation are essential to balance LLM inference with real-time control processes. These domain-specific optimizations are vital to ensure LLMs meet real-world latency and reliability requirements on resource-constrained devices.

\begin{figure}[ht]
    \vspace{2mm}
    \centering    \includegraphics[width=0.8\textwidth]{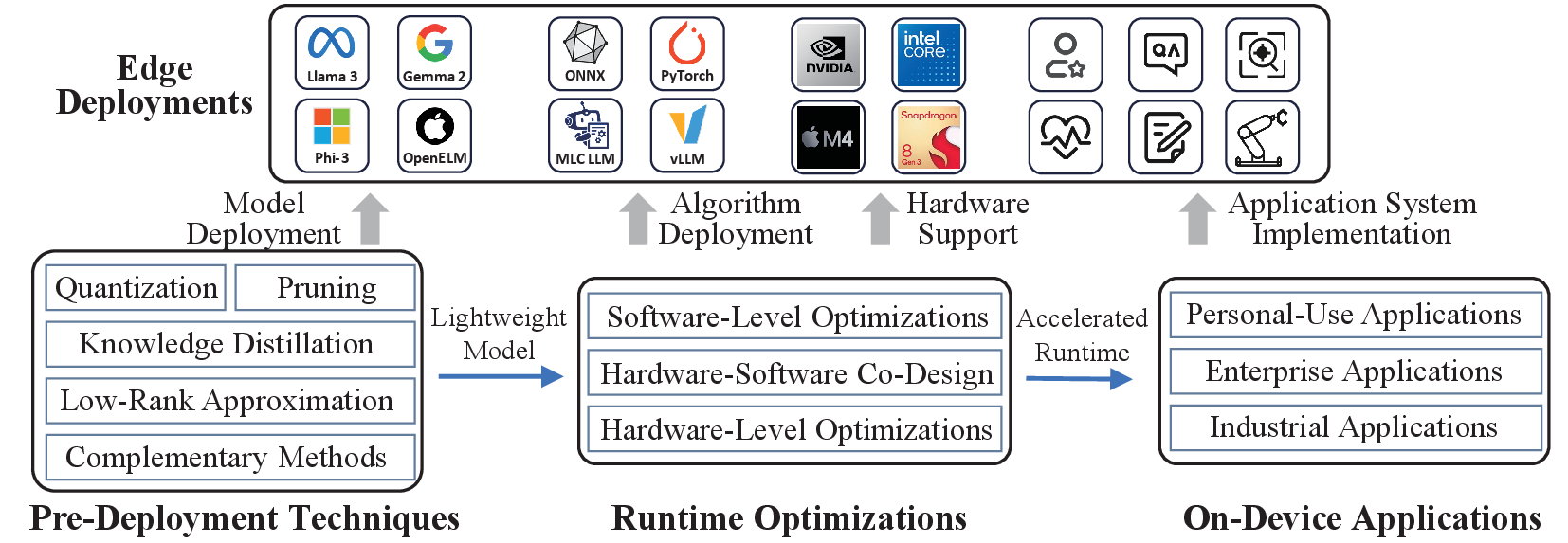}
    \caption{~\modbluetext{Edge LLMs deployment, optimization and application pipeline.}}
    \label{fig: pipeline_figure}
\end{figure}

~\modbluetext{To address these challenges, we have devised a comprehensive optimization pipeline that integrates techniques across the entire lifecycle of edge-based LLM deployment, as illustrated in Fig.~\ref{fig: pipeline_figure}. Starting with pre-deployment methods such as quantization, pruning, and knowledge distillation, the pipeline enables the creation of compact, resource-efficient models that preserve performance while reducing computational demands. These models are then deployed on edge devices, where runtime optimizations—spanning software-level strategies, hardware-level enhancements, and hardware-software co-design—ensure seamless adaptation to heterogeneous environments. Finally, the optimized models power a variety of on-device applications, from personal assistants to enterprise systems and industrial solutions, showcasing the practical impact of edge LLMs. This unified process effectively addresses key deployment challenges, demonstrating how offline compression and real-time optimizations together enable diverse real-world applications.}

Following this pipeline, this survey aims to provide a comprehensive exploration of the key areas involved in enabling LLMs on edge and mobile devices, including three crucial aspects, as summarized in Fig.~\ref{fig: taxonomy_figure}.
Specifically,
\begin{itemize}
\item \textbf{\nameoffline}. It focuses on compression models to reduce size and ease deployment on edge devices. Traditional methods like quantization, pruning, knowledge distillation, and low-rank approximation face unique challenges with LLMs due to their scale, Transformer architecture, and diverse tasks~\cite{abnar2020quantifying,ji2021distribution}. These challenges have motivated novel compression methods tailored for LLMs. Quantization reduces LLM size by representing weights and activations with fewer bits~\cite{linAWQActivationawareWeight2024,shenAgileQuantActivationGuidedQuantization2024}. Pruning removes unnecessary attention heads or other Transformer components, either structurally or unstructurally~\cite{xiaStructuredPruningLearns2022,kurticOptimalBERTSurgeon2022}. Knowledge distillation transfers knowledge to smaller models~\cite{liangLessMoreTaskaware2023,jiangLionAdversarialDistillation2023}. Low-rank approximation exploits matrix redundancy for efficient compression~\cite{liLoSparseStructuredCompression2023,hsuLanguageModelCompression2021}. Complementary methods, such as advanced pre-training strategies, data curation, and architectural optimization, further enhance compression effectiveness~\cite{hughesPhi2SurprisingPower2023,team2024gemma,mehta2024openelm}.

\item \textbf{\nameonline}. It introduces inference optimization techniques that improves the LLM performance on resource-constrained edge devices. Key strategies include software-level optimizations, hardware-software co-design, and hardware-level optimizations. Software-level optimizations encompass resource-aware scheduling strategies for cloud-edge collaboration~\cite{borzunov2023distributed,zhao2024lingualinked,sun2024spectr}, single-device inference scenarios~\cite{zeng2024consistentee,goyal2020powerBERT,shengFlexGenHighThroughputGenerative2023}, and lightweight frameworks for efficient memory management and tensor operations~\cite{kwonEfficientMemoryManagement2023,song2024powerinfer}. Hardware-software co-design integrates software algorithms with specific hardware capabilities, facilitating efficient hardware profiling and enabling the implementation of hardware-aware inference algorithms~\cite{guoOliVeAcceleratingLarge2023,wangSpAttenEfficientSparse2021}. Hardware-level optimizations introduce commonly used edge hardware devices, highlighting their innovations in on-device LLM inference~\cite{QualcommSnapdragon8Gen3MobilePlatform,yuan2024mobilefirmware}.

\item \modbluetext{\textbf{\nameapplication}}. It showcases the practical impact of on-device LLMs across personal, enterprise, and industrial domains. In personal applications, they power AI assistants for tasks such as daily management, healthcare monitoring, and companionship, offering privacy-preserving and low-latency interactions~\cite{wenAutoDroidLLMpoweredTask2024,lu2024multimodal,huang2023innermonologue}. In enterprise settings, on-device LLMs enhance productivity through message completion, meeting summarization, and secure local processing of sensitive data~\cite{zhu2024textrewriting,tian2024dialoguesummarization,li2024sheetcopilot}. In industrial scenarios, they enable real-time and local processing capabilities like autonomous driving, fault localization, and anomaly detection, improving efficiency and safety in complex environments~\cite{tian2024drivevlm,kang2024quantitativeAutoFL,gu2024anomalygpt}.

\end{itemize}

\definecolor{hidden-draw}{rgb}{0,0,0}
\definecolor{blue}{RGB}{214, 225, 244}
\definecolor{orange}{RGB}{252,222,208}
\tikzstyle{my-box}=[
    rectangle,
    draw=hidden-draw,
    rounded corners,
    text opacity=1,
    minimum height=1.2em,
    minimum width=5em,
    inner sep=2pt,
    align=center,
    fill opacity=.5,
    line width=0.8pt,
]
\tikzstyle{leaf}=[my-box, minimum height=1.5em,
    fill=hidden-pink!80, text=black, align=left,font=\normalsize,
    inner xsep=2pt,
    inner ysep=4pt,
    line width=0.8pt,
]
\begin{figure}[ht]
    \centering
    \resizebox{\textwidth}{!}{
        \begin{forest}
            forked edges,
            for tree={
                grow=east,
                reversed=true,
                anchor=base west,
                parent anchor=east,
                child anchor=west,
                base=center,
                font=\large,
                rectangle,
                draw=hidden-draw,
                rounded corners,
                align=left,
                text centered,
                minimum width=4em,
                edge+={darkgray, line width=1pt},
                s sep=3pt,
                inner xsep=2pt,
                inner ysep=3pt,
                line width=0.8pt,
                ver/.style={rotate=90, child anchor=north, parent anchor=south, anchor=center},
            },
            where level=1{text width=18em,font=\normalsize,}{},
            where level=2{text width=18em,font=\normalsize,}{},
            where level=3{text width=23em,font=\normalsize,}{},
            [
                \textbf{Efficient LLMs Methods}, ver
                [
                    \makecell{\textbf{Pre-Deployment Techniques}  (\S \ref{Sec: \nameoffline})\\(Research details in Fig. \ref{fig: fig_sec3_offline_overview})}, fill=blue
                    [
                        \textbf{Quantization}  (\S \ref{SubSec: Quantization}), fill=blue
                        [
                            \textbf{Weight-Only Quantization} (\S \ref{SubSubSec: Weight-Only Quantization}), fill=blue
                        ]
                        [
                            \textbf{Weight-Activation Co-Quantization} (\S \ref{SubSubSec: Weight-Activation Co-Quantization}), fill=blue
                        ]
                    ]
                    [
                        \textbf{Pruning}  (\S \ref{SubSec: Pruning}), fill=blue
                        [
                            \textbf{Structured Pruning} (\S \ref{SubSubSec: Structured Pruning}), fill=blue
                        ]
                        [
                            \textbf{Unstructured Pruning} (\S \ref{SubSubSec: Unstructured Pruning}), fill=blue
                        ]
                    ]
                    [
                        \textbf{Knowledge Distillation}  (\S \ref{SubSec: Knowledge Distillation}), fill=blue
                        [
                            \textbf{White-Box Knowledge Distillation} (\S \ref{SubSubSec: White-Box Knowledge Distillation}), fill=blue
                        ]
                        [
                            \textbf{Black-Box Knowledge Distillation} (\S \ref{SubSubSec: Black-Box Knowledge Distillation}), fill=blue
                        ]
                    ]
                    [
                        \textbf{Low-Rank Approximation}  (\S \ref{SubSec: Low-Rank Approximation}), fill=blue
                    ]
                    [
                        \textbf{Complementary Methods}  (\S \ref{SubSec: Complementary Methods}), fill=blue
                    ]
                ]
                [
                    \makecell{\textbf{Runtime Optimizations}  (\S \ref{Sec: \nameonline})\\(More details in Tables \ref{tab: tab_collaboration_literature}-\ref{tab: tab_hardware})}, fill=orange, 
                     [
                        \textbf{Software-Level Optimizations} (\S \ref{SubSec: Software-Level Optimizations}),  fill=orange
                        [   
                            \makecell{\textbf{Cloud and Multi-Edge Collaboration} (\S \ref{SubSubSec: Cloud and Multi-Edge Collaboration})\\(Research details in Table \ref{tab: tab_collaboration_literature})}, fill=orange
                        ]
                        [
                             \makecell{\textbf{Single-Device Resource Scheduling} (\S \ref{SubSubSec: Single-Device Resource Scheduling})\\(Research details in Table \ref{tab: tab_single_literature})},  fill=orange
                        ]
                        [
                            \makecell{\textbf{Framework-Level Optimizations} (\S \ref{SubSubSec: Framework-Level Optimizations})\\(Research details in Table \ref{tab: tab_frameworks})},  fill=orange
                        ]
                    ]
                    [
                        \makecell{\textbf{Hardware-Software Co-Design} (\S \ref{SubSec: Hardware-Software Co-Design})\\(Research details in Table \ref{tab: tab_hardware_software_codesign})},  fill=orange
                        [
                            \textbf{Hardware-Aware Sparsity} (\S \ref{SubSubSec: Hardware-Aware Sparsity}), fill=orange
                        ]
                        [
                            \textbf{Hardware-Optimized Arithmetic Formats} (\S \ref{SubSubSec: Hardware-Optimized Arithmetic Formats}), fill=orange
                        ]
                    ]
                    [  
                        \makecell{\textbf{Hardware-Level Optimizations} (\S \ref{SubSec: Hardware-Level Optimizations})\\(Research details in Table \ref{tab: tab_hardware})},  fill=orange
                        [   
                            \textbf{CPUs: Foundation of AI Workflow} (\S \ref{SubSubSec: CPUs: Foundation of AI Workflow}), fill=orange
                        ]
                        [
                            \textbf{GPUs: Parallel Acceleration} (\S \ref{SubSubSec: GPUs: Parallel Acceleration}), fill=orange
                        ]
                        [
                            \textbf{NPUs: Neural Network Optimization} (\S \ref{SubSubSec: NPUs: Neural Network Optimization}), fill=orange
                        ]
                    ]
                ]
                [
                    \makecell{\textbf{On-Device Applications} (\S \ref{Sec: \nameapplication})\\(Research details in Fig. \ref{fig: fig_sec5_application})}, fill=green!10
                      [
                        \textbf{Personal-Use Applications} (\S \ref{SubSec: Personal-Use Applications}),  fill=green!10
                         [   
                            \textbf{Personal Agents} (\S \ref{SubSubSec: Personal Agents}), fill=green!10
                        ]
                        [
                            \textbf{Healthcare Assistants} (\S \ref{SubSubSec: Healthcare Assistants}), fill=green!10
                        ]
                        [
                            \textbf{Companion Robots} (\S \ref{SubSubSec: Companion Robots}), fill=green!10
                        ]
                    ]
                     [
                        \textbf{Enterprise Applications} (\S \ref{SubSec: Enterprise Applications}),  fill=green!10
                         [   
                            \textbf{Message Completion} (\S \ref{SubSubSec: Message Completion}), fill=green!10
                        ]
                        [
                            \textbf{Meeting Summarization} (\S \ref{SubSubSec: Meeting Summarization}), fill=green!10
                        ]
                        [
                            \textbf{Computer Operation} (\S \ref{SubSubSec: Computer Operation}), fill=green!10
                        ]
                    ]
                     [
                        \textbf{Industrial Applications} (\S \ref{SubSec: Industrial Applications})),  fill=green!10
                         [   
                            \textbf{Autonomous Driving} (\S \ref{SubSubSec: Autonomous Driving}), fill=green!10
                        ]
                        [
                            \textbf{Fault Localization} (\S \ref{SubSubSec: Fault Localization}), fill=green!10
                        ]
                        [
                            \textbf{Anomaly Detection} (\S \ref{SubSubSec: Anomaly Detection}), fill=green!10
                        ]
                    ]
                ]
            ]
        \end{forest} 
        }
    \captionsetup{font={},justification=raggedright}
    \caption{\modbluetext{A comprehensive overview of on-device LLMs, with detailed research categorized into: Pre-Deployment Techniques (\S \ref{Sec: \nameoffline}, research details in Fig. \ref{fig: fig_sec3_offline_overview}), Runtime Optimizations (\S \ref{Sec: \nameapplication}, details distributed across Tables \ref{tab: tab_collaboration_literature}-\ref{tab: tab_hardware}), and On-Device Applications (\S \ref{Sec: \nameapplication}, research details in Fig. \ref{fig: fig_sec5_application}). Each branch represents a key research direction with detailed methodology and implementation discussions in corresponding sections.}}
    \label{fig: taxonomy_figure}
\end{figure}

By employing these innovative techniques and methodologies, developers can harness the benefits of reduced model sizes and improved computational efficiency, facilitating the seamless integration of LLMs on edge devices. This advancement not only improves edge computing performance but also broadens the applicability of LLMs in resource-constrained environments, potentially revolutionizing the landscape of edge AI applications.

The remainder of this paper is structured as follows: 
Section~\ref{Sec: Background and Related Work} examines the widening gap between LLM complexity and edge device capabilities, reviews related work on efficient LLMs and edge computing, and analyzes research trends in on-device LLM optimizations, establishing the context for our survey.
Section~\ref{Sec: \nameoffline} and Section~\ref{Sec: \nameonline} provide a comprehensive examination of the state-of-the-art approaches for offline pre-deployment techniques and online runtime optimizations, respectively. 
Section~\ref{Sec: \nameapplication} delves into the on-device applications of LLMs, highlighting their vast potential. 
Section~\ref{Sec: Future Directions and Open Challenges} discusses future directions and open challenges in the field of on-device LLMs, while Section~\ref{Sec: Conclusion} concludes the survey, summarizing the key takeaways and insights gained.

\section{Background and Related Work} \label{Sec: Background and Related Work}
\begin{figure}[ht]
    \centering
    \includegraphics[width=0.8\textwidth]{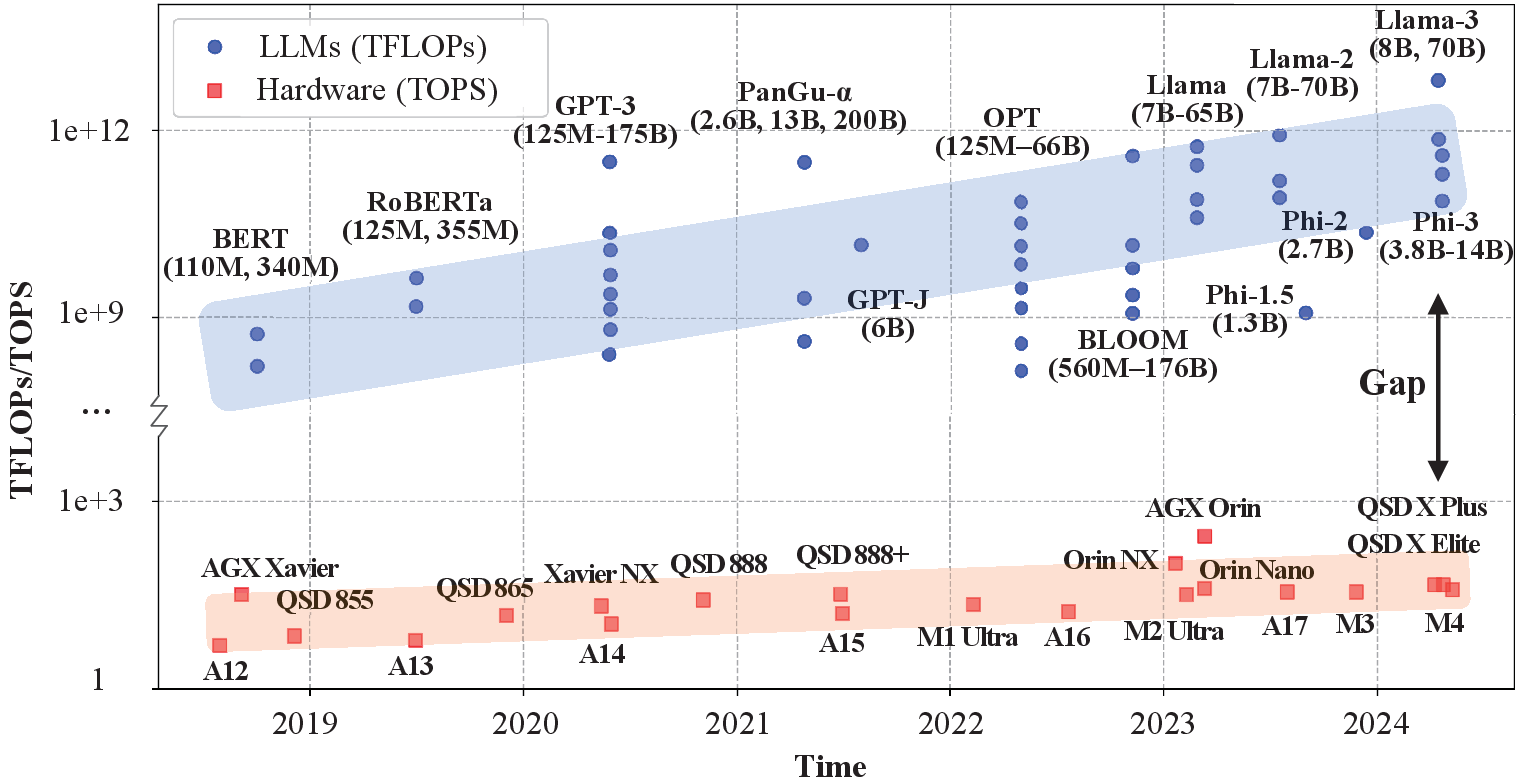}
    \caption{LLMs (TFLOPs) vs Edge Devices (TOPS) Over Time. (QSD: Qualcomm Snapdragon.)}
    \label{fig: gap_figure}
\end{figure}

The rapid advancement of LLMs and the increasing demand for edge computing have led to a growing interest in deploying these powerful AI models on resource-constrained devices~\cite{linAWQActivationawareWeight2024}. However, this endeavor is hindered by a significant disparity between the computational complexity of LLMs and the capabilities of edge devices. Fig.~\ref{fig: gap_figure} illustrates this widening gap, showing the evolution of estimated LLM pre-training FLOPs~\cite{brown2020languageModelsAreFewShotLearners,ruan2024observationalScalingLaws} (measured in TFLOPs) compared to the AI performance of edge devices~\cite{JetsonModulesSupport, SnapdragonSeriesMobile, AIPerformanceNPU} (measured in TOPS) over time. 
The pre-training FLOPs for LLMs are estimated using the widely accepted heuristic \(C \approx 6ND\), where \(N\) denotes the model's parameter count and \(D\) represents the total number of tokens used during pre-training~\cite{kaplan2020scalingLaws}. 

This graph clearly illustrates the widening gap between the rapidly increasing computational complexity of LLMs and the relatively slow improvement in edge device capabilities. While LLMs have experienced a steep rise in estimated pre-training FLOPs, the AI performance of edge devices has improved at a much slower pace. This growing disparity underscores the critical need for researching efficient LLM deployment and implementation methods, which is precisely the focus of our survey. 

While prior surveys in edge computing research have been conducted on efficient learning techniques on Deep Neural Network (DNN) architectures, such as Convolutional Neural Networks (CNNs) and Recurrent Neural Networks (RNNs)~\cite{wangConvergenceEdgeComputing2020a,lalapuraRecurrentNeuralNetworks2022b}, they have not adequately addressed the unique challenges posed by LLMs, including their larger model sizes and the complexities introduced by attention mechanisms. While research has addressed resource management~\cite{zhangResourceManagementMobile2023a} and security enhancements~\cite{wang2024end} in edge environments, these studies primarily focus on general deep learning, neglecting the specific needs of transformer-based LLMs in mobile edge computing.

Complementing these efforts, research in the realm of NLP has also made significant strides.
~\citet{xuSurveyModelCompression2023} review methods for enhancing the efficiency of model compression and acceleration for pre-trained LLMs. ~\citet{hedderichSurveyRecentApproaches2021} survey approaches for improving performance in low-resource NLP settings. ~\citet{wanEfficientLargeLanguage2024} provide a comprehensive review of efficient LLMs research, organizing the literature into model-centric, data-centric, and framework-centric approaches. ~\citet{trevisoEfficientMethodsNatural2023} synthesize methods for conducting NLP under constraints of limited data, time, storage, or energy, emphasizing the trade-offs between performance and resource consumption.
However, these works do not specifically address the challenges of deploying LLMs in edge environments. As a result, there is a critical need for focused research in this area.

Our survey uniquely provides a comprehensive analysis of LLMs in edge environments. The two most relevant surveys are Mobile Edge Intelligence for LLMs~\cite{qu2024mobileedgeintelligence}, which focuses primarily on collaborative resource management across different computing nodes, and Personal LLM Agents~\cite{li2024personal}, which explores the applications and scenarios of LLM assistants. However, the former does not address framework- and hardware-level optimizations for edge devices, and the latter lacks a systematic analysis of runtime optimizations on edge devices.
To bridge this gap, we offer a holistic, top-down perspective on LLMs for edge devices, encompassing the entire optimization pipeline from offline pre-deployment model design techniques to online runtime inference optimizations and on-device LLM-based applications across various sectors. Our analysis covers model architectures, compression strategies, software-level optimizations, hardware-software co-design, and hardware enhancements for Transformer-based architectures at the edge. Additionally, we examine on-device application systems designed to maximize LLM performance under resource constraints. This multi-faceted approach distinguishes our survey, addressing challenges and solutions across the entire optimization pipeline for edge-deployed LLMs.

\def\pubwid{36.496}
\def\pubhit{23.395}

\begin{figure}[ht]
    \centering
    \begin{tikzpicture}
        \node[anchor=north west, inner sep=0] (image) at (0,0) {\includegraphics[width=1\textwidth]{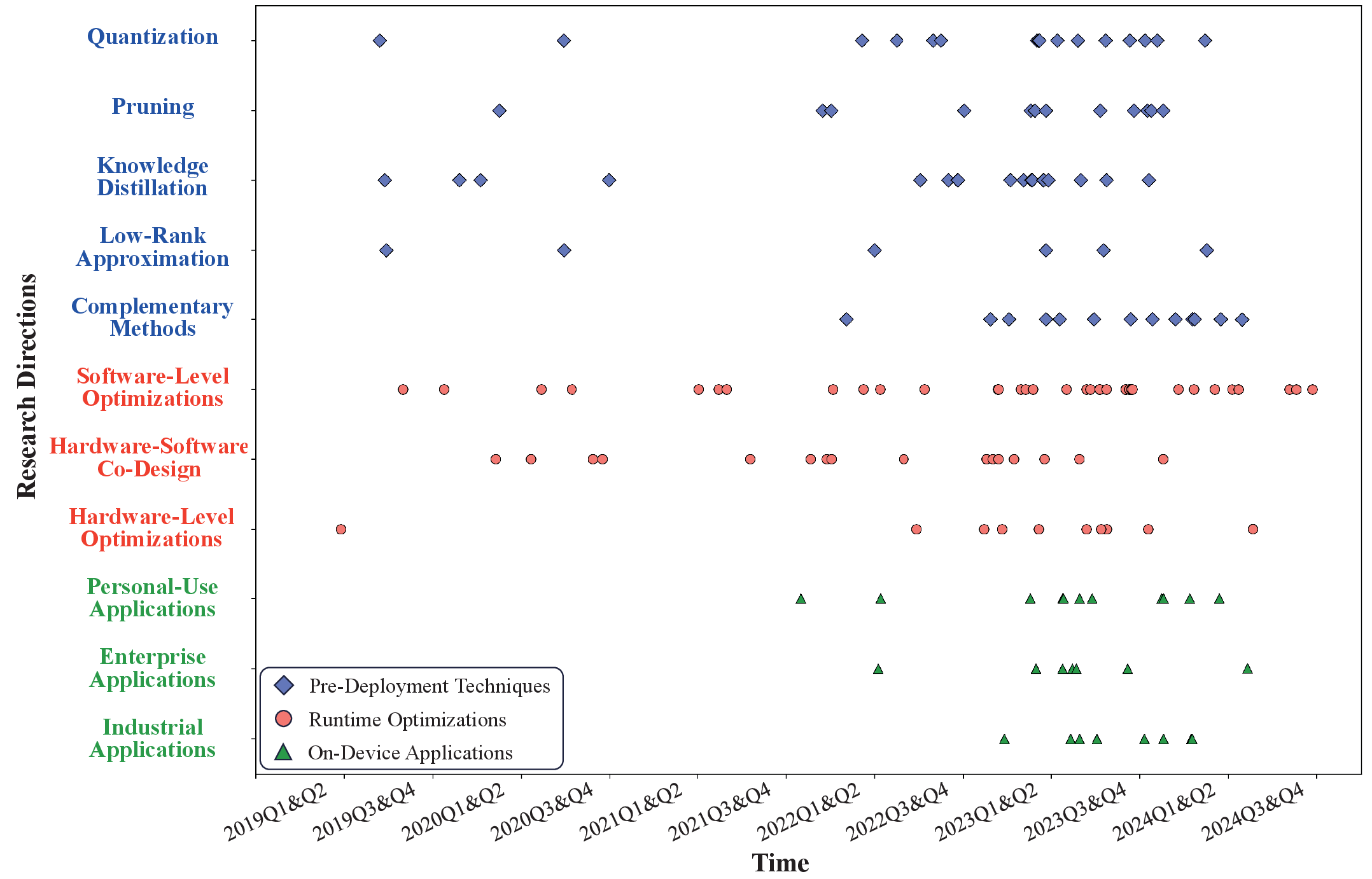}}; 
        \begin{scope}[x={(image.north east)}, y={(image.south west)}] 
            \node[font=\tiny] at (10.11/\pubwid,0.56/\pubhit) {\cite{shenQBERTHessianBased2020a}}; 
            \node[font=\tiny] at (14.98/\pubwid,0.56/\pubhit) {\cite{zhang2020ternarybert}}; 

            \node[font=\tiny] at (23.42/\pubwid,0.56/\pubhit) {\cite{yaoZeroQuantEfficientAffordable2022,dettmersLLMInt88bit}}; 
            \node[font=\tiny] at (24.96/\pubwid,0.56/\pubhit) {\cite{frantarOPTQAccurateQuantization2022}}; 
            \node[font=\tiny] at (24.96/\pubwid,1.73/\pubhit) {\cite{xiaoSmoothQuantAccurateEfficient2023}}; 
            \node[font=\tiny] at (29.55/\pubwid,0.56/\pubhit) {\cite{linAWQActivationawareWeight2024,leeOWQOutlierAwareWeight2024,dettmersSpQRSparseQuantizedRepresentation2023,yuBoostTransformerbasedLanguage2023,shaoOmniQuantOmnidirectionallyCalibrated2023}}; 
             \node[font=\tiny] at (29.55/\pubwid,1.73/\pubhit){\cite{liuQLLMAccurateEfficient2023,shenAgileQuantActivationGuidedQuantization2024,egiazarianextreme,tseng2024quipbetter}};

          \node[font=\tiny] at (32.88/\pubwid,1.09/\pubhit){\cite{zhang2024qhitter}};

             \node[font=\tiny] at (13.3/\pubwid,2.33/\pubhit) {\cite{sanhMovementPruningAdaptive2020}}; 
             \node[font=\tiny] at (21.91/\pubwid,2.33/\pubhit) {\cite{kurticOptimalBERTSurgeon2022}}; 
              \node[font=\tiny] at (22.14/\pubwid,3.56/\pubhit) {\cite{xiaStructuredPruningLearns2022}}; 
             \node[font=\tiny] at (25.67/\pubwid,2.33/\pubhit) {\cite{frantarSparseGPTMassiveLanguage2023}}; 
             \node[font=\tiny] at (27.52/\pubwid,2.33/\pubhit) {\cite{maLLMPrunerStructuralPruning2023,zhang2024loraprunestructuredpruningmeets}}; 
             \node[font=\tiny] at (27.89/\pubwid,3.56/\pubhit) {\cite{sun2024simpleeffectivepruningapproach}}; 
             \node[font=\tiny] at (29.28/\pubwid,3.56/\pubhit) {\cite{xiaShearedLLaMAAccelerating2023}}; 
             \node[font=\tiny] at (30.38/\pubwid,2.33/\pubhit) {\cite{zhang2024plugandplay,ashkboos2024slicegptcompresslargelanguage}}; 
             \node[font=\tiny] at (30.85/\pubwid,3.56/\pubhit) {\cite{an2023fluctuationbasedadaptivestructuredpruning,xu2024besapruning}}; 

            \node[font=\tiny] at (10.25/\pubwid,4.18/\pubhit) {\cite{jiaoTinyBERTDistillingBERT2020}};
             \node[font=\tiny] at (12.24/\pubwid,4.18/\pubhit) {\cite{wangMiniLMDeepSelfAttention2020}}; 
             \node[font=\tiny] at (12.79/\pubwid,5.4/\pubhit) {\cite{sunMobileBERTCompactTaskAgnostic2020}}; 
             \node[font=\tiny] at (16.25/\pubwid,4.18/\pubhit) {\cite{wangMiniLMv2MultiHeadSelfAttention2021}}; 
             \node[font=\tiny] at (24.98/\pubwid,4.18/\pubhit) {\cite{liangLessMoreTaskaware2023,shridharDistillingReasoningCapabilities2023}};
            \node[font=\tiny] at (24.98/\pubwid,5.4/\pubhit) {\cite{hoLargeLanguageModels2023,chenDISCODistillingCounterfactuals2023}}; 

             \node[font=\tiny] at (28.31/\pubwid,4.18/\pubhit) {\cite{wuLaMiniLMDiverseHerd2024,hsiehDistillingStepbyStepOutperforming2023,zhangLiftingCurseCapacity2023,jiangLionAdversarialDistillation2023}}; 
             
             \node[font=\tiny] at (28.31/\pubwid,5.4/\pubhit) {\cite{guMiniLLMKnowledgeDistillation2023,padmanabhanPropagatingKnowledgeUpdates,liSymbolicChainofThoughtDistillation2023,kimTokenScaledLogitDistillation,chenMCCKDMultiCoTConsistent2023}}; 

              \node[font=\tiny] at (31.47/\pubwid,4.8/\pubhit) {\cite{wan2024knowledgefusion}}; 

             \node[font=\tiny] at (10.27/\pubwid,6.07/\pubhit) {\cite{lanALBERTLiteBERT2019}};
             \node[font=\tiny] at (15.01/\pubwid,6.07/\pubhit) {\cite{chenDRONEDataawareLowrank2021}};
             \node[font=\tiny] at (23.27/\pubwid,6.07/\pubhit) {\cite{hsuLanguageModelCompression2021}};
             \node[font=\tiny] at (27.87/\pubwid,6.07/\pubhit) {\cite{liLoSparseStructuredCompression2023}};
             \node[font=\tiny] at (29.37/\pubwid,6.07/\pubhit) {\cite{sahaMatrixCompressionRandomized2023}};
             \node[font=\tiny] at (32.14/\pubwid,6.07/\pubhit) {\cite{jiFeaturebasedLowRankCompression2024,chavanSurgicalFeatureSpaceDecomposition2024}};

              \node[font=\tiny] at (22.53/\pubwid,7.9/\pubhit) {\cite{zhang2022opt}};
              \node[font=\tiny] at (26.62/\pubwid,7.9/\pubhit) {\cite{touvron2023llama,biderman2023pythia}};
              \node[font=\tiny] at (28.05/\pubwid,9.14/\pubhit) {\cite{li2023textbooks,touvron2023llama2}};
            \node[font=\tiny] at (29.00/\pubwid,7.9/\pubhit) {\cite{bai2023qwen}};
              \node[font=\tiny] at (31.6/\pubwid,7.9/\pubhit) {\cite{hughesPhi2SurprisingPower2023,liu2024mobilellm,team2024gemma,dubey2024llama3herd}};
              \node[font=\tiny] at (31.7/\pubwid,9.14/\pubhit){\cite{abdin2024phi3,mehta2024openelm,yang2024qwen2,GoogleGemma2}};

            \node[font=\tiny] at (10.74/\pubwid,9.77/\pubhit) {\cite{onnxruntime}};
              \node[font=\tiny] at (11.87/\pubwid,9.77/\pubhit) {\cite{goyal2020powerBERT}};
            \node[font=\tiny] at (14.41/\pubwid,9.77/\pubhit) {\cite{intelneuralcompressor}};
              \node[font=\tiny] at (15.19/\pubwid,9.77/\pubhit) {\cite{kim2021lengthAdaptiveTransformer}};
            \node[font=\tiny] at (18.63/\pubwid,9.77/\pubhit)          {\cite{kim2022learnedTokenPruning}};
              \node[font=\tiny] at (19.26/\pubwid,10.97/\pubhit) {\cite{zhuLeeBERTLearnedEarly2021,niuDNNFusionAcceleratingDeep2021}};

            \node[font=\tiny] at (21.66/\pubwid,9.77/\pubhit) {\cite{liuFastBERTSelfdistillingBERT2020,zhouBERTLosesPatience2020}};
            \node[font=\tiny] at (21.66/\pubwid,10.97/\pubhit) {\cite{guoSTITurbochargeNLP2023,kongAcceleratingInferencePretrained2022}};

            \node[font=\tiny] at (27.0/\pubwid,9.77/\pubhit) {\cite{llamacpp,shengFlexGenHighThroughputGenerative2023,mlc-llm,wang2023tabi,chevalier2023adapting,executorch,kwonEfficientMemoryManagement2023,tensorrtllm}};

            \node[font=\tiny] at (26.8/\pubwid,10.97/\pubhit) {\cite{jiang2023llmlingua,bae2023fastFREE,sun2024spectr,zhao2024lingualinked,mlx2023,zeng2024consistentee,borzunov2023distributed,alizadehLLMFlashEfficient2024,song2024powerinfer}};
            
            \node[font=\tiny] at (32.35/\pubwid,9.77/\pubhit) {\cite{laskaridisMELTingPointMobile2024,niuSmartMemLayoutTransformation2024}};
            \node[font=\tiny] at (32.35/\pubwid,10.97/\pubhit) {\cite{lee2024autonomous,he2024largelanguagemodelsinference,hu2024edge}};

            \node[font=\tiny] at (34.52/\pubwid,9.77/\pubhit) {\cite{guo2024easter,zhang2024beyondtheCloud}};
            \node[font=\tiny] at (34.88/\pubwid,10.97/\pubhit) {\cite{xu2024EdgeLLMSpeculative}};

            \node[font=\tiny] at (13.21/\pubwid,11.68/\pubhit) {\cite{zadeh2020gobo}};
            \node[font=\tiny] at (14.18/\pubwid,12.77/\pubhit) {\cite{tambe2020AdaptivFloat}};

            \node[font=\tiny] at (15.93/\pubwid,11.68/\pubhit) {\cite{tambeEdgeBERTSentenceLevelEnergy2021}};
            \node[font=\tiny] at (15.93/\pubwid,12.77/\pubhit) {\cite{wangSpAttenEfficientSparse2021}};

            \node[font=\tiny] at (19.97/\pubwid,11.68/\pubhit) {\cite{lu2021sanger}};

            \node[font=\tiny] at (21.56/\pubwid,11.68/\pubhit) {\cite{tu202228nmTranCIM}};
            
            \node[font=\tiny] at (22.1/\pubwid,12.77/\pubhit) {\cite{zadeh2022mokey,zhou2022transpim}};
            \node[font=\tiny] at (24.08/\pubwid,11.68/\pubhit) {\cite{guo2022ant}};
         \node[font=\tiny] at (27.41/\pubwid,11.68/\pubhit) {\cite{tambe2212nm182023,tu202316MulTCIM,tuliAccelTranSparsityAwareAccelerator2023,sridharan2023xformer}};

         \node[font=\tiny] at (27.41/\pubwid,12.77/\pubhit) {\cite{guoOliVeAcceleratingLarge2023,fanTaskFusionEfficientTransfer2023,yuan2024mobilefirmware}};
            
         \node[font=\tiny] at (30.98/\pubwid,11.68/\pubhit) {\cite{kim20CTransformer6182024}};

       \node[font=\tiny] at (9.1/\pubwid,13.58/\pubhit) {\cite{raspberrypi4b}};
         \node[font=\tiny] at (24.43/\pubwid,13.58/\pubhit) {\cite{inteli913900news}};
         \node[font=\tiny] at (26.23/\pubwid,13.58/\pubhit) {\cite{nvidiaGeForceGraphicsCards}};
      \node[font=\tiny] at (26.71/\pubwid,14.66/\pubhit) {\cite{nvidiaLaptopCompare}};
       \node[font=\tiny] at (27.66/\pubwid,13.58/\pubhit) {\cite{JetsonModulesSupport}};
         \node[font=\tiny] at (28.91/\pubwid,13.58/\pubhit) {\cite{applemseries}};
         \node[font=\tiny] at (29.44/\pubwid,14.66/\pubhit) {\cite{appleaseries,QualcommSnapdragon8Gen3MobilePlatform}};
         \node[font=\tiny] at (30.55/\pubwid,13.58/\pubhit) {\cite{samsungExynos2400news}};
         \node[font=\tiny] at (33.39/\pubwid,13.58/\pubhit) {\cite{googleTensorG4}};

         \node[font=\tiny] at (21.31/\pubwid,15.35/\pubhit) {\cite{rawassizadehODSearchFastResource2022}};
         \node[font=\tiny] at (23.43/\pubwid,15.35/\pubhit) {\cite{huang2023innermonologue}};
         \node[font=\tiny] at (27.33/\pubwid,15.35/\pubhit) {\cite{xiang2024languageEmbodied}};
         \node[font=\tiny] at (28.91/\pubwid,15.35/\pubhit) {\cite{hong20233dllm,xu2024mentallm}};
         \node[font=\tiny] at (28.31/\pubwid,16.58/\pubhit) {\cite{wenAutoDroidLLMpoweredTask2024,labrak2024biomistral}};
         \node[font=\tiny] at (31.06/\pubwid,15.35/\pubhit) {\cite{ma2024comprehensive,yang2024mentallama}};
         \node[font=\tiny] at (31.68/\pubwid,16.58/\pubhit) {\cite{zhang2024llamatouch}};
         \node[font=\tiny] at (33.15/\pubwid,15.98/\pubhit) {\cite{lu2024multimodal}};
         
         \node[font=\tiny] at (23.39/\pubwid,17.27/\pubhit) {\cite{caoMobiVQAEfficientOnDevice2022}};

         \node[font=\tiny] at (27.57/\pubwid,17.27/\pubhit) {\cite{li2024sheetcopilot,kim2024languageRCI}};
         
         \node[font=\tiny] at (28.56/\pubwid,18.4/\pubhit) {\cite{gurrealWebAgent,chan2024chateval,zhu2024textrewriting}};

         \node[font=\tiny] at (29.99/\pubwid,17.27/\pubhit) {\cite{GeminiNanoWeb}};

         \node[font=\tiny] at (33.22/\pubwid,17.27/\pubhit) {\cite{tian2024dialoguesummarization}};

         \node[font=\tiny] at (26.74/\pubwid,19.16/\pubhit) {\cite{jeong2023winclip}};
         \node[font=\tiny] at (28.65/\pubwid,19.16/\pubhit) {\cite{kang2024quantitativeAutoFL,gu2024anomalygpt}};

         \node[font=\tiny] at (29.23/\pubwid,20.24/\pubhit) {\cite{chen2024drivingWithLLMs,yang2024largeLLMAO}};

         \node[font=\tiny] at (30.80/\pubwid,19.16/\pubhit) {\cite{pan2024vlp,tian2024drivevlm}};

          \node[font=\tiny] at (31.86/\pubwid,20.24/\pubhit) {\cite{zhu2024llmsALFA,hossain2024deepToggle}};
         
        \end{scope}
    \end{tikzpicture}
    \caption{Temporal Distribution of On-Device LLM Research.}
    \label{fig: publication_timeline}
\end{figure}

Fig.~\ref{fig: publication_timeline} shows the evolution of on-device LLM research from 2019 to 2024, categorized into pre-deployment techniques (blue), runtime optimizations (purple), and on-device applications (green). Offline pre-deployment techniques, such as quantization, pruning, knowledge distillation, and low-rank approximation have been consistently researched throughout the period. 
Online runtime optimizations, including software-level optimizations, hardware-software co-design, and hardware-level optimizations, have gained traction from 2022. The emergence of on-device applications for personal, enterprise, and industrial use cases is particularly notable in the latter half of the timeline, indicating a growing trend towards edge AI and mobile LLM deployments.
This highlights the rapid advancement and diversification of LLM deployment for resource-constrained environments, emphasizing the growing importance of efficient on-device AI. Our survey provides a comprehensive analysis of these trends, offering a foundation for future research and practice in this field.
\section{\nameoffline} \label{Sec: \nameoffline}
The proliferation of LLMs has engendered a burgeoning demand for their deployment on mobile and edge devices, driven by the imperative for enhanced privacy, reduced latency, and improved service availability in connectivity-constrained environments. This paradigm shift towards edge computing for LLMs. However, it presents significant challenges due to the LLMs' inherent computational complexity and substantial memory requirements~\cite{ji2021distribution}. Consequently, offline pre-deployment model design techniques have emerged as a pivotal strategy, aiming to substantially reduce the computational and memory footprint of LLMs while preserving their performance integrity. These techniques, applied prior to the models' deployment on target edge devices, facilitate efficient execution in resource-constrained environments.
\def\offoverwid{28.000}
\def\offoverhit{19.399}
\begin{figure}[ht]
    \centering
    \begin{tikzpicture}
        \node[anchor=north west, inner sep=0] (image) at (0,0) {\includegraphics[width=0.88\textwidth]{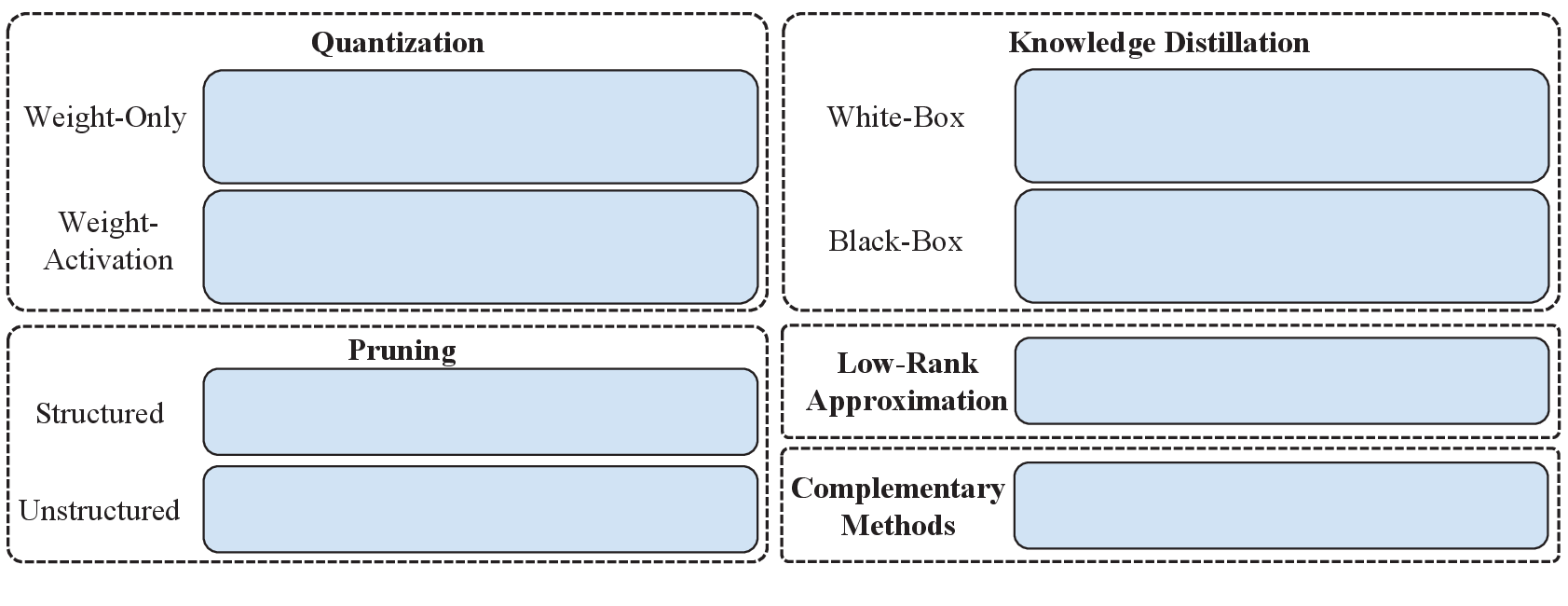}}; 
        \begin{scope}[x={(image.north east)}, y={(image.south west)}] 
            \node[font=\tiny] at (8.66/\offoverwid,4.10/\offoverhit) {            \makecell{AWQ~\cite{linAWQActivationawareWeight2024}, LLM.int8()~\cite{dettmersLLMInt88bit}, GPTQ~\cite{frantarOPTQAccurateQuantization2022},\\AQLM~\cite{egiazarianextreme}, QuIP\#~\cite{tseng2024quipbetter}, SpQR~\cite{dettmersSpQRSparseQuantizedRepresentation2023}, \\GPUSQ-TLM~\cite{yuBoostTransformerbasedLanguage2023}, OWQ~\cite{leeOWQOutlierAwareWeight2024}}}; 

            \node[font=\tiny] at (8.66/\offoverwid,7.98/\offoverhit) {            \makecell{ZeroQuant~\cite{yaoZeroQuantEfficientAffordable2022}, SmoothQuant~\cite{xiaoSmoothQuantAccurateEfficient2023}, \\ OmniQuant~\cite{shaoOmniQuantOmnidirectionallyCalibrated2023}, Agile-Quant~\cite{shenAgileQuantActivationGuidedQuantization2024},  \\QLLM~\cite{liuQLLMAccurateEfficient2023}, Q-Hitter~\cite{zhang2024qhitter}, QBERT~\cite{shenQBERTHessianBased2020a}}}; 

            \node[font=\tiny] at (8.66/\offoverwid,13.3/\offoverhit) {            \makecell{CoFi~\cite{xiaStructuredPruningLearns2022}, LLM-Pruner~\cite{maLLMPrunerStructuralPruning2023}, FLAP~\cite{an2023fluctuationbasedadaptivestructuredpruning}, \\Sheared LLaMA~\cite{xiaShearedLLaMAAccelerating2023}, SliceGPT~\cite{ashkboos2024slicegptcompresslargelanguage}}}; 

            \node[font=\tiny] at (8.66/\offoverwid,16.3/\offoverhit) {            \makecell{Movement Pruning~\cite{sanhMovementPruningAdaptive2020}, oBERT~\cite{kurticOptimalBERTSurgeon2022}, \\SparseGPT~\cite{frantarSparseGPTMassiveLanguage2023}, Plug-and-Play~\cite{zhang2024plugandplay}}}; 

            \node[font=\tiny] at (22.90/\offoverwid,4.10/\offoverhit) {            \makecell{MiniLM~\cite{wangMiniLMDeepSelfAttention2020,wangMiniLMv2MultiHeadSelfAttention2021}, KPTD~\cite{padmanabhanPropagatingKnowledgeUpdates}, \\FUSELLM~\cite{wan2024knowledgefusion}, MobileBERT~\cite{sunMobileBERTCompactTaskAgnostic2020}, \\TED~\cite{liangLessMoreTaskaware2023}, TSLD~\cite{kimTokenScaledLogitDistillation},  MiniMoE~\cite{zhangLiftingCurseCapacity2023}}}; 

            \node[font=\tiny] at (22.90/\offoverwid,7.98/\offoverhit) {            \makecell{Distilling Step-by-Step~\cite{hsiehDistillingStepbyStepOutperforming2023}, Lion~\cite{jiangLionAdversarialDistillation2023},\\Fine-tune-CoT~\cite{hoLargeLanguageModels2023}, LaMini-LM~\cite{wuLaMiniLMDiverseHerd2024}, \\DISCO~\cite{chenDISCODistillingCounterfactuals2023}, SCoTD~\cite{liSymbolicChainofThoughtDistillation2023}, MCC-KD~\cite{chenMCCKDMultiCoTConsistent2023}}}; 

            \node[font=\tiny] at (22.90/\offoverwid,12.32/\offoverhit) {            \makecell{ALBERT~\cite{lanALBERTLiteBERT2019}, FWSVD~\cite{hsuLanguageModelCompression2021}, Boalco~\cite{jiFeaturebasedLowRankCompression2024}, \\SFSD~\cite{chavanSurgicalFeatureSpaceDecomposition2024}, DRONE~\cite{chenDRONEDataawareLowrank2021}, LoSparse~\cite{liLoSparseStructuredCompression2023}}}; 

            \node[font=\tiny] at (22.90/\offoverwid,16.26/\offoverhit) {            \makecell{Data Preprocessing~\cite{li2023textbooks}, RoPE~\cite{su2024roformerrotarypositionembedding},\\GQA/MQA~\cite{ainslie2023gqa}, Layer-wise Scaling~\cite{mehta2020delight}}};

        \end{scope}
    \end{tikzpicture}
    \caption{\modbluetext{An Overview of Offline Pre-deployment Model Design Techniques and Literature.}}
    \label{fig: fig_sec3_offline_overview}
\end{figure}

 \modbluetext{As illustrated in Fig.~\ref{fig: fig_sec3_offline_overview}, these pre-deployment techniques encompass five primary categories: quantization, pruning, knowledge distillation, low-rank approximation, and complementary methods.}

\subsection{Quantization} \label{SubSec: Quantization}
Quantization is a compression technique that reduces the precision of numerical values in models, providing significant deployment advantages for edge devices. However, applying conventional methods to LLMs is challenging due to the architectural complexity of Transformer-based models, which rely heavily on attention mechanisms and high-dimensional representations~\cite{ji2021distribution}. These characteristics result in precision-sensitive tasks, and the high dynamic activation ranges in these models exacerbate quantization difficulties, often leading to performance degradation~\cite{bondarenko2021understanding}. To address these challenges, specialized quantization methods have been developed, typically focused on two main areas:
\begin{itemize}
\item Weight quantization: Reduces the precision of model weights.
\item Activation quantization: Reduces the precision of intermediate activations.
\end{itemize}
As shown in Fig.~\ref{fig: model_compression_figure} (a), it can be categorized into weight-only quantization and weight-activation co-quantization.

\newcommand{\heighta}{24.3mm}
\newcommand{\heightb}{28mm}
\begin{figure}[ht]
    \centering
    \begin{minipage}{\textwidth}
        \centering
        \includegraphics[width=0.8\textwidth]{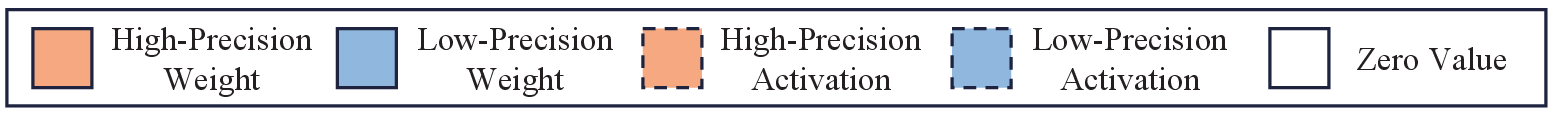}
    \end{minipage}
    \begin{minipage}{\textwidth}
        \centering
  \subfloat[Quantization]{\includegraphics[height=\heighta]{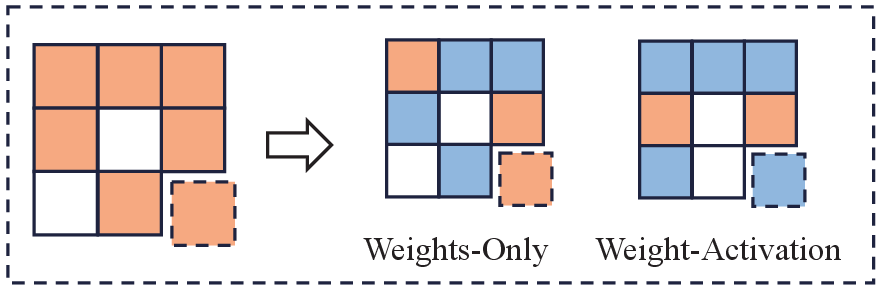}\vspace{-2mm}}
  \subfloat[Pruning]{\includegraphics[height=\heighta]{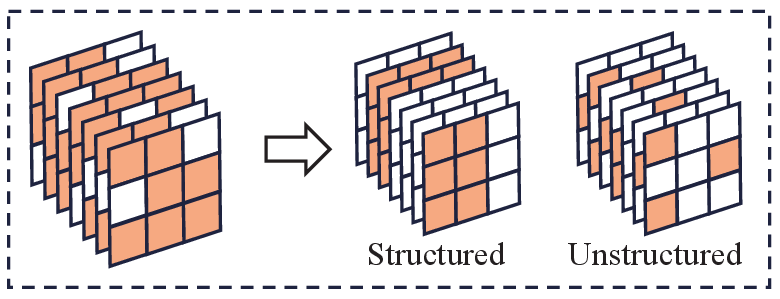}\vspace{-2mm}}
  \end{minipage}
  \newline
      \begin{minipage}{\textwidth}
        \centering
  \subfloat[Knowledge Distillation]{\includegraphics[height=\heightb]{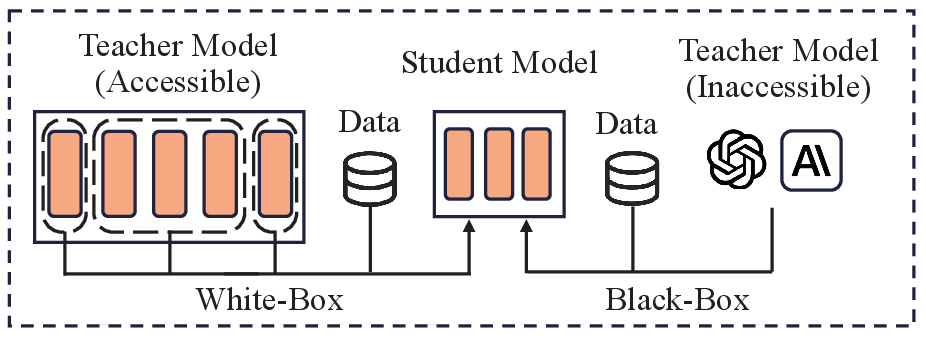}\vspace{-2mm}}	
  \subfloat[Low-Rank Approximation]{\includegraphics[height=\heightb]{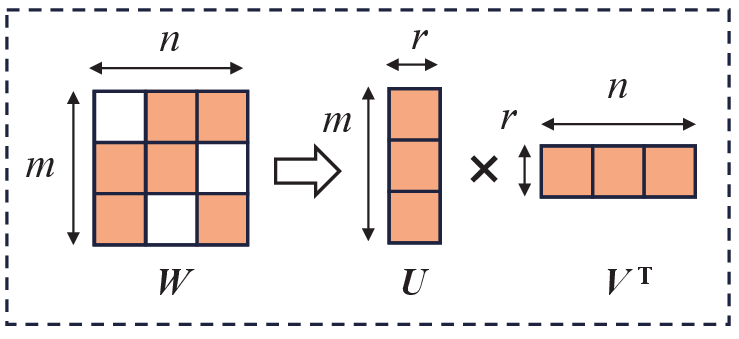}\vspace{-2mm}}	
  \end{minipage}
\caption{Comparative Schematics of Offline Compression Methods for LLMs.}
\label{fig: model_compression_figure}
\end{figure}

~\subsubsection{Weight-Only Quantization} \label{SubSubSec: Weight-Only Quantization}
Weight-only quantization reduces the precision of model weights from high-precision data types (e.g., 32-bit floating point) to lower-precision ones (e.g., 8-bit integers). This reduces memory usage and can speed up inference on resource-constrained devices. For instance, ~\citet{dettmersLLMInt88bit} propose LLM.int8(), which reduces memory requirements during inference without sacrificing performance. Similarly, ~\citet{frantarOPTQAccurateQuantization2022} introduce GPTQ, a post-training method that compresses LLM weights to lower bits, addressing layer-wise quantization challenges with the OBQ method~\cite{frantarOptimalBrainCompression2022}. Moreover, ~\citet{linAWQActivationawareWeight2024} propose AWQ, which preserves salient weights at high precision while quantizing others, optimizing computational complexity and energy consumption. Other weight-only quantization methods, such as AQLM~\cite{egiazarianextreme}, QuIP\#~\cite{tseng2024quipbetter}, and GPUSQ-TLM~\cite{yuBoostTransformerbasedLanguage2023} also contribute to the advancements in this field by exploring different quantization strategies and optimization techniques.

Despite these advances, managing quantization errors in outlier weights remains a challenge. Techniques such as SpQR~\cite{dettmersSpQRSparseQuantizedRepresentation2023} and OWQ~\cite{leeOWQOutlierAwareWeight2024} address this issue by storing outlier weights at higher precision, improving compression efficiency without compromising performance.

~\subsubsection{Weight-Activation Co-Quantization} \label{SubSubSec: Weight-Activation Co-Quantization}
While weight-only quantization offers benefits, it may leave activation values uncompressed. Weight-activation co-quantization, which quantifies both weights and activations, provides further compression. For example, ZeroQuant~\cite{yaoZeroQuantEfficientAffordable2022} combines group-wise weight and token-wise activation quantization. SmoothQuant~\cite{xiaoSmoothQuantAccurateEfficient2023} offering lossless 8-bit quantization by using per-channel scaling transformations. 
Furthermore, Agile-Quant~\cite{shenAgileQuantActivationGuidedQuantization2024} and Q-Hitter~\cite{zhang2024qhitter} employ activation-aware techniques to balance performance and real-time inference speed. In addition to the advancements discussed, other established technologies, including QBERT~\cite{shenQBERTHessianBased2020a} and TernaryBERT~\cite{zhang2020ternarybert} have also played a role in the development of weight-activation co-quantization for edge environments.

Despite these advancements, handling outlier issues in co-quantization remains a challenge. Methods like QLLM~\cite{liuQLLMAccurateEfficient2023} and OmniQuant~\cite{shaoOmniQuantOmnidirectionallyCalibrated2023} address activation and weight outliers by leveraging adaptive calibration and learnable transformations, respectively.

In summary, quantization techniques for LLMs strike a balance between model compression, performance, and computational complexity. Weight-only methods, such as LLM.int8()~\cite{dettmersLLMInt88bit}, are ideal for quick deployment with moderate compression, while weight-activation co-quantization approaches like ZeroQuant~\cite{yaoZeroQuantEfficientAffordable2022} offer higher compression at the cost of increased complexity and potential accuracy loss.

\subsection{Pruning} \label{SubSec: Pruning}
Pruning is a key technique for optimizing large language models (LLMs) by reducing the number of parameters, leading to smaller model sizes and faster inference. However, pruning in LLMs is challenging due to the complexity of their architecture and the varying significance of components such as attention heads. Conventional pruning methods, effective in CNNs, face limitations when applied to LLMs~\cite{li2020efficientTransformerPruning}. As shown in Fig.~\ref{fig: model_compression_figure} (b), specialized pruning techniques for LLMs are generally categorized into structured and unstructured pruning, each with distinct trade-offs. 

\subsubsection{Structured Pruning} \label{SubSubSec: Structured Pruning}
Structured pruning reduces the size of neural networks by removing entire structural components, such as neurons, channels, or layers. For example, CoFi~\cite{xiaStructuredPruningLearns2022} uses multiple pruning masks at different granularities to simultaneously remove layers and attention heads. LLM-Pruner~\cite{maLLMPrunerStructuralPruning2023} employs gradient-based pruning, removing non-critical units while preserving model performance, and incorporates LoRA~\cite{huLoRALowRankAdaptation2021} to recover performance post-pruning. LoRAPrune~\cite{zhang2024loraprunestructuredpruningmeets} leverages LoRA’s weights and gradients for importance estimation. Sheared LLaMA~\cite{xiaShearedLLaMAAccelerating2023} prunes specific layers and dimensions, using dynamic batch loading for domain-specific loss metrics, ideal for resource-constrained edge devices. SliceGPT~\cite{ashkboos2024slicegptcompresslargelanguage} projects transformer block signal matrices onto principal components, removing redundant columns or rows to reduce size. FLAP~\cite{an2023fluctuationbasedadaptivestructuredpruning} formulates importance metrics, enabling adaptive search for the optimal compressed model and implementing compensation mechanisms to mitigate performance loss.

\subsubsection{Unstructured Pruning}  \label{SubSubSec: Unstructured Pruning}
Unstructured pruning removes individual weights or neurons, resulting in sparse models that are harder to optimize. 
Movement Pruning~\cite{sanhMovementPruningAdaptive2020} adapts pruning decisions based on weight dynamics during fine-tuning, preserving important weights that exhibit significant movement. oBERT~\cite{kurticOptimalBERTSurgeon2022} introduces a second-order pruning method supporting both unstructured and block pruning.  SparseGPT~\cite{frantarSparseGPTMassiveLanguage2023} treats pruning as a sparse regression problem, allowing for one-shot pruning without retraining. Plug-and-Play~\cite{zhang2024plugandplay} integrates activation-based importance to prune weights selectively, further improving pruning robustness in large-scale models. Wanda~\cite{sun2024simpleeffectivepruningapproach} prunes weights with the smallest magnitudes multiplied by the corresponding input activations, on a per-output basis. BESA~\cite{xu2024besapruning} targets the overall pruning error with respect to individual transformer blocks and allocates layer-specific sparsity in a differentiable manner. 

Overall, pruning techniques for LLMs offer various strategies for balancing size reduction and performance retention. Structured pruning methods like CoFi~\cite{xiaStructuredPruningLearns2022} and LLM-Pruner~\cite{maLLMPrunerStructuralPruning2023} provide controlled reductions, preserving architecture integrity, while unstructured methods such as Movement Pruning~\cite{sanhMovementPruningAdaptive2020} and oBERT~\cite{kurticOptimalBERTSurgeon2022} offer greater flexibility but may result in irregular, sparse models. In practice, the choice of method depends on the specific deployment scenario, considering the trade-offs between model size, computational efficiency, and task performance.

\subsection{Knowledge Distillation} \label{SubSec: Knowledge Distillation}
Knowledge distillation transfers knowledge from complex teacher models to simpler student models to create computationally efficient alternatives without sacrificing performance. This process reduces model size, computational costs, and deployment requirements, while enhancing the student model's diversity and stability. However, distilling knowledge from large teacher models, such as LLMs, remains challenging due to the difficulty in transferring internal representations and the complexity of attention mechanisms in Transformers~\cite{ji2021distribution}. As illustrated in Fig.~\ref{fig: model_compression_figure} (c), distillation methods for LLMs are categorized into white-box and black-box approaches, each tailored to handle the scale and intricacies of LLMs.

\subsubsection{White-Box Knowledge Distillation}  \label{SubSubSec: White-Box Knowledge Distillation}
White-box knowledge distillation leverages access to the teacher model's architecture and parameters, using internal features and logits for knowledge transfer. MiniLM~\cite{wangMiniLMDeepSelfAttention2020} extracts knowledge from the final Transformer~\cite{vaswani2017attention} layer, alleviating the complexity of layer-to-layer mapping, while MiniLMv2~\cite{wangMiniLMv2MultiHeadSelfAttention2021} extends this for task-agnostic compression. Other significant contributions include MobileBERT~\cite{sunMobileBERTCompactTaskAgnostic2020}, and TinyBERT~\cite{jiaoTinyBERTDistillingBERT2020} collectively laying the foundation in this area.

To mitigate capacity mismatches between teacher and student models, techniques such as reverse Kullback-Leibler Divergence in MiniLLM~\cite{guMiniLLMKnowledgeDistillation2023} and layer-wise alignment in TED~\cite{liangLessMoreTaskaware2023} improve the effectiveness of distillation. KPTD~\cite{padmanabhanPropagatingKnowledgeUpdates} and TSLD~\cite{kimTokenScaledLogitDistillation} further refine the process through entity-based transfer and token-scaled logit distillation, respectively. MiniMoE~\cite{zhangLiftingCurseCapacity2023} addresses the capacity gap by employing a Mixture-of-Experts (MoE) model, advancing the scalability of white-box distillation for LLMs. FUSELLM~\cite{wan2024knowledgefusion} integrates the capabilities of existing LLMs and transfers them into a single one, thereby elevating the capabilities of the target model. These advancements contribute to the evolving landscape of white-box knowledge distillation.

\subsubsection{Black-Box Knowledge Distillation}   \label{SubSubSec: Black-Box Knowledge Distillation}
Black-box knowledge distillation focuses solely on the outputs of the teacher model, bypassing internal model details. This approach is valuable when teacher models are proprietary or deployed via APIs. Chain-of-Thought (CoT) distillation aims to transfer reasoning abilities from LLMs to smaller student models~\cite{hsiehDistillingStepbyStepOutperforming2023,liSymbolicChainofThoughtDistillation2023}. For example, Distilling Step-by-Step~\cite{hsiehDistillingStepbyStepOutperforming2023} uses LLM rationales to supervise task-specific models, improving dataset quality and model performance. Fine-tune-CoT~\cite{hoLargeLanguageModels2023} successfully transfers reasoning from large models (over 100B parameters) to students with as few as 0.3B. SCoTD~\cite{liSymbolicChainofThoughtDistillation2023} further enhances performance in both supervised and few-shot tasks. Other methods for CoT distillation include Socratic CoT~\cite{shridharDistillingReasoningCapabilities2023} and MCC-KD~\cite{chenMCCKDMultiCoTConsistent2023}.

Black-box distillation also includes methods that focus on instruction-following, a crucial ability for LLMs in real-world applications. Lion~\cite{jiangLionAdversarialDistillation2023} employs adversarial distillation to generate complex instructions, producing a 13B-parameter model comparable to ChatGPT~\cite{IntroducingChatGPTOpenAI}. DISCO~\cite{chenDISCODistillingCounterfactuals2023} distills counterfactual data to enhance robustness, while LaMini-LM~\cite{wuLaMiniLMDiverseHerd2024} uses diverse instruction sets to compress large models effectively.

In conclusion, white-box methods, such as MiniLM~\cite{wangMiniLMDeepSelfAttention2020}, excel in scenarios where model internals are accessible. Conversely, black-box methods, such as Lion~\cite{jiangLionAdversarialDistillation2023}, are valuable in industrial or proprietary contexts where access to model internals is not accessible.

\subsection{Low-Rank Approximation} \label{SubSec: Low-Rank Approximation}
Matrix factorization techniques, such as principal component analysis and regularized matrix factorization, have been pivotal in improving the generalization and interoperability of CNNs and RNNs. These methods reduce high-dimensional data to lower-dimensional spaces, enhancing model performance~\cite{ji2021distribution}. However, the large-scale parameters of Transformer-based LLMs pose challenges to traditional factorization approaches, as their computational complexity and unique structural elements, such as multi-head attention and feed-forward networks, require specialized adaptations~\cite{gengattention}. To address these, low-rank approximation has emerged as a promising strategy within the Transformer framework~\cite{chenDRONEDataawareLowrank2021,chavanSurgicalFeatureSpaceDecomposition2024}.

As illustrated in Fig.~\ref{fig: model_compression_figure} (d), this technique approximates a high-dimensional matrix \(W_{m\times n}\) with the product of two lower-rank matrices, \(U_{m\times r}\) and \((V^T)_{r\times n}\), where \(r\) is much smaller than \(m\) and \(n\). For example, ALBERT~\cite{lanALBERTLiteBERT2019} applied low-rank approximation to vocabulary embeddings, decoupling hidden layer size from vocabulary size. FWSVD~\cite{hsuLanguageModelCompression2021} enhances singular value decomposition by incorporating Fisher information to weight parameter importance, while DRONE~\cite{chenDRONEDataawareLowrank2021} optimizes weight matrix compression by leveraging data distribution. Innovations such as LoSparse~\cite{liLoSparseStructuredCompression2023} introduce a method for separating coherent and incoherent neuron components, outperforming traditional pruning approaches. Additionally, inherent low-rank characteristics in LLMs have led to Bayesian optimization-based feature compression techniques~\cite{jiFeaturebasedLowRankCompression2024}. SFSD~\cite{chavanSurgicalFeatureSpaceDecomposition2024} refine feature space approximation, and fast randomized algorithms using Gaussian sketches facilitate efficient low-rank factorization on consumer-grade hardware~\cite{sahaMatrixCompressionRandomized2023}. 

In summary, low-rank approximation effectively reduces parameters in LLMs, making it ideal for optimizing large-scale matrices like embedding layers and attention weights~\cite{yan2024federa}. By minimizing redundancy, it significantly reduces storage and computation costs while maintaining performance.

\subsection{Complementary Methods}  \label{SubSec: Complementary Methods}
Recent research in the field of LLMs has witnessed a paradigm shift towards developing innovative approaches specifically tailored for compact models, typically with around 10B parameters or less. As illustrated in Fig.~\ref{fig: other_methods_figure}, these technological advancements including:
\begin{itemize}
\item \modbluetext{Data Preprocessing}: Meticulous curation of high-quality training data~\cite{gunasekar2023textbooks}.
\item Grouped query and Multi-query Attention: Optimization of attention mechanisms~\cite{ainslie2023gqa}.
\item Rotary Position Embedding (RoPE): Advanced positional information encoding~\cite{su2024roformerrotarypositionembedding}.
\item Layer-wise Scaling: Strategic distribution of parameters across model layers~\cite{mehta2020delight}.
\end{itemize}

\newcommand{\widtha}{0.37\textwidth}
\newcommand{\widthb}{0.368\textwidth}
\begin{figure}[ht]
    \centering
    \raisebox{4mm}{
     \subfloat[Data Preprocessing]{\includegraphics[width=\widtha]{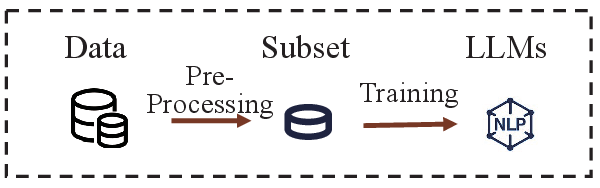}\vspace{-2mm}}}
  \raisebox{-3.6mm}{
    \subfloat[Grouped and Multi-query Attention]{\includegraphics[width=\widtha]{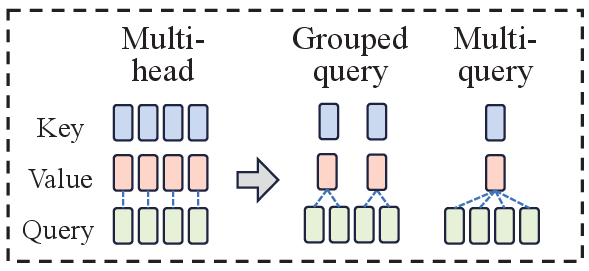}\vspace{-2mm}}}
\vspace{-8.5mm} 
\\
  \raisebox{-0.4mm}{
    \subfloat[RoPE]{\includegraphics[width=\widtha]{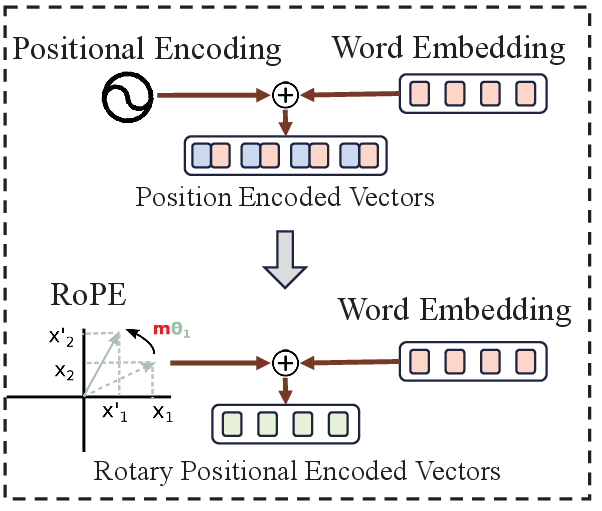}\vspace{-2mm}}}
\raisebox{-1mm}{
    \subfloat[Layer-wise Scaling]{\includegraphics[width=\widthb]{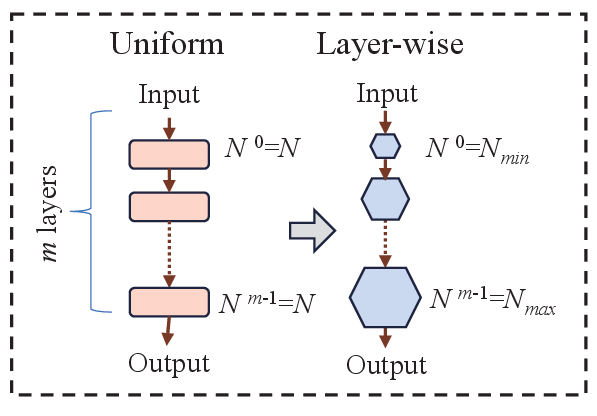}\vspace{-2mm}}}
\caption{Illustrations of complementary methods for designing efficient LLMs.}
\label{fig: other_methods_figure}
\end{figure}

The integration of these techniques has resulted in compact LLMs to excel on resource-constrained edge devices, supporting high-performance AI applications in diverse scenarios. For instance, Meta’s LLaMA series~\cite{touvron2023llama,touvron2023llama2,dubey2024llama3herd, IntroducingLlama3.2} have achieved remarkable performance gains through efficient pre-training on carefully curated datasets and leverages grouped query attention and RoPE to improve inference speed, reduce cache size, and expand context lengths. Similarly, Microsoft’s Phi series~\cite{gunasekar2023textbooks,hughesPhi2SurprisingPower2023,abdin2024phi3} showcases the impact of data preprocessing, using high-quality synthetic and filtered datasets to outperform larger models across benchmarks. Google’s Gemma models~\cite{team2024gemma, GoogleGemma2} combine multi-query attention and RoPE to balance efficiency and accuracy, while Apple’s OpenELM family~\cite{mehta2024openelm} employs layer-wise scaling and grouped
query attention to optimize performance on low-powered devices. Other notable compact LLMs, such as Pythia~\cite{biderman2023pythia}, OPT~\cite{zhang2022opt}, Qwen~\cite{bai2023qwen,yang2024qwen2}, and MobileLLM~\cite{liu2024mobilellm} also benefit from similar optimizations, thereby reinforcing the growing potential of small-scale models for edge deployment.

While these methods significantly reduce parameter counts and resource demands, their development and optimization are predominantly executed by large organizations with substantial computational resources and domain expertise~\cite{mehta2024openelm}. However, individual researchers can still benefit by using pre-trained compact LLMs from Hugging Face\footnote{Hugging Face: \url{https://huggingface.co/}}. These models, often with optimized weights and configurations, can be deployed on edge devices for efficient inference, lowering the entry barrier and enabling broader access to advanced AI capabilities in resource-limited settings~\cite{touvron2023llama,biderman2023pythia}.

\subsection{Comparative Analysis and Selection}
The efficacy of offline pre-deployment techniques for LLMs depends on model architecture, deployment constraints, and performance requirements. The subsequent comparison provides an analysis to guide practitioners in selecting the most appropriate technique for their use case.

The selection of appropriate pre-deployment techniques should be informed by a comprehensive analysis of the deployment environment, hardware constraints, performance requirements, and LLM characteristics. For example, scenarios with strict memory constraints but moderate computational resources may benefit from weight-only quantization (e.g., LLM.int8()~\cite{dettmersLLMInt88bit}) and structured pruning (e.g., CoFi~\cite{xiaStructuredPruningLearns2022}). Conversely, applications requiring maximum compression with some tolerance for accuracy loss might employ weight-activation co-quantization (e.g., SmoothQuant~\cite{xiaoSmoothQuantAccurateEfficient2023}) and aggressive unstructured pruning (e.g., Movement Pruning~\cite{sanhMovementPruningAdaptive2020}). Knowledge 
distillation techniques such as MiniLM~\cite{wangMiniLMDeepSelfAttention2020} or Lion~\cite{jiangLionAdversarialDistillation2023} are suitable for transferring knowledge from larger to smaller models, with the choice between white-box or black-box approaches depending on teacher model accessibility. For models with large, redundant matrices, low-rank approximation methods like ALBERT~\cite{lanALBERTLiteBERT2019} or LoSparse~\cite{liLoSparseStructuredCompression2023} can effectively reduce model size while preserving performance. In practice, combining multiple techniques often yields better results. ZeroQuant~\cite{yaoZeroQuantEfficientAffordable2022} demonstrates the efficacy of integrating quantization with knowledge distillation. NVIDIA's Nemotron-4 4B~\cite{nvidia2024NemotronMini4b} exhibit high performance for on-device inference, which was pruned and distilled from Nemotron-4 15B~\cite{parmar2024nemotron4-15b}. DeepSeek-R1~\cite{guo2025deepseekr1} distills knowledge from compact models like Llama 3~\cite{dubey2024llama3herd}, forming an edge-optimized 1.5B model further refined via large-scale reinforcement learning for mathematics, code generation, and complex reasoning, significantly enhancing inference capabilities.

When feasible, developing inherently efficient architectures such as LLaMA~\cite{touvron2023llama} or Gemma~\cite{team2024gemma} may provide optimal performance within strict resource constraints. These approaches, typically developed by major tech companies with access to high-quality data and ample computing resources, demonstrate that careful architecture design and training strategies can yield highly efficient models without extensive post-training compression.

\section{\nameonline} \label{Sec: \nameonline}
\begin{figure}[ht]
    \centering
    \includegraphics[width=0.64\textwidth]{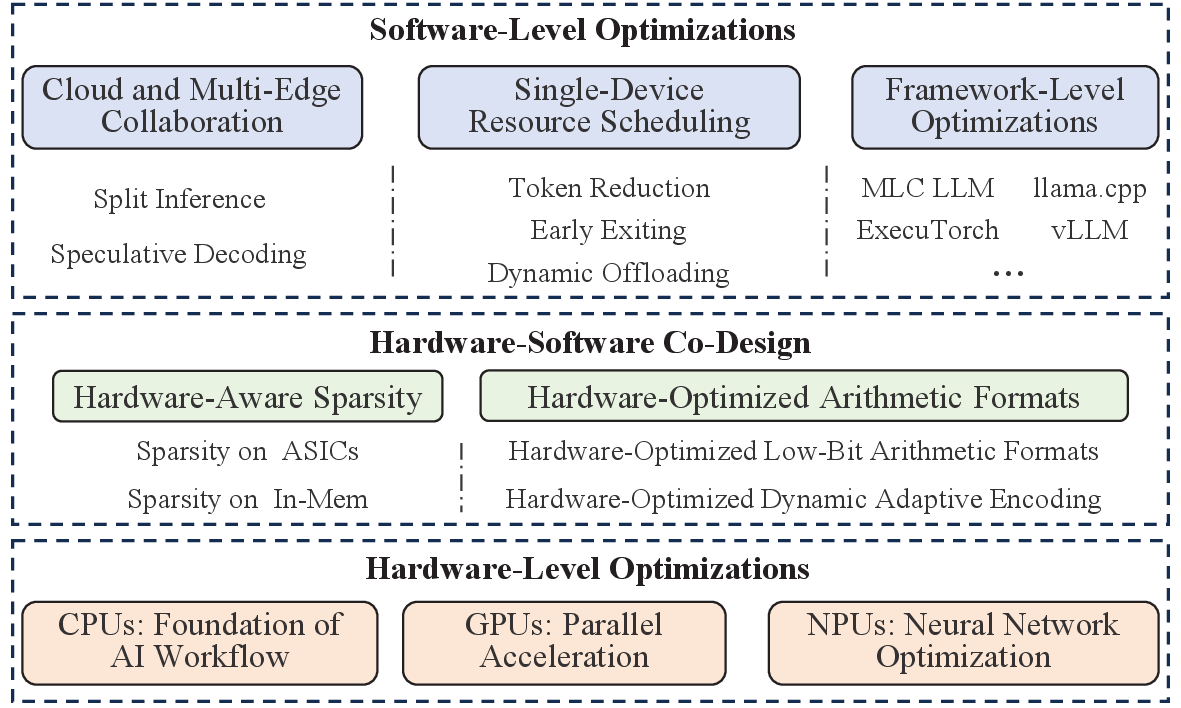}
    \caption{\modbluetext{A top-down view of runtime inference optimizations. (In-Mem: Computing/Processing in Memory.)}}
    \label{fig: runtime_figure}
\end{figure}

\modbluetext{We previously discussed offline optimization techniques in the pre-deployment phase, including model compression techniques and pre-training compact models for edge devices. This section shifts to runtime optimizations for efficient inference of LLMs directly on edge devices. As shown in Fig.~\ref{fig: runtime_figure}, these optimizations are categorized into three areas: first, \textbf{Software-Level Optimizations}, which focuses on algorithmic strategies, resource scheduling, and framework optimizations independent of hardware; second, \textbf{Hardware-Software Co-Design}, which involves co-optimized solutions that leverage specific hardware features by adapting techniques such as sparsity and quantization to be hardware-friendly; and third, \textbf{Hardware-Level Optimizations}, which highlights innovations in hardware designed to improve the performance of LLMs.}

\subsection{Software-Level Optimizations} \label{SubSec: Software-Level Optimizations}
\modbluetext{Software-level strategies can be categorized into Cloud and Multi-Edge Collaboration, Single-Device Resource Scheduling, and Framework-Level Optimizations, all of which focus on optimizing software algorithms, systems, frameworks or engines without modifications tailored to specific hardware. }

\subsubsection{Cloud and Multi-Edge Collaboration} \label{SubSubSec: Cloud and Multi-Edge Collaboration}
\begin{figure}[ht]
    \centering
    \begin{minipage}{\textwidth}
        \centering
            \includegraphics[width=0.76\textwidth]{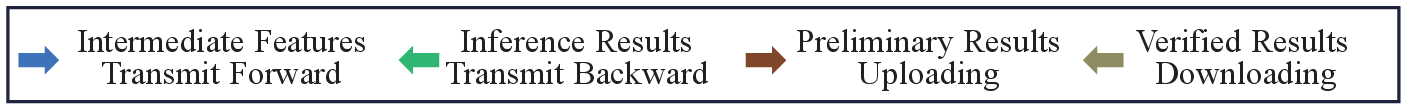}  
    \end{minipage}
    \begin{minipage}{\textwidth}
        \centering
        \vspace{1mm}
  \subfloat[Split Inference]{\includegraphics[width=0.443\textwidth]{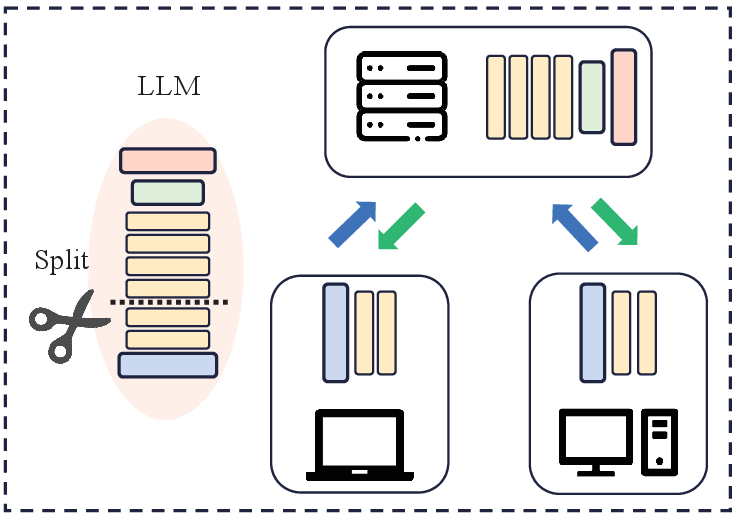}\vspace{0mm}}	
  \hspace{2mm}
  \subfloat[Speculative Decoding]{\includegraphics[width=0.396\textwidth]{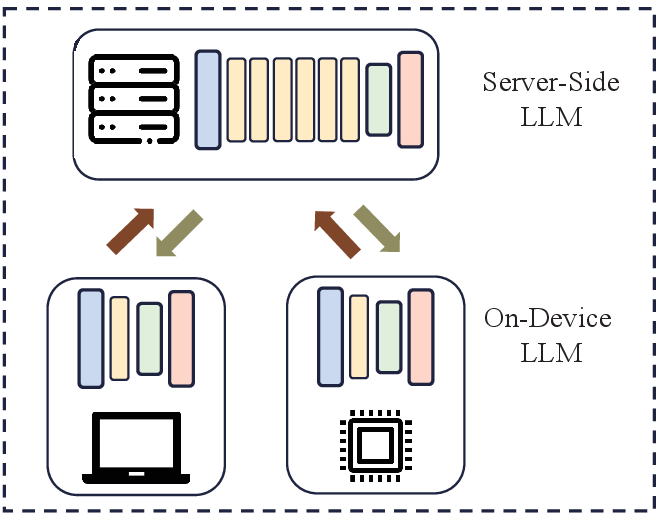}\vspace{0.3mm}}
  \end{minipage}
    \caption{\modbluetext{Illustrations of cloud and multi-edge collaboration for on-device LLMs.}}
    \label{fig: fig_cross_device}
\end{figure}

Collaborative computing across cloud and edge devices plays a critical role in enhancing LLM performance by distributing computational workloads efficiently~\cite{du2024distributed,10734312}. Although it has seen extensive use in tasks like video analytics~\cite{zhang2024vulcan,lu2023multiview,jain2020spatula,lu2022turbo,padmanabhan2023gemel,ga19mobisys,wang19hotedgevideo}, multi-device and cloud-edge collaboration for LLM computation remains in its infancy and generally follows one of two approaches: split inference or speculative decoding, as illustrated in Fig.~\ref{fig: fig_cross_device}.

~

~\modbluetext{~\textbf{Split inference} optimizes resource utilization and accelerates inference by partitioning computation across cloud and edge devices. PETALS~\cite{borzunov2023distributed} aggregates idle computational resources from diverse sources to ensure robust performance under dynamic network conditions. Voltage~\cite{hu2024edge} improves throughput by distributing transformer layers across edge devices, achieving linear speed-ups through parallel computation. EASTER~\cite{guo2024easter} addresses reliability challenges with adaptive partition strategies that maintain performance despite device failures. LinguaLinked~\cite{zhao2024lingualinked} enhances split inference for mobile scenarios by aligning model segments with trusted device capabilities, enabling efficient data exchange. Approaches such as rewardless guidance~\cite{he2024largelanguagemodelsinference} and Hepti~\cite{lee2024autonomous} further optimize offloading decisions and dynamically adapt workloads to network and resource conditions. Building on these advancements, ~\citet{zhang2024beyondtheCloud} propose a tree-search framework to efficiently manage request batching and resource allocation, demonstrating the potential for adaptive and efficient cloud-edge collaboration.}

\modbluetext{\textbf{Speculative decoding}~\cite{leviathan2023fastinferencespeculativedecoding} addresses a key limitation of split inference, where the transmission of intermediate features often incurs significant communication overhead. In contrast, speculative decoding uploads preliminary results from edge devices and downloads only the verified results from a cloud-side LLM. This streamlined communication mechanism offers a practical alternative for scenarios where bandwidth or latency constraints are critical. For example, SpecTr~\cite{sun2024spectr} uses lightweight models on edge devices to propose token drafts, enabling parallel validation by larger cloud models. Tabi~\cite{wang2023tabi} uses calibrated confidence scores to decide whether to upload tokens to the cloud for verification. Built atop speculative decoding, EdgeLLM~\cite{xu2024EdgeLLMSpeculative} features a branch navigation and self-adaptive fallback strategy, enabling fast and accurate token generation. }

\modbluetext{To provide a concise overview of the advancements in cloud and multi-edge collaboration, Table~\ref{tab: tab_collaboration_literature} summarizes the primary literature. The table highlights the key methodologies, objectives, and application scenarios of each work, offering a comparative perspective on the progress in split inference and speculative decoding. By integrating these approaches, future research can address challenges such as latency, reliability, and resource constraints more effectively.}

~\newcommand{\widcollit}{8cm}
\begin{table}[ht]\footnotesize
\centering
\caption{\modbluetext{Summary of the primary literature in cloud and multi-edge collaboration.}}
\label{tab: tab_collaboration_literature}
\begin{tabular}{ >{\centering\arraybackslash}m{1.6cm} >{\centering\arraybackslash}m{2.4cm} p{\widcollit} }
\toprule
\textbf{Techniques} & \multicolumn{1}{c}{\textbf{Works}} & \multicolumn{1}{c}{\textbf{Features}} \\
\midrule
\multirow{10}{*}{\makecell{Split\\Inference}} & PETALS~\cite{borzunovPetalsCollaborativeInference2023} & \begin{tabular}[c]{@{}p{\widcollit}@{}} Utilizes distributed idle computational resources to ensure robust performance under dynamic network conditions.\end{tabular} \\
    \noalign{\vspace{0.5ex}}
& Voltage~\cite{hu2024edge} & \begin{tabular}[c]{@{}p{\widcollit}@{}} Achieves linear speed-ups by distributing transformer layers across edge devices, leveraging parallel computation. \end{tabular}\\
    \noalign{\vspace{0.5ex}}
& EASTER~\cite{guo2024easter} &\begin{tabular}[c]{@{}p{\widcollit}@{}}  Implements adaptive partition strategies to sustain performance despite device failures.\end{tabular}\\
    \noalign{\vspace{0.5ex}}
& LinguaLinked~\cite{zhao2024lingualinked} & \begin{tabular}[c]{@{}p{\widcollit}@{}} Optimizes split inference by aligning model segments with trusted mobile device capabilities for efficient data exchange.\end{tabular} \\
    \noalign{\vspace{0.5ex}}
& Hepti~\cite{lee2024autonomous}& \begin{tabular}[c]{@{}p{\widcollit}@{}} Dynamically adapts workload distribution and optimizes offloading decisions based on network and resource conditions. \end{tabular}\\
\midrule
\multirow{5}{*}{\makecell{Speculative\\Decoding}} & SpecTr~\cite{sun2024spectr} & \begin{tabular}[c]{@{}p{\widcollit}@{}} Employs lightweight models on edge devices to generate token drafts, enabling parallel validation by larger cloud models. \end{tabular}\\
    \noalign{\vspace{0.5ex}}
& Tabi~\cite{wang2023tabi} & \begin{tabular}[c]{@{}p{\widcollit}@{}} Utilizes calibrated confidence scores to decide token uploads for the cloud verification, balancing efficiency and accuracy.\end{tabular} \\
    \noalign{\vspace{0.5ex}}
& EdgeLLM~\cite{xu2024EdgeLLMSpeculative} & \begin{tabular}[c]{@{}p{\widcollit}@{}}  Features a branch navigation and self-adaptive fallback strategy, enabling fast and accurate token generation.\end{tabular} \\

\bottomrule
\end{tabular}
\end{table}


\subsubsection{Single-Device Resource Scheduling} \label{SubSubSec: Single-Device Resource Scheduling}
\begin{figure}[ht]
    \centering
    \begin{minipage}{\textwidth}
        \centering
            \includegraphics[width=0.56\textwidth]{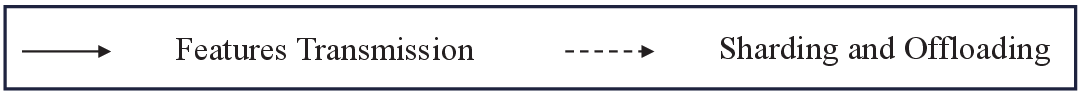}  
    \end{minipage}
    \begin{minipage}{\textwidth}
        \centering
        \vspace{1mm}
  \subfloat[Token Reduction]{\includegraphics[width=0.234\textwidth]{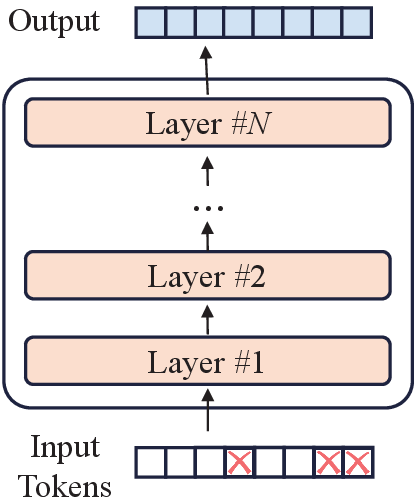}\vspace{0mm}}	\hspace{6mm}
  \subfloat[Early Exiting]{\includegraphics[width=0.240\textwidth]{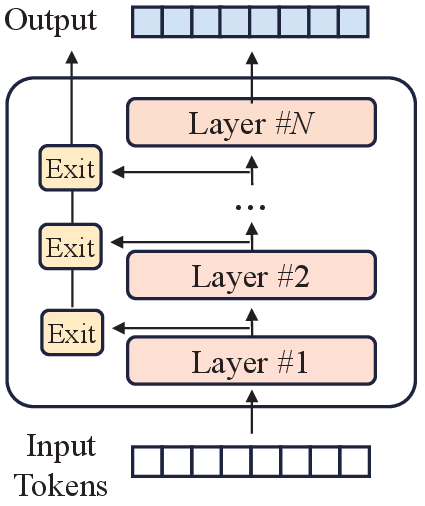}\vspace{0mm}}\hspace{6mm}
  \subfloat[Dynamic Offloading]{\includegraphics[width=0.32\textwidth]{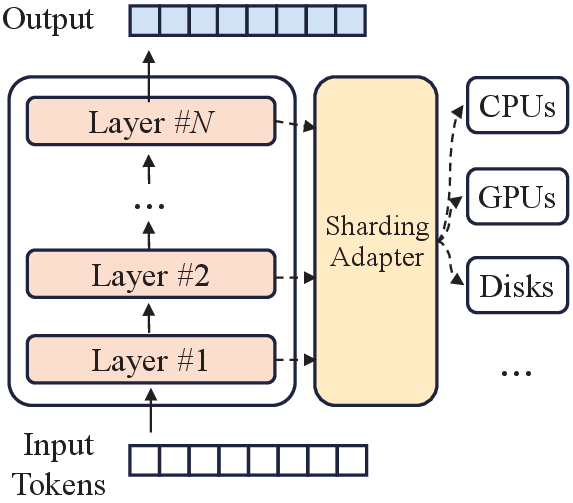}\vspace{0mm}}\hspace{6mm}
  \end{minipage}
    \caption{\modbluetext{Illustrations of single-device resource scheduling.}}
    \label{fig: fig_single_device}
\end{figure}

~\modbluetext{Efficient scheduling for single-device inference builds upon a layered approach, where different techniques tackle optimization challenges from complementary perspectives. As illustrated in Fig.~\ref{fig: fig_single_device}, recent advancements can be categorized into token reduction, early exiting, and dynamic offloading.}

~

~\modbluetext{~\textbf{Token reduction} focuses on reducing the computational load by selectively eliminating unnecessary tokens or inputs. For example, PoWER-BERT~\cite{goyal2020powerBERT} eliminates non-critical word vectors. Length-Adaptive Transformer~\cite{kim2021lengthAdaptiveTransformer} adjusts input sequence lengths dynamically using a dropout variant called LengthDrop. LTP~\cite{kim2022learnedTokenPruning} prunes less significant tokens based on attention scores. Additionally, LLMLingua~\cite{jiang2023llmlingua} compresses prompts iteratively at the token level, and AutoCompressors~\cite{chevalier2023adapting} reduce long context windows into more compact summary vectors. These techniques collectively aim to minimize the input size, ensuring that only the most important data is processed, thereby optimizing memory usage and reducing the computational burden on edge devices.}

\textbf{Early exiting} terminates inference once a predefined confidence threshold is met, thereby reducing unnecessary computations during the model's forward pass. This technique can further benefit from the reduced input size. For example, PABEE~\cite{zhouBERTLosesPatience2020} enhances the efficiency and robustness of pre-trained language models by integrating internal-classifiers at each layer. MPEE~\cite{kongAcceleratingInferencePretrained2022} combines horizontal and vertical early exit strategies for adaptive inference. FREE~\cite{bae2023fastFREE} proposes a shallow-deep module to synchronize the decoding process of the current token with previously stacked early-exit tokens. ConsistentEE~\cite{zeng2024consistentee} integrates reinforcement learning to determine the optimal early exit points, balancing inference speed with result accuracy. Techniques like LeeBERT~\cite{zhuLeeBERTLearnedEarly2021} and FastBERT~\cite{liuFastBERTSelfdistillingBERT2020} enable adaptive computation via early exiting with layer-wise confidence evaluation, ensuring efficient resource utilization while maintaining output quality.

~\textbf{Dynamic offloading} further optimizes resource utilization by distributing tasks across different processing units. Techniques like STI~\cite{guoSTITurbochargeNLP2023} introduce elastic pipelining, where model components are dynamically sharded and assigned to available resources. FlexGen~\cite{shengFlexGenHighThroughputGenerative2023} further refines this concept by leveraging a hybrid memory model, utilizing GPU, CPU, and disk memory with intelligent I/O scheduling for offloading computationally intensive tasks. LLM in a Flash~\cite{alizadehLLMFlashEfficient2024} stores the model parameters in flash memory and dynamically loads them into DRAM for inference. They ensure balanced workloads, ensuring that edge devices can efficiently handle large-scale LLM inference with minimal bottlenecks.

\newcommand{\widsinglit}{7.4cm}
\begin{table}[ht]\footnotesize
\centering
\caption{\modbluetext{Summary of the primary literature in single-device resource scheduling.}}
\label{tab: tab_single_literature}
\begin{tabular}{ >{\centering\arraybackslash}m{2.1cm} >{\centering\arraybackslash}m{2.8cm} p{\widsinglit} }
\toprule
\textbf{Techniques} & \multicolumn{1}{c}{\textbf{Works}} & \multicolumn{1}{c}{\textbf{Features}} \\
\midrule
\multirow{10}{*}{\makecell{Token Reduction}} & PoWER-BERT
~\cite{goyal2020powerBERT} & \begin{tabular}[c]{@{}p{\widsinglit}@{}}Eliminates non-critical word vectors to reduce computation while preserving accuracy.\end{tabular} \\
    \noalign{\vspace{0.5ex}}
&  Length-Adaptive Transformer
~\cite{kim2021lengthAdaptiveTransformer} & \begin{tabular}[c]{@{}p{\widsinglit}@{}}Dynamically adjusts input sequence lengths using a dropout-based technique called LengthDrop~\cite{kim2021lengthAdaptiveTransformer}.\end{tabular}\\
    \noalign{\vspace{0.5ex}}
& LTP~\cite{kim2022learnedTokenPruning} &\begin{tabular}[c]{@{}p{\widsinglit}@{}}Prunes less significant tokens by leveraging attention scores to optimize inference efficiency.\end{tabular}\\
    \noalign{\vspace{0.5ex}}
& LLMLingua~\cite{jiang2023llmlingua} & \begin{tabular}[c]{@{}p{\widsinglit}@{}}Compresses prompts iteratively at the token level to streamline input processing.\end{tabular} \\
    \noalign{\vspace{0.5ex}}
& AutoCompressors~\cite{chevalier2023adapting}& \begin{tabular}[c]{@{}p{\widsinglit}@{}}Reduces long context windows into more compact summary vectors for efficient processing. \end{tabular}\\
\midrule
\multirow{10}{*}{\makecell{Early Exiting}} & PABEE
~\cite{zhouBERTLosesPatience2020} & \begin{tabular}[c]{@{}p{\widsinglit}@{}}Improves the efficiency and robustness of pre-trained language models by integrating internal classifiers at each layer. \end{tabular}\\
    \noalign{\vspace{0.5ex}}
& MPEE~\cite{kongAcceleratingInferencePretrained2022} & \begin{tabular}[c]{@{}p{\widsinglit}@{}}Combines horizontal and vertical early exit strategies, enabling a more adaptive and efficient inference process.\end{tabular} \\

    \noalign{\vspace{0.5ex}}
& ConsistentEE~\cite{zeng2024consistentee} & \begin{tabular}[c]{@{}p{\widsinglit}@{}}Utilizes reinforcement learning to identify optimal early exit points, balancing inference speed with output accuracy.\end{tabular} \\

    \noalign{\vspace{0.5ex}}
& FREE~\cite{bae2023fastFREE} & \begin{tabular}[c]{@{}p{\widsinglit}@{}}Introduces a shallow-deep module to sync current token decoding with earlier early-exit tokens.\end{tabular} \\

\midrule
\multirow{7}{*}{\makecell{Dynamic Offloading}} & STI~\cite{guoSTITurbochargeNLP2023} & \begin{tabular}[c]{@{}p{\widsinglit}@{}}Implements elastic pipelining, dynamically sharding model components and assigning them to available resources.\end{tabular}\\
    \noalign{\vspace{0.5ex}}
& FlexGen~\cite{shengFlexGenHighThroughputGenerative2023} & \begin{tabular}[c]{@{}p{\widsinglit}@{}}Leverages a hybrid memory architecture, combining GPU, CPU, and disk memory with intelligent I/O scheduling to handle computationally intensive tasks.\end{tabular} \\

    \noalign{\vspace{0.5ex}}
& LLM in a Flash~\cite{alizadehLLMFlashEfficient2024} & \begin{tabular}[c]{@{}p{\widsinglit}@{}}Stores model parameters in flash memory and dynamically loads them into DRAM for efficient inference.\end{tabular} \\

\bottomrule
\end{tabular}
\end{table}


Recent studies, such as MELTing Point~\cite{laskaridisMELTingPointMobile2024}, offer insights beyond traditional benchmarking by evaluating performance, memory, and energy demands across various model sizes and devices. These analyses reveal critical bottlenecks in computational efficiency, quality of experience, and accuracy, providing a foundation for advancements in algorithms and hardware. As summarized in Table~\ref{tab: tab_single_literature}, they enhance understanding of single-device resource scheduling optimization, addressing challenges in computational efficiency and resource allocation for edge inference tasks.

\subsubsection{Framework-Level Optimizations} \label{SubSubSec: Framework-Level Optimizations}
To meet the performance and portability demands of edge computing, framework-level optimizations focus on lightweight frameworks, libraries, and engines specifically designed. For instance, PyTorch, widely adopted for its flexibility and ecosystem support, has also extended its capabilities to edge computing with ExecuTorch~\cite{executorch}, enabling efficient LLM inference through optimized execution plans for low-latency scenarios. 

Building on this, DNNFusion~\cite{niuDNNFusionAcceleratingDeep2021} optimizes mobile execution with advanced operator fusion, combining graph rewriting and fusion plan generation. SmartMem~\cite{niuSmartMemLayoutTransformation2024} reduces memory overhead by eliminating redundant layout transformations and selecting optimal memory layouts. PowerInfer~\cite{song2024powerinfer} enhances runtime efficiency by preloading frequently activated neurons onto the GPU and processing less active neurons on the CPU to minimize memory and data transfer overhead. Furthermore, vLLM~\cite{kwonEfficientMemoryManagement2023} introduces PagedAttention, inspired by virtual memory management, to segment attention key-value caches into blocks, enabling efficient memory sharing across sequences and requests. This approach improves memory efficiency while supporting quantization methods and optimized GPU kernels, making vLLM versatile for edge use.

~\modbluetext{The framework-level optimization ecosystem, as summarized in Table~\ref{tab: tab_frameworks}, includes tools tailored for LLM deployment on edge devices. The table categorizes frameworks by their descriptions, platform compatibility, and references to related works, helping readers identify suitable options and learn from previous implementations.}

\newcommand{\widdescription}{6.3cm}
\newcommand{\widframeworkworks}{2.435cm}
\newcommand{\widframeworknames}{1.64cm}

\begin{table}[ht]\scriptsize
\centering
\renewcommand{\arraystretch}{0.8}
\captionsetup{font={},justification=raggedright}
\caption{\modbluetext{Comparison of popular edge LLM frameworks, including examples of related works that have optimized or experimentally deployed these frameworks.}}
\label{tab: tab_frameworks}
\begin{tabular}{m{\widframeworknames} m{\widdescription} >{\centering\arraybackslash}m{0.92cm} >{\centering\arraybackslash}m{0.98cm} m{\widframeworkworks}}
\toprule
\centering{\textbf{Name}} & \centering{\textbf{Description}} & \textbf{Mobile Support} & \textbf{Desktop Support} & \makecell{\textbf{Related Works}} \\ 
\midrule
MLC LLM~\cite{mlc-llm} 
& Comprehensive LLM deployment framework leveraging machine learning compilation techniques. 
& \greencheck 
& \greencheck 
& OmniQuant~\cite{shaoOmniQuantOmnidirectionallyCalibrated2023}, MobileLLM~\cite{liu2024mobilellm}, MELTing Point~\cite{laskaridisMELTingPointMobile2024}. \\ 
\noalign{\vspace{0.5ex}}

llama.cpp~\cite{llamacpp} 
& Efficient C/C++ implementation for LLM inference. 
& \greencheck 
& \greencheck 
& EdgeLLM~\cite{xu2024EdgeLLMSpeculative}, MELTing Point~\cite{laskaridisMELTingPointMobile2024}. \\ 
\noalign{\vspace{0.5ex}}

ONNX Runtime \cite{onnxruntime}
& A cross-platform machine-learning model accelerator, with a flexible interface to integrate hardware-specific libraries. 
& \greencheck 
& \greencheck 
& EASTER~\cite{guo2024easter}. \\ 
\noalign{\vspace{0.5ex}}

ExecuTorch~\cite{executorch} 
& On-device LLM across mobile and edge for PyTorch. 
& \greencheck 
& \greencheck 
& Llama 3.2~\cite{IntroducingLlama3.2}. \\ 
\noalign{\vspace{0.5ex}}

vLLM~\cite{kwonEfficientMemoryManagement2023} 
& Utilize key-value cache stored in non-contiguous paged memory to efficiently inference. 
& \redcross 
& \greencheck 
& AWQ~\cite{linAWQActivationawareWeight2024}, GPTQ~\cite{frantarOPTQAccurateQuantization2022}, QLLM~\cite{liuQLLMAccurateEfficient2023}. \\ 
\noalign{\vspace{0.5ex}}

\makecell[l]{Neural\\ Compressor \cite{intelneuralcompressor}} \vspace{-1.5ex}
& State-of-the-art low-bit quantization and sparsity for runtime LLMs. 
& \redcross 
& \greencheck 
& SmoothQuant~\cite{xiaoSmoothQuantAccurateEfficient2023}, AWQ~\cite{linAWQActivationawareWeight2024}, GPTQ~\cite{frantarOPTQAccurateQuantization2022}, SparseGPT~\cite{frantarSparseGPTMassiveLanguage2023}. \\ 
\noalign{\vspace{0.5ex}}

TensorRT-LLM~\cite{tensorrtllm} 
& Advanced library for optimizing LLM inference powered by NVIDIA. 
& \redcross 
& \greencheck 
& SmoothQuant~\cite{xiaoSmoothQuantAccurateEfficient2023}, AWQ~\cite{linAWQActivationawareWeight2024}. \\ 
\noalign{\vspace{0.5ex}}

MLX~\cite{mlx2023} 
& Apple-developed array framework tailored for machine learning research on Apple silicon. 
& \redcross 
& \greencheck 
& OpenELM~\cite{mehta2024openelm}. \\ 

\bottomrule
\end{tabular}
\end{table}

\subsection{Hardware-Software Co-Design} \label{SubSec: Hardware-Software Co-Design}
\modbluetext{Hardware-software co-design is a cross-disciplinary approach that integrates hardware and software optimization to enhance the performance, energy efficiency, and scalability of machine learning models. Unlike isolated optimization strategies, co-design aligns algorithmic innovations with hardware-specific capabilities, addressing the unique constraints of edge devices, such as limited power, memory, and computational resources. This synergy is essential for achieving high-efficiency inference in LLMs on resource-constrained platforms.}

~\modbluetext{To provide a comprehensive overview of co-design optimizations, Table~\ref{tab: tab_hardware_software_codesign} compares software features and hardware platforms, highlighting the inference speedup and energy efficiency achieved by various approaches. Techniques are categorized into two primary areas: hardware-aware sparsity and hardware-optimized arithmetic formats. These classifications reflect different optimization focuses, enabling a clearer understanding of their respective contributions.}

~\newcommand{\widswdesign}{4cm}
\newcommand{\widhwdesign}{2.4cm}
\begin{table}[t]\scriptsize
\centering
\captionsetup{font={},justification=raggedright}
\caption{\modbluetext{Comparison of edge LLM-based co-design optimization techniques. (In-Mem:  Computing/Processing in Memory. SW: Software. HW: Hardware.)}}
  \label{tab: tab_hardware_software_codesign}
\begin{tabular}{ >{\centering\arraybackslash}m{1.3cm} >{\centering\arraybackslash}m{2.05cm} >{\centering\arraybackslash}m{\widswdesign} >{\centering\arraybackslash}m{1.6cm}  >{\centering\arraybackslash}m{2.6cm}}
\toprule
    \textbf{Type}& \textbf{Name}  &\textbf{SW Features} &\textbf{HW Chips}  & \textbf{Inference Speedup\textsuperscript{\textasteriskcentered} or Energy Efficiency\textsuperscript{\textdagger}} \\
\midrule
\multirow{24}{=}[0ex]{\centering \textbf{Hardware-Aware Sparsity}}
&SpAtten\cite{wangSpAttenEfficientSparse2021}  & \begin{tabular}[c]{@{}p{\widswdesign}@{}} Token-based sparsity and multi-head quantization.\end{tabular} & \makecell{ASIC 40nm\\simulator}
& \makecell{1095$\times$\textsuperscript{\textasteriskcentered}, 406$\times$\textsuperscript{\textdagger}\\Jetson Nano GPU}
\\
                \noalign{\vspace{0.5ex}}

&Sanger\cite{lu2021sanger} & \begin{tabular}[c]{@{}p{\widswdesign}@{}}Threshold-based pruning prediction.\end{tabular} &\makecell{ASIC 55nm\\simulator} &\makecell{1.47$\times$\textsuperscript{\textasteriskcentered} SpAtten} 
\\
                \noalign{\vspace{0.5ex}}

&EdgeBERT\cite{tambeEdgeBERTSentenceLevelEnergy2021} &\begin{tabular}[c]{@{}p{\widswdesign}@{}}Early exit prediction with pruning-aware optimizations\end{tabular}	& \makecell{ASIC 12nm\\simulator} &\makecell{53$\times$\textsuperscript{\textdagger} Jetson\\ Tegra X2 GPU}
\\ 
                \noalign{\vspace{0.5ex}}

&TaskFusion \cite{fanTaskFusionEfficientTransfer2023}  & \begin{tabular}[c]{@{}p{\widswdesign}@{}} Hardware-aware sub-task pruning with combined sparsity\end{tabular}& \makecell{ASIC 22nm\\simulator} &  \makecell{8.21$\times$\textsuperscript{\textasteriskcentered}, 19.83$\times$\textsuperscript{\textdagger}\\Jetson Nano GPU}
\\
                \noalign{\vspace{0.5ex}}

&AccelTran\cite{tuliAccelTranSparsityAwareAccelerator2023} &\begin{tabular}[c]{@{}p{\widswdesign}@{}}Activation sparsity via dynamic cycle simulation.\end{tabular}  & \makecell{ASIC 14nm\\simulator}
& \makecell{280$\times$\textsuperscript{\textasteriskcentered} , 236$\times$\textsuperscript{\textdagger}\\Apple M1 GPU} 
\\
                \noalign{\vspace{0.5ex}}

&4.60mm$^{2}$ STP\cite{tambe202322STP} &\begin{tabular}[c]{@{}p{\widswdesign}@{}}Entropy-based early exit and mixed-precision sparsity.\end{tabular} & \makecell{ASIC 12nm\\ tapeout} 
& \makecell{18.1 TFLOPS/W (FP4)~\textsuperscript{\textdagger}} 
\\
                \noalign{\vspace{0.5ex}}

&C-Transformer\cite{kim20CTransformer6182024} &\begin{tabular}[c]{@{}p{\widswdesign}@{}}Hybrid pruning for spiking/non-spiking Transformers.\end{tabular} & \makecell{ASIC 28nm\\tapeout}
& \makecell{47.8 TOPS/W (INT8)\textsuperscript{\textdagger}}
\\
                \noalign{\vspace{0.5ex}}
&TransPIM\cite{zhou2022transpim} &\begin{tabular}[c]{@{}p{\widswdesign}@{}}Token-based dataflow optimization\end{tabular}& \makecell{In-Mem 65nm\\simulator} 
& \makecell{114.9$\times$\textsuperscript{\textasteriskcentered}, 666.6$\times$\textsuperscript{\textdagger}\\RTX 2080 Ti GPU}
\\
                \noalign{\vspace{0.5ex}}
&X-Former\cite{sridharan2023xformer}  &\begin{tabular}[c]{@{}p{\widswdesign}@{}}Attention pruning with projection-based sparsity\end{tabular} & \makecell{In-Mem 32nm\\simulator} 
& \makecell{19.6$\times$\textsuperscript{\textasteriskcentered}, 13$\times$\textsuperscript{\textdagger}\\GTX 1060 GPU}
\\
                \noalign{\vspace{0.5ex}}
&TranCIM\cite{tu202228nmTranCIM} & \begin{tabular}[c]{@{}p{\widswdesign}@{}}Dynamic reconfiguration of streaming networks.\end{tabular} & \makecell{In-Mem 28nm\\tapeout}
& \makecell{20.5 TOPS/W (INT8)~\textsuperscript{\textdagger}} 
\\
                \noalign{\vspace{0.5ex}}
&MulTCIM\cite{tu202316MulTCIM} &\begin{tabular}[c]{@{}p{\widswdesign}@{}}Token pruning with reshaped attention matrices.\end{tabular}	
& \makecell{In-Mem 28nm\\tapeout}
& \makecell{101.1 TOPS/W (INT8)~\textsuperscript{\textdagger}} 
\\
\midrule
\multirow{12}{=}[0ex]{\centering \textbf{Hardware-Optimized Arithmetic Formats}}
&GOBO\cite{zadeh2020gobo} &  \begin{tabular}[c]{@{}p{\widswdesign}@{}}Map weights to a dictionary of 3-/4-bit indexed representative floats.\end{tabular}& \makecell{ASIC 65nm\\simulator} & \makecell{7$\times$\textsuperscript{\textasteriskcentered}, 3$\times$\textsuperscript{\textdagger} Tensor\\Cores-based accelerator} \\
                \noalign{\vspace{0.5ex}}
&Mokey\cite{zadeh2022mokey} & \begin{tabular}[c]{@{}p{\widswdesign}@{}}Quantizing values to 4-bit index counting instead of floating-point multiplications and accumulations.
\end{tabular} & \makecell{ASIC 65nm\\simulator}& 9$\times$\textsuperscript{\textdagger} GOBO\\
                \noalign{\vspace{0.5ex}}

&OliVe\cite{guoOliVeAcceleratingLarge2023} & \begin{tabular}[c]{@{}p{\widswdesign}@{}}Abfloat Format: An outlier-victim pair quantization strategy designed for sparse outlier distributions.\end{tabular} & \makecell{ASIC 22nm\\simulator}
& 4.5$\times$\textsuperscript{\textasteriskcentered}, 4$\times$\textsuperscript{\textdagger} GOBO\\
                \noalign{\vspace{0.5ex}}

&AdaptivFloat\cite{tambe2020AdaptivFloat} &\begin{tabular}[c]{@{}p{\widswdesign}@{}}HFINT Format: Dynamically adjusts exponent bias across network layers to optimize precision.
\end{tabular}  & \makecell{ASIC 16nm\\simulator}
& \makecell{1.09$\times$\textsuperscript{\textdagger} INT8-based\\accelerator}
\\
                \noalign{\vspace{0.5ex}}

&ANT\cite{guo2022ant}   & \begin{tabular}[c]{@{}p{\widswdesign}@{}}Flint Format: Adaptive bit-width selection between float and int.
\end{tabular} & \makecell{ASIC 28nm\\simulator}
& \makecell{4$\times$\textsuperscript{\textasteriskcentered}, 3.33$\times$\textsuperscript{\textdagger}\\AdaptivFloat}
\\

\bottomrule
\end{tabular}
\end{table}

\subsubsection{Hardware-Aware Sparsity} \label{SubSubSec: Hardware-Aware Sparsity} 
~\modbluetext{Hardware-aware sparsity focuses on integrating hardware-specific considerations into model design. This strategy can be further categorized into two primary approaches: sparsity on ASICs and sparsity on in-memory accelerators.}

~\modbluetext{\textbf{Sparsity on ASICs.} 
ASIC-based accelerators achieve fine-grained control over model sparsity through custom circuit designs. Simulator-based methods explore early-stage designs: SpAtten~\cite{wangSpAttenEfficientSparse2021} employs token pruning and a top-k ranking engine to prioritize token and head importance. Sanger~\cite{lu2021sanger} synergistically co-designs software that prunes attention matrices into dynamic structured patterns and hardware with a reconfigurable architecture. EdgeBERT~\cite{tambeEdgeBERTSentenceLevelEnergy2021} dynamically scales voltage and frequency, prunes networks, and quantizes floating points. TaskFusion~\cite{fanTaskFusionEfficientTransfer2023} combines weight and activation sparsity with hardware-aware sub-task inference algorithms. AccelTran~\cite{tuliAccelTranSparsityAwareAccelerator2023} enhances activation sparsity dynamically using a cycle-accurate simulator. Real-world tapeouts validate these designs: STP~\cite{tambe202322STP} optimizes latency and energy through mixed-precision computation and fine-grained power management. Similarly, C-Transformer~\cite{kim20CTransformer6182024} integrates spiking and non-spiking Transformers, achieving high sparsity and hardware utilization, illustrating the potential of hybrid architectures.}

~\modbluetext{~\textbf{Sparsity on In-Memory Accelerators.} 
In-memory accelerators address data movement challenges by embedding computation within memory arrays. Processing-in-Memory (PIM) augments conventional memory hierarchies with compute units near memory, while Computing-in-Memory (CIM) embeds compute capabilities directly into memory cells for fine-grained operations. PIM-based sparsity designs like TransPIM~\cite{zhou2022transpim} use token-based dataflows and high-bandwidth memory to minimize communication overhead, and X-Former~\cite{sridharan2023xformer} combines a software-level attention engine with nonvolatile memory and CMOS tiles. CIM-based sparsity designs, such as TranCIM~\cite{tu202228nmTranCIM}, dynamically reconfigure streaming networks and bitline-transpose structures to reduce complexity. MulTCIM~\cite{tu202316MulTCIM} addresses hybrid sparsity with runtime token pruning and attention matrix reshaping, improving scalability and efficiency.}

~\modbluetext{\textbf{Comparative Insights.} ASIC accelerators excel in throughput optimization through fine-grained sparsity, whereas in-memory accelerators focus on reducing data movement with token-based designs. Together, these approaches demonstrate the transformative potential of hardware-aware sparsity in edge scenarios.}

\subsubsection{Hardware-Optimized Arithmetic Formats} \label{SubSubSec: Hardware-Optimized Arithmetic Formats} 
~\modbluetext{Hardware-optimized arithmetic formats optimize inference performance by balancing computational accuracy with hardware efficiency. These formats employ two synergistic strategies: low-bit arithmetic and dynamic adaptive encoding.}

~\modbluetext{\textbf{Hardware-Optimized Low-Bit Arithmetic Formats.} Low-bit arithmetic formats reduce the numerical precision of model parameters to fixed low-bit representations, tailored for specialized hardware accelerators. For example, GOBO ~\cite{zadeh2020gobo} reduces 32-bit floating-point parameters to just 3 bits, significantly lowering power consumption. Mokey ~\cite{zadeh2022mokey} quantizes parameters to 4-bit representations, enabling area- and energy-efficient hardware acceleration. OliVe ~\cite{guoOliVeAcceleratingLarge2023} introduces an outlier-victim pair quantization approach, which sacrifices normal values to accommodate outliers, enabling more efficient memory alignment and boosting performance.}

~\modbluetext{\textbf{Hardware-Optimized Dynamic Adaptive Encoding.} Dynamic adaptive encoding adjusts numerical precision based on runtime requirements, providing greater flexibility compared to fixed low-bit formats. AdaptivFloat~\cite{tambe2020AdaptivFloat} dynamically adjusts tensor-wide exponent biases to maintain accuracy at low bit widths. ANT~\cite{guo2022ant} introduces a flexible adaptive data format, Flint, which combines floating-point and integer precision for low-bit quantization with minimal hardware overhead. }

\textbf{Comparative Insights.} Low-bit formats offer energy efficiency and predictable performance but may sacrifice accuracy, whereas adaptive encoding provides precision flexibility at the cost of added complexity. Combining both strategies can yield robust and adaptable hardware-software designs for on-device LLMs.

In conclusion, hardware-software co-design represents a critical avenue for optimizing LLMs on edge devices. However, much of the research remains in the simulation phase, with limited real-world deployment on general-purpose heterogeneous platforms. By summarizing these approaches, we aim to highlight how tailored software optimizations can exploit hardware-specific features to outperform similar algorithms on existing hardware platforms. This discussion serves as a guide for future developments in deploying efficient and scalable LLMs in resource-constrained environments.

\subsection{Hardware-Level Optimizations}  ~\label{SubSec: Hardware-Level Optimizations}
The development of hardware architectures is essential for deploying LLMs on devices, enhancing inference speed and energy efficiency. Table~\ref{tab: tab_hardware} outlines prominent commercial hardware chips, emphasizing AI performance and related studies. The AI performance column highlights key metrics of System-on-Chips (SoCs), CPUs, and GPUs, while the Related Works column helps readers identify suitable hardware and explore relevant optimization strategies from previous implementations.

~\newcommand{\widperformance}{3.6cm}
\newcommand{\widhardworks}{4.2cm}
\newcommand{\widchip}{1.8cm}
\begin{table}[ht]\scriptsize
\centering
\captionsetup{font={},justification=raggedright}
\caption{\modbluetext{Comparison of popular commercial hardware devices for on-device LLMs. (RPi: Raspberry Pi.)}}
  \label{tab: tab_hardware}
\begin{tabular}{>{\centering\arraybackslash}m{1.1cm} >{\centering\arraybackslash}m{2.0cm}  >{\centering\arraybackslash}m{0.8cm} >{\centering\arraybackslash}m{\widperformance} >{\centering\arraybackslash}m{\widhardworks}}
\toprule
\textbf{Hardware} & \textbf{Chip} &\textbf{Type} & \textbf{AI Performance} & \textbf{Related Works}\\ 
\midrule

\multirow{14}{=}[0ex]{\centering \textbf{SoCs (CPU + iGPU + NPU)}}

&\makecell{Samsung\\Exynos}   & Mobile & \begin{tabular}[c]{@{}p{\widperformance}@{}}Exynos 2400 equips with NPU and DSP AI Engine~\cite{samsungExynos2400news} \end{tabular} & \begin{tabular}[c]{@{}p{\widhardworks}@{}}Llama 3.2~\cite{IntroducingLlama3.2}\end{tabular}\\ 
                \noalign{\vspace{0.5ex}}

&\makecell{Google\\Tensor}& Mobile & \begin{tabular}[c]{@{}p{\widperformance}@{}}Tensor G4 can run Gemini Nano \cite{team2023gemini} with multi-modality~\cite{googleTensorG4} \end{tabular} & \begin{tabular}[c]{@{}p{\widhardworks}@{}}MELTing Point~\cite{laskaridisMELTingPointMobile2024}, Llama 3.2~\cite{IntroducingLlama3.2}, Gemini Nano~\cite{team2023gemini}, MobileLLM~\cite{liu2024mobilellm}\end{tabular} \\
                \noalign{\vspace{0.5ex}}

& Apple A & Mobile & \begin{tabular}[c]{@{}p{\widperformance}@{}}A16 Bionic features 16-core Neural Engine with 17 TOPS~\cite{appleaseries}\end{tabular}& \begin{tabular}[c]{@{}p{\widhardworks}@{}}MobileLLM~\cite{liu2024mobilellm}, Llama 3.2~\cite{IntroducingLlama3.2}, MELTing Point~\cite{laskaridisMELTingPointMobile2024}, SpQR~\cite{dettmersSpQRSparseQuantizedRepresentation2023}\end{tabular}\\ 
                \noalign{\vspace{0.5ex}}

&Apple M & Laptop & \begin{tabular}[c]{@{}p{\widperformance}@{}}M2 Ultra features a 32-core Neural Engine with 31.6 TOPS~\cite{applemseries}\end{tabular} & \begin{tabular}[c]{@{}p{\widhardworks}@{}}MELTing Point~\cite{laskaridisMELTingPointMobile2024}, LLM in a Flash~\cite{alizadehLLMFlashEfficient2024}\end{tabular}\\ 
                \noalign{\vspace{0.5ex}}

& \makecell{Qualcomm\\Snapdragon} & Mobile & \begin{tabular}[c]{@{}p{\widperformance}@{}}Snapdragon 8 Gen 3 achieves up to 20 tokens/sec for LLM inference~\cite{QualcommSnapdragon8Gen3MobilePlatform}\end{tabular}& \begin{tabular}[c]{@{}p{\widhardworks}@{}}EdgeLLM~\cite{xu2024EdgeLLMSpeculative}, MobileLLM~\cite{liu2024mobilellm}, MobileBERT~\cite{sunMobileBERTCompactTaskAgnostic2020}, Llama 3.2~\cite{IntroducingLlama3.2}, MELTing Point~\cite{laskaridisMELTingPointMobile2024}, Agile-Quant~\cite{shenAgileQuantActivationGuidedQuantization2024}\end{tabular}\\

\midrule
                \noalign{\vspace{-1ex}}
\multirow{6}{=}[1ex]{\centering \textbf{CPUs}}
& \makecell{Intel Core} & \makecell{Desktop}  & \begin{tabular}[c]{@{}p{\widperformance}@{}}Intel i9-13900k combines Performance and Efficient cores with Deep Learning Boost~\cite{inteli913900news}\end{tabular} & \begin{tabular}[c]{@{}p{\widhardworks}@{}}SmoothQuant~\cite{xiaoSmoothQuantAccurateEfficient2023}, EASTER~\cite{guo2024easter}, AQLM~\cite{egiazarianextreme}, Neural Compressor~\cite{intelneuralcompressor}, FREE~\cite{bae2023fastFREE}\end{tabular} \\ 
                \noalign{\vspace{0.5ex}}
& \makecell{ARM Cortex-A72\\(on RPi 4/4B)} & \makecell{\\IoT\\\\}  & \begin{tabular}[c]{@{}p{\widperformance}@{}} RPi 4B handles lightweight LLM inference efficiently with quantization and token pruning~\cite{shenAgileQuantActivationGuidedQuantization2024} \end{tabular} & \begin{tabular}[c]{@{}p{\widhardworks}@{}}MELTing Point~\cite{laskaridisMELTingPointMobile2024}, Hepti~\cite{lee2024autonomous}, Agile-Quant~\cite{shenAgileQuantActivationGuidedQuantization2024} \end{tabular}\\ 

\midrule

\multirow{10}{=}[1ex]{\centering \textbf{GPUs}}
& \makecell{NVIDIA GeForce\\Laptop GPU} & Laptop  & \begin{tabular}[c]{@{}p{\widperformance}@{}}RTX 4070 Laptop GPU achieves 321 TOPS with 8GB memory~\cite{nvidiaLaptopCompare} \end{tabular} &\begin{tabular}[c]{@{}p{\widhardworks}@{}} AWQ~\cite{linAWQActivationawareWeight2024}\end{tabular} \\ 
                \noalign{\vspace{0.5ex}}

& \makecell{\\NVIDIA GeForce\\Desktop GPU\\} & \makecell{\\Desktop}  & \begin{tabular}[c]{@{}p{\widperformance}@{}}\\RTX 4090 offers 1321 TOPS with 24GB memory~\cite{nvidiaGeForceGraphicsCards} \end{tabular} & \begin{tabular}[c]{@{}p{\widhardworks}@{}}AWQ~\cite{linAWQActivationawareWeight2024}, SpQR~\cite{dettmersSpQRSparseQuantizedRepresentation2023}, OWQ~\cite{leeOWQOutlierAwareWeight2024}, AQLM~\cite{egiazarianextreme}, TensorRT-LLM~\cite{tensorrtllm}, QuIP\#~\cite{tseng2024quipbetter}, LLM in a Flash~\cite{alizadehLLMFlashEfficient2024}, Drone~\cite{chenDRONEDataawareLowrank2021}\end{tabular} \\ 
                \noalign{\vspace{0.5ex}}

& \makecell{NVIDIA Jetson\\Modules GPU} &IoT & \begin{tabular}[c]{@{}p{\widperformance}@{}}Jetson AGX Orin delivers 275 TOPS with 64GB memory~\cite{JetsonModulesSupport} \end{tabular} &\begin{tabular}[c]{@{}p{\widhardworks}@{}} EdgeLLM~\cite{xu2024EdgeLLMSpeculative}, MELTing Point~\cite{laskaridisMELTingPointMobile2024}, STI~\cite{guoSTITurbochargeNLP2023}, Beyond the Cloud~\cite{zhang2024beyondtheCloud}, TensorRT-LLM~\cite{tensorrtllm}\end{tabular}\\

\bottomrule
\end{tabular}
\end{table}

\subsubsection{CPUs: Foundation of AI Workflow} \label{SubSubSec: CPUs: Foundation of AI Workflow}
~\modbluetext{CPUs remain fundamental in AI workflow management, providing essential flexibility and system-level integration. Recent advancements have expanded their LLM potential for LLM inference. For example, Arm Cortex A72 on Raspberry Pi 4B (8GB RAM) achieves low latency inference of LLaMA-7B~\cite{touvron2023llama} when paired with Agile-Quant~\cite{shenAgileQuantActivationGuidedQuantization2024}. Intel i9-13900k~\cite{inteli913900news} enhanced by AQLM~\cite{egiazarianextreme}, provides efficient inference for the LLaMA-2 model~\cite{touvron2023llama2}.}

However, CPUs' architecture, optimized for sequential tasks, struggle with the parallel computations needed for LLM inference, especially in edge scenarios requiring real-time performance and energy efficiency. To address this, modern CPUs are often combined with specialized accelerators (e.g., GPUs and NPUs) in SoCs like Apple’s M-series or A-series SoCs~\cite{applemseries,appleaseries}, Google's Tensor G4~\cite{googleTensorG4}, and Qualcomm Snapdragon chips~\cite{QualcommSnapdragon8Gen3MobilePlatform}. These heterogeneous architectures aim to balance flexibility and performance by integrating general-purpose and parallel processing units. However, managing data flow and power consumption across these components remains a challenge.

\subsubsection{GPUs: Parallel Acceleration} \label{SubSubSec: GPUs: Parallel Acceleration}
To address the inefficiency of CPUs, GPUs have been developed as highly parallel processors to accelerate computationally intensive tasks in edge computing. Equipped with hundreds or thousands of smaller cores, GPUs enable massive parallel computation for LLM inference. Modern edge GPUs, such as the NVIDIA Jetson series~\cite{JetsonModulesSupport}, feature specialized hardware such as Tensor Cores, optimized for matrix operations prevalent in LLM inference tasks. For example, ~\citet{yuan2024mobilefirmware} demonstrates the feasibility of LLM inference on NVIDIA Jetson ORIN NX~\cite{JetsonModulesSupport}.

However, GPUs face significant power consumption challenges in edge scenarios, often necessitating hybrid models that distribute intensive computations between cloud and local resources. This strategy balances high-performance inference with the energy constraints of mobile and edge devices. Additionally, GPUs are frequently integrated with CPUs and NPUs in heterogeneous systems, complicating integration and resource management. Effective scheduling and task offloading among these processors are critical challenges in edge deployment, requiring advanced software frameworks to ensure efficient collaboration.

\subsubsection{NPUs: Neural Network Optimization} \label{SubSubSec: NPUs: Neural Network Optimization}
~\modbluetext{NPUs are specialized accelerators designed to optimize neural network computations, offering substantial improvements in performance and energy efficiency for edge LLM inference. By employing low-precision arithmetic (e.g., INT8) and highly parallelized architectures, NPUs enable real-time inference with minimal power consumption. Notable examples include Apple’s Neural Engine in the M-series or A-series chips~\cite{applemseries,appleaseries} and Qualcomm’s AI engines in Snapdragon processors~\cite{QualcommSnapdragon8Gen3MobilePlatform}, which boost on-device LLM capabilities and reduce cloud dependency.}

~\modbluetext{However, NPUs face key limitations that restrict their broader applicability. They are optimized for a limited set of neural network operators, making them incompatible with many modern LLM architectures. This often forces fallback strategies like CPU-NPU co-processing, which can offset performance gains. Additionally, the rapid evolution of neural architectures further complicates matters, as NPUs struggle to keep up with model diversity and complexity, requiring extensive adaptations or failing to execute some models entirely~\cite{yuan2024mobilefirmware}. In heterogeneous hardware environments, effective integration of NPUs with CPUs and GPUs is essential, as NPUs' specialization may necessitate offloading tasks to general-purpose processors, complicating resource management.}


In conclusion, online runtime inference optimizations enhance LLM performance on edge devices through software-level optimizations, hardware-software co-design, and hardware-level enhancements. These methods complement offline pre-deployment techniques, forming a holistic approach to on-device LLM optimization. Offline techniques reduce computational complexity and memory footprint via pre-training and fine-tuning, while runtime strategies focus on efficient resource utilization, dynamic adaptation, and scalability. Integrating both phases is crucial for high-performance LLMs on resource-constrained platforms.



\section{\nameapplication} \label{Sec: \nameapplication}
Leveraging compact models and runtime optimizations, on-device LLMs enable efficient, low-latency, and privacy-preserving AI in edge environments. As shown in Fig.~\ref{fig: fig_sec5_application}, on-device LLM-based application systems span personal, enterprise, and industrial domains.

\def\appwid{26.997}
\def\apphit{12.601}
\begin{figure}[ht]
    \centering
    \begin{tikzpicture}
        \node[anchor=north west, inner sep=0] (image) at (0,0) {\includegraphics[width=0.92\textwidth]{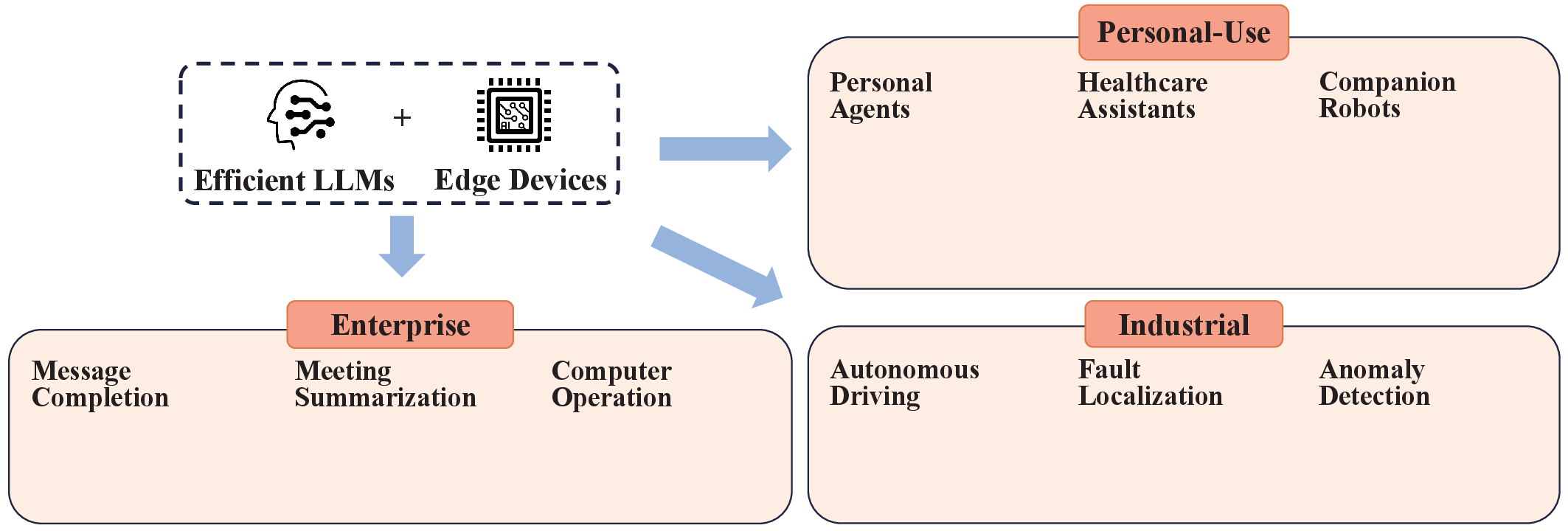}}; 
        \begin{scope}[x={(image.north east)}, y={(image.south west)}] 
             \node[font=\tiny,align=left,text width=2cm] at (16.40/\appwid,3.28/\apphit){AutoDroid~\cite{wenAutoDroidLLMpoweredTask2024}}; 
             \node[font=\tiny,align=left,text width=2cm] at (16.40/\appwid,3.98/\apphit){LlamaTouch~\cite{zhang2024llamatouch}}; 
             \node[font=\tiny,align=left,text width=2cm] at (16.40/\appwid,4.68/\apphit){CoCo-Agent~\cite{ma2024comprehensive}};

             \node[font=\tiny,align=left,text width=2cm] at (20.7/\appwid,3.28/\apphit){BioMistral~\cite{labrak2024biomistral}}; 
             \node[font=\tiny,align=left,text width=2cm] at (20.7/\appwid,3.98/\apphit){PathChat~\cite{lu2024multimodal}}; 
             \node[font=\tiny,align=left,text width=2cm] at (20.7/\appwid,4.68/\apphit){Menta-LLM~\cite{xu2024mentallm}};
             \node[font=\tiny,align=left,text width=2cm] at (20.7/\appwid,5.38/\apphit){MentaLLaMA~\cite{yang2024mentallama}}; 
             \node[font=\tiny,align=left,text width=2cm] at (20.7/\appwid,6.08/\apphit){ODSearch~\cite{rawassizadehODSearchFastResource2022}};

            \node[font=\tiny,align=left,text width=3cm] at (25.9/\appwid,3.28/\apphit){3D-LLM~\cite{hong20233dllm}}; 
             \node[font=\tiny,align=left,text width=3cm] at (25.9/\appwid,3.98/\apphit){E2WM~\cite{xiang2024languageEmbodied}}; 
             \node[font=\tiny,align=left,text width=3cm] at (25.9/\appwid,5.08/\apphit){Inner\\Monologue~\cite{huang2023innermonologue}};

            \node[font=\tiny,align=left,text width=3cm] at (3.7/\appwid,10.08/\apphit) {Gboard~\cite{GeminiNanoWeb}}; 
            \node[font=\tiny,align=left,text width=3cm] at (3.7/\appwid,10.78/\apphit) {Rewriting Agent~\cite{zhu2024textrewriting}}; 
            \node[font=\tiny,align=left,text width=3cm] at (3.7/\appwid,11.49/\apphit) {ChatEval~\cite{chan2024chateval}}; 
            
            \node[font=\tiny,align=left,text width=3cm] at (8.25/\appwid,10.08/\apphit) {MobiVQA~\cite{caoMobiVQAEfficientOnDevice2022}}; 
            \node[font=\tiny,align=left,text width=3cm] at (8.25/\appwid,10.78/\apphit) {DiaSum-MoE~\cite{tian2024dialoguesummarization}}; 

            \node[font=\tiny,align=left,text width=3cm] at (12.6/\appwid,10.08/\apphit) {WebAgent\cite{gurrealWebAgent}}; 
            \node[font=\tiny,align=left,text width=3cm] at (12.6/\appwid,10.78/\apphit) {SheetCopilot\cite{li2024sheetcopilot}}; 
            \node[font=\tiny,align=left,text width=3cm] at (12.6/\appwid,11.49/\apphit) {RCI Agent~\cite{kim2024languageRCI}}; 

             \node[font=\tiny,align=left,text width=2cm] at (16.40/\appwid,10.08/\apphit){DriveVLM~\cite{tian2024drivevlm}}; 
             \node[font=\tiny,align=left,text width=2cm] at (16.40/\appwid,10.78/\apphit){LLM-Driver~\cite{chen2024drivingWithLLMs}}; 
             \node[font=\tiny,align=left,text width=2cm] at (16.40/\appwid,11.49/\apphit){VLP~\cite{pan2024vlp}};

            \node[font=\tiny,align=left,text width=2cm] at (20.7/\appwid,10.08/\apphit){AutoFL~\cite{kang2024quantitativeAutoFL}}; 
             \node[font=\tiny,align=left,text width=2cm] at (20.7/\appwid,10.78/\apphit){LLMAO~\cite{yang2024largeLLMAO}}; 
             \node[font=\tiny,align=left,text width=2cm] at (20.7/\appwid,11.49/\apphit){Toggle~\cite{hossain2024deepToggle}};

            \node[font=\tiny,align=left,text width=3cm] at (25.9/\appwid,10.08/\apphit){WinCLIP~\cite{jeong2023winclip}}; 
             \node[font=\tiny,align=left,text width=3cm] at (25.9/\appwid,10.78/\apphit){AnomalyGPT~\cite{gu2024anomalygpt}}; 
             \node[font=\tiny,align=left,text width=3cm] at (25.9/\appwid,11.49/\apphit){ALFA~\cite{zhu2024llmsALFA}};

        \end{scope}
    \end{tikzpicture}
    \caption{\modbluetext{Illustrations of on-device applications for LLMs.}}
    \label{fig: fig_sec5_application}
\end{figure}

\subsection{Personal-Use Applications}  \label{SubSec: Personal-Use Applications}
On-device LLMs enable ubiquitous AI experiences in edge environments, driving advancements in personal assistants, healthcare assistants, and companion robots.

~\subsubsection{Personal Agents} \label{SubSubSec: Personal Agents}
On-device LLMs are transforming personal digital assistants by providing low-latency, privacy-preserving, and always available support tailored to individual needs. AutoDroid~\cite{wenAutoDroidLLMpoweredTask2024} automates daily tasks such as food ordering and health tracking, leveraging multi-granularity token pruning techniques to operate efficiently within the limited computational budget of mobile devices. Similarly, LlamaTouch~\cite{zhang2024llamatouch} redefines mobile UI task automation evaluation, while CoCo-Agent~\cite{ma2024comprehensive} enhances GUI automation for complex interactions with real-world environments. These advancements streamline personal productivity by automating routine tasks, including email management, photo editing, and scheduling~\cite{wenAutoDroidLLMpoweredTask2024}.

~\subsubsection{Healthcare Assistants} \label{SubSubSec: Healthcare Assistants}
In healthcare, on-device LLMs demonstrate versatility by providing domain-specific support in areas like clinical decision-making and mental health analysis. BioMistral~\cite{labrak2024biomistral}, a 7B biomedical LLM, exemplifies the synergy between domain-specific fine-tuning and quantization to achieve low-latency inference. PathChat~\cite{lu2024multimodal} integrates visual and natural language inputs through a multimodal projector module connecting a vision encoder to the LLaMA-2 model~\cite{touvron2023llama2}, transforming pathology education and research. Mental-LLM~\cite{xu2024mentallm} and MentaLLaMA~\cite{yang2024mentallama} focus on interpretable mental health analysis on social media platforms. On-device LLMs also enable wearable applications such as ODSearch~\cite{rawassizadehODSearchFastResource2022}, which provides a natural language interface for fitness tracker data, delivering near real-time search capabilities.

~\subsubsection{Companion Robots} \label{SubSubSec: Companion Robots}
On-device LLMs are transforming companion robots by enabling rapid, context-aware interactions and excelling in both verbal and non-verbal tasks. These robots can build connections through dialogue, negotiate with users, generate contextually appropriate responses, and execute actions requiring physical interaction. 3D-LLM~\cite{hong20233dllm} integrates 3D environment reasoning and planning, enabling robots to analyze spatial contexts and make informed decisions. E2WM~\cite{xiang2024languageEmbodied} augments multiple compact LLMs (e.g., LLaMA~\cite{touvron2023llama} and OPT~\cite{zhang2022opt}) with embodied knowledge and skills, improving adaptability in dynamic environments. Inner Monologue~\cite{huang2023innermonologue} enhances these capabilities by allowing 1.3B InstructGPT~\cite{ouyang2022trainingFollowingInstructions} to reason over natural language feedback for better planning and execution in embodied tasks.

\subsection{Enterprise Applications} \label{SubSec: Enterprise Applications}
On-device LLMs boost enterprise privacy and cost-efficiency, improving tasks like message completion, meeting summaries, and computer operations.

\subsubsection{Message Completion}  \label{SubSubSec: Message Completion}
On-device LLMs streamline enterprise communication by generating context-aware responses, improving productivity through automated replies. For instance, Google's Gboard~\cite{GeminiNanoWeb} integrates Gemini Nano~\cite{team2023gemini} for real-time, context-sensitive suggestions, utilizing a cloud-edge framework where the cloud handles complex tasks and the edge manages lightweight interactions. ~\citet{zhu2024textrewriting} propose a mobile-centric text rewriting LLM with efficient dataset generation and cascading mechanisms, while ChatEval~\cite{chan2024chateval} autonomously evaluates text quality, aligning closely with human judgment.

\subsubsection{Meeting Summarization} \label{SubSubSec: Meeting Summarization}
In meetings, on-device LLMs are enabling automated summarization of key discussions and decisions, offering a privacy-preserving alternative to cloud-based solutions while maintaining high efficiency. For example, MobiVQA~\cite{caoMobiVQAEfficientOnDevice2022} delivers an efficient on-device visual question-answering system with attention-based early exit and question-aware pruning techniques. ~\citet{tian2024dialoguesummarization} propose a MoE framework for role-oriented dialogue summarization, enhancing contextual nuance in business settings.

\subsubsection{Computer Operation} \label{SubSubSec: Computer Operation}
On-device LLMs facilitate natural language interfaces for computer operations, automating tasks and reducing user cognitive load. WebAgent~\cite{gurrealWebAgent} enables task execution on websites through real-time programming. SheetCopilot~\cite{li2024sheetcopilot} simplifies spreadsheet interactions by translating natural language commands into actionable tasks. The RCI agent~\cite{kim2024languageRCI} uses LLMs to autonomously complete tasks on a computer using a keyboard and mouse, iteratively refining its output to improve performance. These systems leverage cloud-edge collaboration, with the cloud handling intensive planning and the edge executing lightweight, privacy-preserving operations.

\subsection{Industrial Applications} \label{SubSec: Industrial Applications}
On-device LLMs enhance industrial systems by enabling real-time analysis and decision-making in applications like autonomous driving, fault localization, and anomaly detection, while reducing network overhead and improving response times.

\subsubsection{Autonomous Driving} \label{SubSubSec: Autonomous Driving}
On-device LLMs enhance autonomous driving by integrating linguistic understanding with navigation and decision-making. DriveVLM-Dual~\cite{tian2024drivevlm} uses a visual LLM to interpret urban environments and plan routes via natural language, efficiently deployed on NVIDIA Orin platforms~\cite{JetsonModulesSupport}. LLM-Driver~\cite{chen2024drivingWithLLMs} leverages object-level vector inputs for explainable driving actions prediction. Similarly, VLP~\cite{pan2024vlp} strengthens both contextual understanding and memory foundation in self-driving systems. These advancements demonstrate the transformative potential of LLMs in autonomous driving.

\subsubsection{Fault Localization} \label{SubSubSec: Fault Localization}
In software fault localization, LLM-based techniques offer significant improvements over traditional machine learning methods. AutoFL~\cite{kang2024quantitativeAutoFL} generates bug explanations and identifies fault locations by navigating software repositories and mitigating LLM context length constraints. LLMAO~\cite{yang2024largeLLMAO} detects logic and security vulnerabilities without relying on extensive program analysis or test cases, outperforming deep learning-based methods. Toggle~\cite{hossain2024deepToggle} employs tailored prompts and an adjustment module to localize and fix bugs at the token level, enabling fine-grained debugging.

\subsubsection{Anomaly Detection} \label{SubSubSec: Anomaly Detection}
On-device LLMs also advance industrial anomaly detection by leveraging multi-modal capabilities and efficient feature aggregation~\cite{zhouanomalyclip}. WinCLIP~\cite{jeong2023winclip} supports zero-shot and few-shot anomaly classification and segmentation using state word ensembles and prompt templates. AnomalyGPT~\cite{gu2024anomalygpt} employs large vision-language models to identify anomalies directly, removing the need for manual threshold adjustments while showcasing strong few-shot learning performance. ALFA~\cite{zhu2024llmsALFA} addresses zero-shot visual anomaly detection by generating adaptive prompts, reducing semantic ambiguities, and fusing local pixel-level information for precise localization.

In summary, on-device LLMs use both offline and online optimization techniques to enable efficient, private, and responsive edge AI applications. Offline methods like quantization, pruning, and compact model design (e.g., Gemini Nano~\cite{team2023gemini}, BioMistral~\cite{labrak2024biomistral}) reduce model size and computational demands, making them suitable for resource-constrained devices. These are supported by runtime optimizations such as cloud-edge collaboration (e.g., WebAgent~\cite{gurrealWebAgent}), early exiting (e.g., MobiVQA~\cite{caoMobiVQAEfficientOnDevice2022}), and hardware acceleration (e.g., DriveVLM-Dual~\cite{tian2024drivevlm} on NVIDIA Orin~\cite{JetsonModulesSupport}). Together, these strategies highlight the synergy between offline compression and runtime efficiency, driving innovation across personal, enterprise, and industrial applications.

\section{Future Directions and Open Challenges}  \label{Sec: Future Directions and Open Challenges}
The rapid deployment of LLMs on mobile edge devices offers opportunities but faces key limitations. First, transitioning from centralized servers to decentralized edge nodes brings challenges due to device heterogeneity, including variations in computational power, communication latency, and fault tolerance~\cite{lu2022turbo,jiang2021flexible}. Second, many hardware-software co-design techniques show promise in simulations but often fail in real-world heterogeneous deployments~\cite{wang2025feddfa}. Third, on-device LLMs struggle with complex scenarios requiring multi-hop reasoning, such as dynamic multi-agent interactions or real-time adaptation in personalized applications~\cite{xu2023mesen,chen2024exploring,xu2024gestureprint,yang2024maf}. These limitations hinder efficient inference and complicate the design of robust edge LLM-based systems. In the following, we explore key research directions and open challenges to address these limitations:

\textbf{Compact LLM Architecture Development. } 
Traditional Transformer-based models~\cite{vaswani2017attention} are computationally intensive and impractical for edge devices due to their high memory and processing demands. To address edge device heterogeneity and reduce resource consumption, Mamba~\cite{gu2023mamba} introduces linear computational scaling through selective mechanisms, while Jamba~\cite{lieber2024jamba} combines Transformer and Mamba layers for better adaptability. However, deploying these models on edge devices faces challenges~\cite{gu2023mamba,lieber2024jamba}: (1) their selective state space architecture demands substantial computational resources and memory bandwidth; (2) their recurrent nature inherently limits parallelization and hardware acceleration compared to transformers; and (3) optimizing them through quantization and pruning is complex due to interactions between selective state spaces and nonlinear components.

\textbf{Innovative Edge-Cloud Collaboration. } 
Cloud-edge collaboration can mitigate the limitation of insufficient memory and computing power of individual edge devices, which has seen progress with techniques like split inference~\cite{borzunov2023distributed} and speculative decoding~\cite{leviathan2023fastinferencespeculativedecoding}. However, distributed decoding strategies for edge deployment remain underexplored. Techniques such as disaggregated prefill and decoding~\cite{zhong2024distserve}, which separate tasks to optimize resource allocation, show potential for reducing inter-phase interference. However, to achieve efficient edge-cloud collaboration, it is crucial to address the challenges posed by fault tolerance and communication costs.

\textbf{Heterogeneous Deployment with Simulation Techniques.}
\modbluetext{Many hardware-software co-design approaches~\cite{tambeEdgeBERTSentenceLevelEnergy2021,wangSpAttenEfficientSparse2021,guoOliVeAcceleratingLarge2023} for edge LLM inference have been evaluated in simulation or on specific ASIC prototypes, but the limitations are Cannot be used effectively on general-purpose hardware platforms. Future research should focus on improving scalability with heterogeneous hardware platforms, including CPUs, GPUs, and NPUs~\cite{yuan2024mobilefirmware}. By optimizing for real-world scenarios, these methods can achieve faster, more energy-efficient LLM inference while addressing the challenges of dynamic workloads and diverse hardware configurations.}

\textbf{Graph-based LLM Development.}
On-device LLMs face significant limitations in handling graph-related inference tasks, which are critical for complex relationships and multi-hop reasoning in domains such as social networks, biology, and transportation. Recent approaches, including GraphRAG~\cite{edge2024graphrag} for structured data extraction, GraphGPT~\cite{tang2024graphgpt} for graph interpretation through structural knowledge alignment, and GraphWiz~\cite{chen2024graphwiz} can address the multi-hop reasoning and complex scenarios limitation by extracting and interpreting graph-based data. However, the resource constraints of edge devices limit the scalability of these methods.

\textbf{Multi-Agent Collaboration.} 
While many on-device LLM applications, such as personal assistants~\cite{wenAutoDroidLLMpoweredTask2024,zhang2024llamatouch} and companion robots~\cite{huang2023innermonologue,hong20233dllm}, excel in single-agent tasks, they often struggle to operate effectively in dynamic and complex environments. To address this limitation, collaborative intelligence among heterogeneous agents leverages dynamic orchestration frameworks to enable multi-agent coordination, enhancing task performance beyond single-model systems~\cite{ma2024hpipeparallelismheterogeneous,jain2020spatula}. However, deploying such frameworks on edge devices remains challenging due to constraints in communication bandwidth, latency, and energy efficiency~\cite{lu2023multiview,tian2024large}.

\textbf{Continual Learning and Personalization. } 
On-device LLMs encounter significant challenges in real-time adaptation and personalization, especially under resource constraints. Continual learning provides a promising approach by enabling models to dynamically adapt while addressing issues such as catastrophic forgetting~\cite{bhardwaj2022ekya,zhang2024vulcan,lu19sec}. For example, Interactive Continual Learning~\cite{qi2024interactive} utilizes ongoing interactions to enhance model performance in complex scenarios. However, the limited parameter capacity of edge models restricts their ability to acquire new knowledge, and fine-tuning on-device demands substantial computational resources, complicating efficient adaptation~\cite{khani2023recl,padmanabhan2023gemel}. 

\section{Conclusion} \label{Sec: Conclusion}
This survey provides a comprehensive review of recent advancements in enabling on-device LLMs, systematically exploring Offline Pre-Deployment Model Design Techniques, Online Runtime Inference Optimizations, and On-Device LLM-Based Applications. These components form a cohesive optimization pipeline: pre-deployment methods such as quantization and pruning create compact, efficient models; runtime techniques ensure adaptability and performance across heterogeneous environments; and diverse applications showcase the practical impact of edge LLMs. By addressing key challenges in efficiency and scalability, this survey offers valuable insights for researchers and practitioners, paving the way for accessible, sustainable AI solutions that unlock the full potential of LLMs.

\section*{Acknowledgments}
This work was supported in part by the National Natural Science Foundation of China (NSFC) under Grant 92467301, U23A20326, 62293511, and 62372414, the Key Research and Development Program of Zhejiang Province under Grant 2025C01061 and 2025C01012, and in part by the ZJUCSE-Enflame cloud and edge intelligence joint laboratory.

\bibliographystyle{ACM-Reference-Format}
\bibliography{sample-base}


\begin{thebibliography}{234}


\ifx \showCODEN    \undefined \def \showCODEN     #1{\unskip}     \fi
\ifx \showDOI      \undefined \def \showDOI       #1{#1}\fi
\ifx \showISBNx    \undefined \def \showISBNx     #1{\unskip}     \fi
\ifx \showISBNxiii \undefined \def \showISBNxiii  #1{\unskip}     \fi
\ifx \showISSN     \undefined \def \showISSN      #1{\unskip}     \fi
\ifx \showLCCN     \undefined \def \showLCCN      #1{\unskip}     \fi
\ifx \shownote     \undefined \def \shownote      #1{#1}          \fi
\ifx \showarticletitle \undefined \def \showarticletitle #1{#1}   \fi
\ifx \showURL      \undefined \def \showURL       {\relax}        \fi
\providecommand\bibfield[2]{#2}
\providecommand\bibinfo[2]{#2}
\providecommand\natexlab[1]{#1}
\providecommand\showeprint[2][]{arXiv:#2}

\bibitem[Abdin et~al\mbox{.}(2024)]%
        {abdin2024phi3}
\bibfield{author}{\bibinfo{person}{Marah Abdin}, \bibinfo{person}{Sam~Ade Jacobs}, \bibinfo{person}{Ammar~Ahmad Awan}, \bibinfo{person}{Jyoti Aneja}, \bibinfo{person}{Ahmed Awadallah}, \bibinfo{person}{Hany Awadalla}, \bibinfo{person}{Nguyen Bach}, \bibinfo{person}{Amit Bahree}, \bibinfo{person}{Arash Bakhtiari}, \bibinfo{person}{Harkirat Behl}, {et~al\mbox{.}}} \bibinfo{year}{2024}\natexlab{}.
\newblock \showarticletitle{Phi-3 technical report: A highly capable language model locally on your phone}.
\newblock \bibinfo{journal}{\emph{arXiv preprint arXiv:2404.14219}} (\bibinfo{year}{2024}).
\newblock


\bibitem[Abnar and Zuidema(2020)]%
        {abnar2020quantifying}
\bibfield{author}{\bibinfo{person}{Samira Abnar} {and} \bibinfo{person}{Willem Zuidema}.} \bibinfo{year}{2020}\natexlab{}.
\newblock \showarticletitle{Quantifying Attention Flow in Transformers}.
\newblock \bibinfo{journal}{\emph{ACL}} (\bibinfo{year}{2020}).
\newblock


\bibitem[Ainslie et~al\mbox{.}(2023)]%
        {ainslie2023gqa}
\bibfield{author}{\bibinfo{person}{Joshua Ainslie}, \bibinfo{person}{James Lee-Thorp}, \bibinfo{person}{Michiel de Jong}, \bibinfo{person}{Yury Zemlyanskiy}, \bibinfo{person}{Federico Lebron}, {and} \bibinfo{person}{Sumit Sanghai}.} \bibinfo{year}{2023}\natexlab{}.
\newblock \showarticletitle{GQA: Training Generalized Multi-Query Transformer Models from Multi-Head Checkpoints}.
\newblock \bibinfo{journal}{\emph{EMNLP}} (\bibinfo{year}{2023}).
\newblock


\bibitem[Alizadeh et~al\mbox{.}(2024)]%
        {alizadehLLMFlashEfficient2024}
\bibfield{author}{\bibinfo{person}{Keivan Alizadeh}, \bibinfo{person}{Iman Mirzadeh}, \bibinfo{person}{Dmitry Belenko}, \bibinfo{person}{Karen Khatamifard}, \bibinfo{person}{Minsik Cho}, \bibinfo{person}{Carlo~C Del~Mundo}, \bibinfo{person}{Mohammad Rastegari}, {and} \bibinfo{person}{Mehrdad Farajtabar}.} \bibinfo{year}{2024}\natexlab{}.
\newblock \showarticletitle{Llm in a flash: Efficient large language model inference with limited memory}.
\newblock \bibinfo{journal}{\emph{ACL}} (\bibinfo{year}{2024}).
\newblock


\bibitem[An et~al\mbox{.}(2024)]%
        {an2023fluctuationbasedadaptivestructuredpruning}
\bibfield{author}{\bibinfo{person}{Yongqi An}, \bibinfo{person}{Xu Zhao}, \bibinfo{person}{Tao Yu}, \bibinfo{person}{Ming Tang}, {and} \bibinfo{person}{Jinqiao Wang}.} \bibinfo{year}{2024}\natexlab{}.
\newblock \showarticletitle{Fluctuation-based Adaptive Structured Pruning for Large Language Models}.
\newblock \bibinfo{journal}{\emph{AAAI}} (\bibinfo{year}{2024}).
\newblock


\bibitem[Anand et~al\mbox{.}(2024)]%
        {IntroducingLlama3.2}
\bibfield{author}{\bibinfo{person}{Yuvanesh Anand}, \bibinfo{person}{Zach Nussbaum}, \bibinfo{person}{Brandon Duderstadt}, \bibinfo{person}{Benjamin Schmidt}, {and} \bibinfo{person}{Andriy Mulyar}.} \bibinfo{year}{2024}\natexlab{}.
\newblock \bibinfo{title}{Introducing Llama 3.2}.
\newblock \bibinfo{howpublished}{\url{https://www.llama.com/docs/model-cards-and-prompt-formats/llama3_2/}}.
\newblock
\newblock
\shownote{Accessed on December 5, 2024}.


\bibitem[Ananthanarayanan et~al\mbox{.}(2019)]%
        {ga19mobisys}
\bibfield{author}{\bibinfo{person}{Ganesh Ananthanarayanan}, \bibinfo{person}{Victor Bahl}, \bibinfo{person}{Landon Cox}, \bibinfo{person}{Alex Crown}, \bibinfo{person}{Shadi Nogbahi}, {and} \bibinfo{person}{Yuanchao Shu}.} \bibinfo{year}{2019}\natexlab{}.
\newblock \showarticletitle{{Demo: Video Analytics - Killer App for Edge Computing}}. In \bibinfo{booktitle}{\emph{ACM MobiSys}}.
\newblock


\bibitem[Apple(2023a)]%
        {appleaseries}
\bibfield{author}{\bibinfo{person}{Apple}.} \bibinfo{year}{2023}\natexlab{a}.
\newblock \bibinfo{title}{Apple debuts iPhone 15 and iPhone 15 Plus}.
\newblock \bibinfo{howpublished}{\url{https://www.apple.com/newsroom/2023/09/apple-debuts-iphone-15-and-iphone-15-plus/}}.
\newblock
\newblock
\shownote{Accessed on December 3, 2024}.


\bibitem[Apple(2023b)]%
        {applemseries}
\bibfield{author}{\bibinfo{person}{Apple}.} \bibinfo{year}{2023}\natexlab{b}.
\newblock \bibinfo{title}{Apple introduces M2 Ultra}.
\newblock \bibinfo{howpublished}{\url{https://www.apple.com/newsroom/2023/06/apple-introduces-m2-ultra/}}.
\newblock
\newblock
\shownote{Accessed on December 3, 2024}.


\bibitem[Ashkboos et~al\mbox{.}(2024)]%
        {ashkboos2024slicegptcompresslargelanguage}
\bibfield{author}{\bibinfo{person}{Saleh Ashkboos}, \bibinfo{person}{Maximilian~L. Croci}, \bibinfo{person}{Marcelo~Gennari do Nascimento}, \bibinfo{person}{Torsten Hoefler}, {and} \bibinfo{person}{James Hensman}.} \bibinfo{year}{2024}\natexlab{}.
\newblock \showarticletitle{SliceGPT: Compress Large Language Models by Deleting Rows and Columns}.
\newblock \bibinfo{journal}{\emph{ICLR}} (\bibinfo{year}{2024}).
\newblock


\bibitem[Bae et~al\mbox{.}(2023)]%
        {bae2023fastFREE}
\bibfield{author}{\bibinfo{person}{Sangmin Bae}, \bibinfo{person}{Jongwoo Ko}, \bibinfo{person}{Hwanjun Song}, {and} \bibinfo{person}{Se-Young Yun}.} \bibinfo{year}{2023}\natexlab{}.
\newblock \showarticletitle{Fast and Robust Early-Exiting Framework for Autoregressive Language Models with Synchronized Parallel Decoding}.
\newblock \bibinfo{journal}{\emph{EMNLP}} (\bibinfo{year}{2023}).
\newblock


\bibitem[Bai et~al\mbox{.}(2023)]%
        {bai2023qwen}
\bibfield{author}{\bibinfo{person}{Jinze Bai}, \bibinfo{person}{Shuai Bai}, \bibinfo{person}{Yunfei Chu}, \bibinfo{person}{Zeyu Cui}, \bibinfo{person}{Kai Dang}, \bibinfo{person}{Xiaodong Deng}, \bibinfo{person}{Yang Fan}, \bibinfo{person}{Wenbin Ge}, \bibinfo{person}{Yu Han}, \bibinfo{person}{Fei Huang}, {et~al\mbox{.}}} \bibinfo{year}{2023}\natexlab{}.
\newblock \showarticletitle{Qwen technical report}.
\newblock \bibinfo{journal}{\emph{arXiv preprint arXiv:2309.16609}} (\bibinfo{year}{2023}).
\newblock


\bibitem[Bhardwaj et~al\mbox{.}(2022)]%
        {bhardwaj2022ekya}
\bibfield{author}{\bibinfo{person}{Romil Bhardwaj}, \bibinfo{person}{Zhengxu Xia}, \bibinfo{person}{Ganesh Ananthanarayanan}, \bibinfo{person}{Junchen Jiang}, \bibinfo{person}{Yuanchao Shu}, \bibinfo{person}{Nikolaos Karianakis}, \bibinfo{person}{Kevin Hsieh}, \bibinfo{person}{Paramvir Bahl}, {and} \bibinfo{person}{Ion Stoica}.} \bibinfo{year}{2022}\natexlab{}.
\newblock \showarticletitle{Ekya: Continuous learning of video analytics models on edge compute servers}.
\newblock \bibinfo{journal}{\emph{USENIX NSDI}} (\bibinfo{year}{2022}).
\newblock


\bibitem[Biderman et~al\mbox{.}(2023)]%
        {biderman2023pythia}
\bibfield{author}{\bibinfo{person}{Stella Biderman}, \bibinfo{person}{Hailey Schoelkopf}, \bibinfo{person}{Quentin~Gregory Anthony}, \bibinfo{person}{Herbie Bradley}, \bibinfo{person}{Kyle O’Brien}, \bibinfo{person}{Eric Hallahan}, \bibinfo{person}{Mohammad~Aflah Khan}, \bibinfo{person}{Shivanshu Purohit}, \bibinfo{person}{USVSN~Sai Prashanth}, \bibinfo{person}{Edward Raff}, {et~al\mbox{.}}} \bibinfo{year}{2023}\natexlab{}.
\newblock \showarticletitle{Pythia: A suite for analyzing large language models across training and scaling}.
\newblock \bibinfo{journal}{\emph{ICML}} (\bibinfo{year}{2023}).
\newblock


\bibitem[Bondarenko et~al\mbox{.}(2021)]%
        {bondarenko2021understanding}
\bibfield{author}{\bibinfo{person}{Yelysei Bondarenko}, \bibinfo{person}{Markus Nagel}, {and} \bibinfo{person}{Tijmen Blankevoort}.} \bibinfo{year}{2021}\natexlab{}.
\newblock \showarticletitle{Understanding and Overcoming the Challenges of Efficient Transformer Quantization}.
\newblock \bibinfo{journal}{\emph{EMNLP}} (\bibinfo{year}{2021}).
\newblock


\bibitem[Borzunov et~al\mbox{.}(2023a)]%
        {borzunovPetalsCollaborativeInference2023}
\bibfield{author}{\bibinfo{person}{Alexander Borzunov}, \bibinfo{person}{Dmitry Baranchuk}, \bibinfo{person}{Tim Dettmers}, \bibinfo{person}{Maksim Riabinin}, \bibinfo{person}{Younes Belkada}, \bibinfo{person}{Artem Chumachenko}, \bibinfo{person}{Pavel Samygin}, {and} \bibinfo{person}{Colin Raffel}.} \bibinfo{year}{2023}\natexlab{a}.
\newblock \showarticletitle{Petals: Collaborative Inference and Fine-tuning of Large Models}.
\newblock \bibinfo{journal}{\emph{ACL}} (\bibinfo{year}{2023}).
\newblock


\bibitem[Borzunov et~al\mbox{.}(2023b)]%
        {borzunov2023distributed}
\bibfield{author}{\bibinfo{person}{Alexander Borzunov}, \bibinfo{person}{Max Ryabinin}, \bibinfo{person}{Artem Chumachenko}, \bibinfo{person}{Dmitry Baranchuk}, \bibinfo{person}{Tim Dettmers}, \bibinfo{person}{Younes Belkada}, \bibinfo{person}{Pavel Samygin}, {and} \bibinfo{person}{Colin Raffel}.} \bibinfo{year}{2023}\natexlab{b}.
\newblock \showarticletitle{Distributed inference and fine-tuning of large language models over the internet}.
\newblock \bibinfo{journal}{\emph{NeurIPS}} (\bibinfo{year}{2023}).
\newblock


\bibitem[Brown et~al\mbox{.}(2020)]%
        {brown2020languageModelsAreFewShotLearners}
\bibfield{author}{\bibinfo{person}{Tom Brown}, \bibinfo{person}{Benjamin Mann}, \bibinfo{person}{Nick Ryder}, \bibinfo{person}{Melanie Subbiah}, \bibinfo{person}{Jared~D Kaplan}, \bibinfo{person}{Prafulla Dhariwal}, \bibinfo{person}{Arvind Neelakantan}, \bibinfo{person}{Pranav Shyam}, \bibinfo{person}{Girish Sastry}, \bibinfo{person}{Amanda Askell}, {et~al\mbox{.}}} \bibinfo{year}{2020}\natexlab{}.
\newblock \showarticletitle{Language models are few-shot learners}.
\newblock \bibinfo{journal}{\emph{NeurIPS}} (\bibinfo{year}{2020}).
\newblock


\bibitem[Cao et~al\mbox{.}(2022)]%
        {caoMobiVQAEfficientOnDevice2022}
\bibfield{author}{\bibinfo{person}{Qingqing Cao}, \bibinfo{person}{Prerna Khanna}, \bibinfo{person}{Nicholas~D Lane}, {and} \bibinfo{person}{Aruna Balasubramanian}.} \bibinfo{year}{2022}\natexlab{}.
\newblock \showarticletitle{MobiVQA: Efficient On-Device Visual Question Answering}.
\newblock \bibinfo{journal}{\emph{ACM UbiComp}} (\bibinfo{year}{2022}).
\newblock


\bibitem[Chan et~al\mbox{.}(2024)]%
        {chan2024chateval}
\bibfield{author}{\bibinfo{person}{Chi-Min Chan}, \bibinfo{person}{Weize Chen}, \bibinfo{person}{Yusheng Su}, \bibinfo{person}{Jianxuan Yu}, \bibinfo{person}{Wei Xue}, \bibinfo{person}{Shanghang Zhang}, \bibinfo{person}{Jie Fu}, {and} \bibinfo{person}{Zhiyuan Liu}.} \bibinfo{year}{2024}\natexlab{}.
\newblock \showarticletitle{ChatEval: Towards Better {LLM}-based Evaluators through Multi-Agent Debate}.
\newblock \bibinfo{journal}{\emph{ICLR}} (\bibinfo{year}{2024}).
\newblock


\bibitem[Chavan et~al\mbox{.}(2024)]%
        {chavanSurgicalFeatureSpaceDecomposition2024}
\bibfield{author}{\bibinfo{person}{Arnav Chavan}, \bibinfo{person}{Nahush Lele}, {and} \bibinfo{person}{Deepak Gupta}.} \bibinfo{year}{2024}\natexlab{}.
\newblock \showarticletitle{Surgical Feature-Space Decomposition of LLMs: Why, When and How?}
\newblock \bibinfo{journal}{\emph{ACL}} (\bibinfo{year}{2024}).
\newblock


\bibitem[Chen et~al\mbox{.}(2023b)]%
        {chenMCCKDMultiCoTConsistent2023}
\bibfield{author}{\bibinfo{person}{Hongzhan Chen}, \bibinfo{person}{Siyue Wu}, \bibinfo{person}{Xiaojun Quan}, \bibinfo{person}{Rui Wang}, \bibinfo{person}{Ming Yan}, {and} \bibinfo{person}{Ji Zhang}.} \bibinfo{year}{2023}\natexlab{b}.
\newblock \showarticletitle{MCC-KD: Multi-CoT Consistent Knowledge Distillation}.
\newblock \bibinfo{journal}{\emph{EMNLP}} (\bibinfo{year}{2023}).
\newblock


\bibitem[Chen et~al\mbox{.}(2024b)]%
        {chen2024drivingWithLLMs}
\bibfield{author}{\bibinfo{person}{Long Chen}, \bibinfo{person}{Oleg Sinavski}, \bibinfo{person}{Jan H{\"u}nermann}, \bibinfo{person}{Alice Karnsund}, \bibinfo{person}{Andrew~James Willmott}, \bibinfo{person}{Danny Birch}, \bibinfo{person}{Daniel Maund}, {and} \bibinfo{person}{Jamie Shotton}.} \bibinfo{year}{2024}\natexlab{b}.
\newblock \showarticletitle{Driving with llms: Fusing object-level vector modality for explainable autonomous driving}.
\newblock \bibinfo{journal}{\emph{IEEE ICRA}} (\bibinfo{year}{2024}).
\newblock


\bibitem[Chen et~al\mbox{.}(2024a)]%
        {chen2024graphwiz}
\bibfield{author}{\bibinfo{person}{Nuo Chen}, \bibinfo{person}{Yuhan Li}, \bibinfo{person}{Jianheng Tang}, {and} \bibinfo{person}{Jia Li}.} \bibinfo{year}{2024}\natexlab{a}.
\newblock \showarticletitle{Graphwiz: An instruction-following language model for graph computational problems}.
\newblock \bibinfo{journal}{\emph{ACM SIGKDD}} (\bibinfo{year}{2024}), \bibinfo{pages}{353--364}.
\newblock


\bibitem[Chen et~al\mbox{.}(2021)]%
        {chenDRONEDataawareLowrank2021}
\bibfield{author}{\bibinfo{person}{Patrick Chen}, \bibinfo{person}{Hsiang-Fu Yu}, \bibinfo{person}{Inderjit Dhillon}, {and} \bibinfo{person}{Cho-Jui Hsieh}.} \bibinfo{year}{2021}\natexlab{}.
\newblock \showarticletitle{DRONE: Data-aware Low-rank Compression for Large NLP Models}.
\newblock \bibinfo{journal}{\emph{NeurIPS}} (\bibinfo{year}{2021}).
\newblock


\bibitem[Chen et~al\mbox{.}(2024c)]%
        {chen2024exploring}
\bibfield{author}{\bibinfo{person}{Tao Chen}, \bibinfo{person}{Yongjie Yang}, \bibinfo{person}{Xiaoran Fan}, \bibinfo{person}{Xiuzhen Guo}, \bibinfo{person}{Jie Xiong}, {and} \bibinfo{person}{Longfei Shangguan}.} \bibinfo{year}{2024}\natexlab{c}.
\newblock \showarticletitle{Exploring the Feasibility of Remote Cardiac Auscultation Using Earphones}.
\newblock \bibinfo{journal}{\emph{ACM MobiCom}} (\bibinfo{year}{2024}).
\newblock


\bibitem[Chen et~al\mbox{.}(2023c)]%
        {chen2023confidant}
\bibfield{author}{\bibinfo{person}{Yuhao Chen}, \bibinfo{person}{Yuxuan Yan}, \bibinfo{person}{Qianqian Yang}, \bibinfo{person}{Yuanchao Shu}, \bibinfo{person}{Shibo He}, {and} \bibinfo{person}{Jiming Chen}.} \bibinfo{year}{2023}\natexlab{c}.
\newblock \showarticletitle{Confidant: Customizing Transformer-based LLMs via Collaborative Edge Training}.
\newblock \bibinfo{journal}{\emph{arXiv preprint arXiv:2311.13381}} (\bibinfo{year}{2023}).
\newblock


\bibitem[Chen et~al\mbox{.}(2023a)]%
        {chenDISCODistillingCounterfactuals2023}
\bibfield{author}{\bibinfo{person}{Zeming Chen}, \bibinfo{person}{Qiyue Gao}, \bibinfo{person}{Antoine Bosselut}, \bibinfo{person}{Ashish Sabharwal}, {and} \bibinfo{person}{Kyle Richardson}.} \bibinfo{year}{2023}\natexlab{a}.
\newblock \showarticletitle{DISCO: Distilling Counterfactuals with Large Language Models}.
\newblock \bibinfo{journal}{\emph{ACL}} (\bibinfo{year}{2023}).
\newblock


\bibitem[Chevalier et~al\mbox{.}(2023)]%
        {chevalier2023adapting}
\bibfield{author}{\bibinfo{person}{Alexis Chevalier}, \bibinfo{person}{Alexander Wettig}, \bibinfo{person}{Anirudh Ajith}, {and} \bibinfo{person}{Danqi Chen}.} \bibinfo{year}{2023}\natexlab{}.
\newblock \showarticletitle{Adapting Language Models to Compress Contexts}.
\newblock \bibinfo{journal}{\emph{EMNLP}} (\bibinfo{year}{2023}).
\newblock


\bibitem[CPU-Monkey(2024)]%
        {AIPerformanceNPU}
\bibfield{author}{\bibinfo{person}{CPU-Monkey}.} \bibinfo{year}{2024}\natexlab{}.
\newblock \bibinfo{title}{AI Performance (NPU) CPU Benchmark List}.
\newblock \bibinfo{howpublished}{\url{https://www.cpu-monkey.com/en/cpu\_benchmark-ai\_benchmark}}.
\newblock
\newblock
\shownote{Accessed on July 18, 2024}.


\bibitem[Dai et~al\mbox{.}(2025)]%
        {dai25hotmobile}
\bibfield{author}{\bibinfo{person}{Yubin Dai}, \bibinfo{person}{Bin Qian}, \bibinfo{person}{Yangkun Liu}, \bibinfo{person}{Yuxuan Yan}, {and} \bibinfo{person}{Yuanchao Shu}.} \bibinfo{year}{2025}\natexlab{}.
\newblock \showarticletitle{{Eros: Real-time Dense Mapping Made Easy on Mobile Devices}}. In \bibinfo{booktitle}{\emph{ACM HotMobile}}.
\newblock


\bibitem[Dettmers et~al\mbox{.}(2022)]%
        {dettmersLLMInt88bit}
\bibfield{author}{\bibinfo{person}{Tim Dettmers}, \bibinfo{person}{Mike Lewis}, \bibinfo{person}{Younes Belkada}, {and} \bibinfo{person}{Luke Zettlemoyer}.} \bibinfo{year}{2022}\natexlab{}.
\newblock \showarticletitle{LLM. int8 () 8-bit matrix multiplication for transformers at scale}.
\newblock \bibinfo{journal}{\emph{NeurIPS}} (\bibinfo{year}{2022}).
\newblock


\bibitem[Dettmers et~al\mbox{.}(2023)]%
        {dettmersSpQRSparseQuantizedRepresentation2023}
\bibfield{author}{\bibinfo{person}{Tim Dettmers}, \bibinfo{person}{Ruslan~A. Svirschevski}, \bibinfo{person}{Vage Egiazarian}, \bibinfo{person}{Denis Kuznedelev}, \bibinfo{person}{Elias Frantar}, \bibinfo{person}{Saleh Ashkboos}, \bibinfo{person}{Alexander Borzunov}, \bibinfo{person}{Torsten Hoefler}, {and} \bibinfo{person}{Dan Alistarh}.} \bibinfo{year}{2023}\natexlab{}.
\newblock \showarticletitle{SpQR: A Sparse-Quantized Representation for Near-Lossless LLM Weight Compression}.
\newblock \bibinfo{journal}{\emph{ICLR}} (\bibinfo{year}{2023}).
\newblock


\bibitem[Developer(2024)]%
        {JetsonModulesSupport}
\bibfield{author}{\bibinfo{person}{NVIDIA Developer}.} \bibinfo{year}{2024}\natexlab{}.
\newblock \bibinfo{title}{Jetson Modules, Support, Ecosystem, and Lineup}.
\newblock \bibinfo{howpublished}{\url{https://developer.nvidia.com/embedded/jetson-modules}}.
\newblock
\newblock
\shownote{Accessed on July 16, 2024}.


\bibitem[Dong et~al\mbox{.}(2022)]%
        {dong2022surveyNaturalLanguageGeneration}
\bibfield{author}{\bibinfo{person}{Chenhe Dong}, \bibinfo{person}{Yinghui Li}, \bibinfo{person}{Haifan Gong}, \bibinfo{person}{Miaoxin Chen}, \bibinfo{person}{Junxin Li}, \bibinfo{person}{Ying Shen}, {and} \bibinfo{person}{Min Yang}.} \bibinfo{year}{2022}\natexlab{}.
\newblock \showarticletitle{A survey of natural language generation}.
\newblock \bibinfo{journal}{\emph{Comput. Surveys}} (\bibinfo{year}{2022}).
\newblock


\bibitem[Dong et~al\mbox{.}(2019)]%
        {dong2019naturalLanguageUnderstandingGeneration}
\bibfield{author}{\bibinfo{person}{Li Dong}, \bibinfo{person}{Nan Yang}, \bibinfo{person}{Wenhui Wang}, \bibinfo{person}{Furu Wei}, \bibinfo{person}{Xiaodong Liu}, \bibinfo{person}{Yu Wang}, \bibinfo{person}{Jianfeng Gao}, \bibinfo{person}{Ming Zhou}, {and} \bibinfo{person}{Hsiao-Wuen Hon}.} \bibinfo{year}{2019}\natexlab{}.
\newblock \showarticletitle{Unified language model pre-training for natural language understanding and generation}.
\newblock \bibinfo{journal}{\emph{NeurIPS}} (\bibinfo{year}{2019}).
\newblock


\bibitem[Du et~al\mbox{.}(2024)]%
        {du2024distributed}
\bibfield{author}{\bibinfo{person}{Jun Du}, \bibinfo{person}{Tianyi Lin}, \bibinfo{person}{Chunxiao Jiang}, \bibinfo{person}{Qianqian Yang}, \bibinfo{person}{C~Faouzi Bader}, {and} \bibinfo{person}{Zhu Han}.} \bibinfo{year}{2024}\natexlab{}.
\newblock \showarticletitle{Distributed Foundation Models for Multi-Modal Learning in 6G Wireless Networks}.
\newblock \bibinfo{journal}{\emph{IEEE Wireless Communications}} (\bibinfo{year}{2024}).
\newblock


\bibitem[Dubey et~al\mbox{.}(2024)]%
        {dubey2024llama3herd}
\bibfield{author}{\bibinfo{person}{Abhimanyu Dubey}, \bibinfo{person}{Abhinav Jauhri}, \bibinfo{person}{Abhinav Pandey}, \bibinfo{person}{Abhishek Kadian}, \bibinfo{person}{Ahmad Al-Dahle}, \bibinfo{person}{Aiesha Letman}, \bibinfo{person}{Akhil Mathur}, \bibinfo{person}{Alan Schelten}, \bibinfo{person}{Amy Yang}, \bibinfo{person}{Angela Fan}, {et~al\mbox{.}}} \bibinfo{year}{2024}\natexlab{}.
\newblock \showarticletitle{The llama 3 herd of models}.
\newblock \bibinfo{journal}{\emph{arXiv preprint arXiv:2407.21783}} (\bibinfo{year}{2024}).
\newblock


\bibitem[Edge et~al\mbox{.}(2024)]%
        {edge2024graphrag}
\bibfield{author}{\bibinfo{person}{Darren Edge}, \bibinfo{person}{Ha Trinh}, \bibinfo{person}{Newman Cheng}, \bibinfo{person}{Joshua Bradley}, \bibinfo{person}{Alex Chao}, \bibinfo{person}{Apurva Mody}, \bibinfo{person}{Steven Truitt}, {and} \bibinfo{person}{Jonathan Larson}.} \bibinfo{year}{2024}\natexlab{}.
\newblock \showarticletitle{From local to global: A graph rag approach to query-focused summarization}.
\newblock \bibinfo{journal}{\emph{arXiv preprint arXiv:2404.16130}} (\bibinfo{year}{2024}).
\newblock


\bibitem[Egiazarian et~al\mbox{.}(2024)]%
        {egiazarianextreme}
\bibfield{author}{\bibinfo{person}{Vage Egiazarian}, \bibinfo{person}{Andrei Panferov}, \bibinfo{person}{Denis Kuznedelev}, \bibinfo{person}{Elias Frantar}, \bibinfo{person}{Artem Babenko}, {and} \bibinfo{person}{Dan Alistarh}.} \bibinfo{year}{2024}\natexlab{}.
\newblock \showarticletitle{Extreme Compression of Large Language Models via Additive Quantization}.
\newblock \bibinfo{journal}{\emph{ICML}} (\bibinfo{year}{2024}).
\newblock


\bibitem[Exyno(2024)]%
        {samsungExynos2400news}
\bibfield{author}{\bibinfo{person}{Samsung Exyno}.} \bibinfo{year}{2024}\natexlab{}.
\newblock \bibinfo{title}{Exynos 2400 with 10-core CPU gets detailed after Galaxy S24 launch}.
\newblock \bibinfo{howpublished}{\url{https://www.sammobile.com/news/exynos-2400-10-core-cpu-amd-rdna3-xclipse-940-gpu-specs-detailed/}}.
\newblock
\newblock
\shownote{Accessed on December 11, 2024}.


\bibitem[Fan et~al\mbox{.}(2023)]%
        {fanTaskFusionEfficientTransfer2023}
\bibfield{author}{\bibinfo{person}{Zichen Fan}, \bibinfo{person}{Qirui Zhang}, \bibinfo{person}{Pierre Abillama}, \bibinfo{person}{Sara Shoouri}, \bibinfo{person}{Changwoo Lee}, \bibinfo{person}{David Blaauw}, \bibinfo{person}{Hun-Seok Kim}, {and} \bibinfo{person}{Dennis Sylvester}.} \bibinfo{year}{2023}\natexlab{}.
\newblock \showarticletitle{TaskFusion: An Efficient Transfer Learning Architecture with Dual Delta Sparsity for Multi-Task Natural Language Processing}.
\newblock \bibinfo{journal}{\emph{ACM/IEEE ISCA}} (\bibinfo{year}{2023}).
\newblock


\bibitem[Frantar and Alistarh(2022)]%
        {frantarOptimalBrainCompression2022}
\bibfield{author}{\bibinfo{person}{Elias Frantar} {and} \bibinfo{person}{Dan Alistarh}.} \bibinfo{year}{2022}\natexlab{}.
\newblock \showarticletitle{Optimal Brain Compression: A Framework for Accurate Post-Training Quantization and Pruning}.
\newblock \bibinfo{journal}{\emph{NeurIPS}} (\bibinfo{year}{2022}).
\newblock


\bibitem[Frantar and Alistarh(2023)]%
        {frantarSparseGPTMassiveLanguage2023}
\bibfield{author}{\bibinfo{person}{Elias Frantar} {and} \bibinfo{person}{Dan Alistarh}.} \bibinfo{year}{2023}\natexlab{}.
\newblock \showarticletitle{SparseGPT: Massive Language Models Can Be Accurately Pruned in One-Shot}.
\newblock \bibinfo{journal}{\emph{ICML}} (\bibinfo{year}{2023}).
\newblock


\bibitem[Frantar et~al\mbox{.}(2022)]%
        {frantarOPTQAccurateQuantization2022}
\bibfield{author}{\bibinfo{person}{Elias Frantar}, \bibinfo{person}{Saleh Ashkboos}, \bibinfo{person}{Torsten Hoefler}, {and} \bibinfo{person}{Dan Alistarh}.} \bibinfo{year}{2022}\natexlab{}.
\newblock \showarticletitle{OPTQ: Accurate Quantization for Generative Pre-trained Transformers}.
\newblock \bibinfo{journal}{\emph{ICLR}} (\bibinfo{year}{2022}).
\newblock


\bibitem[GeForce(2023a)]%
        {nvidiaLaptopCompare}
\bibfield{author}{\bibinfo{person}{NVIDIA GeForce}.} \bibinfo{year}{2023}\natexlab{a}.
\newblock \bibinfo{title}{Compare Gaming Laptops: GeForce RTX 40 Series}.
\newblock \bibinfo{howpublished}{\url{https://www.nvidia.com/en-us/geforce/laptops/compare/}}.
\newblock
\newblock
\shownote{Accessed on December 6, 2024}.


\bibitem[GeForce(2023b)]%
        {nvidiaGeForceGraphicsCards}
\bibfield{author}{\bibinfo{person}{NVIDIA GeForce}.} \bibinfo{year}{2023}\natexlab{b}.
\newblock \bibinfo{title}{Compare GeForce Graphics Cards}.
\newblock \bibinfo{howpublished}{\url{https://www.nvidia.com/en-us/geforce/graphics-cards/compare/}}.
\newblock
\newblock
\shownote{Accessed on December 6, 2024}.


\bibitem[Geng et~al\mbox{.}(2021)]%
        {gengattention}
\bibfield{author}{\bibinfo{person}{Zhengyang Geng}, \bibinfo{person}{Meng-Hao Guo}, \bibinfo{person}{Hongxu Chen}, \bibinfo{person}{Xia Li}, \bibinfo{person}{Ke Wei}, {and} \bibinfo{person}{Zhouchen Lin}.} \bibinfo{year}{2021}\natexlab{}.
\newblock \showarticletitle{Is Attention Better Than Matrix Decomposition?}
\newblock \bibinfo{journal}{\emph{ICLR}} (\bibinfo{year}{2021}).
\newblock


\bibitem[Gerganov(2023)]%
        {llamacpp}
\bibfield{author}{\bibinfo{person}{Georgi Gerganov}.} \bibinfo{year}{2023}\natexlab{}.
\newblock \bibinfo{title}{llama.cpp}.
\newblock \bibinfo{howpublished}{\url{https://github.com/ggerganov/llama.cpp}}.
\newblock
\newblock
\shownote{Accessed on June 8, 2024}.


\bibitem[Google(2024)]%
        {googleTensorG4}
\bibfield{author}{\bibinfo{person}{Google}.} \bibinfo{year}{2024}\natexlab{}.
\newblock \bibinfo{title}{Google Tensor: the brains behind Pixel phones.}
\newblock \bibinfo{howpublished}{\url{https://store.google.com/intl/en_in/ideas/articles/google-tensor-pixel-smartphone/}}.
\newblock
\newblock
\shownote{Accessed on December 6, 2024}.


\bibitem[Goyal et~al\mbox{.}(2020)]%
        {goyal2020powerBERT}
\bibfield{author}{\bibinfo{person}{Saurabh Goyal}, \bibinfo{person}{Anamitra~Roy Choudhury}, \bibinfo{person}{Saurabh Raje}, \bibinfo{person}{Venkatesan Chakaravarthy}, \bibinfo{person}{Yogish Sabharwal}, {and} \bibinfo{person}{Ashish Verma}.} \bibinfo{year}{2020}\natexlab{}.
\newblock \showarticletitle{Power-bert: Accelerating bert inference via progressive word-vector elimination}.
\newblock \bibinfo{journal}{\emph{ICML}} (\bibinfo{year}{2020}).
\newblock


\bibitem[Gu and Dao(2023)]%
        {gu2023mamba}
\bibfield{author}{\bibinfo{person}{Albert Gu} {and} \bibinfo{person}{Tri Dao}.} \bibinfo{year}{2023}\natexlab{}.
\newblock \showarticletitle{Mamba: Linear-time sequence modeling with selective state spaces}.
\newblock \bibinfo{journal}{\emph{arXiv preprint arXiv:2312.00752}} (\bibinfo{year}{2023}).
\newblock


\bibitem[Gu et~al\mbox{.}(2023)]%
        {guMiniLLMKnowledgeDistillation2023}
\bibfield{author}{\bibinfo{person}{Yuxian Gu}, \bibinfo{person}{Li Dong}, \bibinfo{person}{Furu Wei}, {and} \bibinfo{person}{Minlie Huang}.} \bibinfo{year}{2023}\natexlab{}.
\newblock \showarticletitle{MiniLLM: Knowledge Distillation of Large Language Models}.
\newblock \bibinfo{journal}{\emph{ICLR}} (\bibinfo{year}{2023}).
\newblock


\bibitem[Gu et~al\mbox{.}(2024)]%
        {gu2024anomalygpt}
\bibfield{author}{\bibinfo{person}{Zhaopeng Gu}, \bibinfo{person}{Bingke Zhu}, \bibinfo{person}{Guibo Zhu}, \bibinfo{person}{Yingying Chen}, \bibinfo{person}{Ming Tang}, {and} \bibinfo{person}{Jinqiao Wang}.} \bibinfo{year}{2024}\natexlab{}.
\newblock \showarticletitle{Anomalygpt: Detecting industrial anomalies using large vision-language models}.
\newblock \bibinfo{journal}{\emph{AAAI}} (\bibinfo{year}{2024}).
\newblock


\bibitem[Gunasekar et~al\mbox{.}(2023)]%
        {gunasekar2023textbooks}
\bibfield{author}{\bibinfo{person}{Suriya Gunasekar}, \bibinfo{person}{Yi Zhang}, \bibinfo{person}{Jyoti Aneja}, \bibinfo{person}{Caio C{\'e}sar~Teodoro Mendes}, \bibinfo{person}{Allie Del~Giorno}, \bibinfo{person}{Sivakanth Gopi}, \bibinfo{person}{Mojan Javaheripi}, \bibinfo{person}{Piero Kauffmann}, \bibinfo{person}{Gustavo de Rosa}, \bibinfo{person}{Olli Saarikivi}, {et~al\mbox{.}}} \bibinfo{year}{2023}\natexlab{}.
\newblock \showarticletitle{Textbooks are all you need}.
\newblock \bibinfo{journal}{\emph{arXiv preprint arXiv:2306.11644}} (\bibinfo{year}{2023}).
\newblock


\bibitem[Guo et~al\mbox{.}(2023b)]%
        {guoOliVeAcceleratingLarge2023}
\bibfield{author}{\bibinfo{person}{Cong Guo}, \bibinfo{person}{Jiaming Tang}, \bibinfo{person}{Weiming Hu}, \bibinfo{person}{Jingwen Leng}, \bibinfo{person}{Chen Zhang}, \bibinfo{person}{Fan Yang}, \bibinfo{person}{Yunxin Liu}, \bibinfo{person}{Minyi Guo}, {and} \bibinfo{person}{Yuhao Zhu}.} \bibinfo{year}{2023}\natexlab{b}.
\newblock \showarticletitle{OliVe: Accelerating Large Language Models via Hardware-friendly Outlier-Victim Pair Quantization}.
\newblock \bibinfo{journal}{\emph{ACM/IEEE ISCA}} (\bibinfo{year}{2023}).
\newblock


\bibitem[Guo et~al\mbox{.}(2022)]%
        {guo2022ant}
\bibfield{author}{\bibinfo{person}{Cong Guo}, \bibinfo{person}{Chen Zhang}, \bibinfo{person}{Jingwen Leng}, \bibinfo{person}{Zihan Liu}, \bibinfo{person}{Fan Yang}, \bibinfo{person}{Yunxin Liu}, \bibinfo{person}{Minyi Guo}, {and} \bibinfo{person}{Yuhao Zhu}.} \bibinfo{year}{2022}\natexlab{}.
\newblock \showarticletitle{Ant: Exploiting adaptive numerical data type for low-bit deep neural network quantization}.
\newblock \bibinfo{journal}{\emph{IEEE/ACM MICRO}} (\bibinfo{year}{2022}).
\newblock


\bibitem[Guo et~al\mbox{.}(2025)]%
        {guo2025deepseekr1}
\bibfield{author}{\bibinfo{person}{Daya Guo}, \bibinfo{person}{Dejian Yang}, \bibinfo{person}{Haowei Zhang}, \bibinfo{person}{Junxiao Song}, \bibinfo{person}{Ruoyu Zhang}, \bibinfo{person}{Runxin Xu}, \bibinfo{person}{Qihao Zhu}, \bibinfo{person}{Shirong Ma}, \bibinfo{person}{Peiyi Wang}, \bibinfo{person}{Xiao Bi}, {et~al\mbox{.}}} \bibinfo{year}{2025}\natexlab{}.
\newblock \showarticletitle{Deepseek-r1: Incentivizing reasoning capability in llms via reinforcement learning}.
\newblock \bibinfo{journal}{\emph{arXiv preprint arXiv:2501.12948}} (\bibinfo{year}{2025}).
\newblock


\bibitem[Guo et~al\mbox{.}(2023a)]%
        {guoSTITurbochargeNLP2023}
\bibfield{author}{\bibinfo{person}{Liwei Guo}, \bibinfo{person}{Wonkyo Choe}, {and} \bibinfo{person}{Felix~Xiaozhu Lin}.} \bibinfo{year}{2023}\natexlab{a}.
\newblock \showarticletitle{STI: Turbocharge NLP Inference at the Edge via Elastic Pipelining}.
\newblock \bibinfo{journal}{\emph{ACM ASPLOS}} (\bibinfo{year}{2023}).
\newblock


\bibitem[Guo et~al\mbox{.}(2024)]%
        {guo2024easter}
\bibfield{author}{\bibinfo{person}{Xiaotian Guo}, \bibinfo{person}{Quan Jiang}, \bibinfo{person}{Yixian Shen}, \bibinfo{person}{Andy~D Pimentel}, {and} \bibinfo{person}{Todor Stefanov}.} \bibinfo{year}{2024}\natexlab{}.
\newblock \showarticletitle{EASTER: Learning to Split Transformers at the Edge Robustly}.
\newblock \bibinfo{journal}{\emph{IEEE Transactions on Computer-Aided Design of Integrated Circuits and Systems}} (\bibinfo{year}{2024}).
\newblock


\bibitem[Gur et~al\mbox{.}(2024)]%
        {gurrealWebAgent}
\bibfield{author}{\bibinfo{person}{Izzeddin Gur}, \bibinfo{person}{Hiroki Furuta}, \bibinfo{person}{Austin~V Huang}, \bibinfo{person}{Mustafa Safdari}, \bibinfo{person}{Yutaka Matsuo}, \bibinfo{person}{Douglas Eck}, {and} \bibinfo{person}{Aleksandra Faust}.} \bibinfo{year}{2024}\natexlab{}.
\newblock \showarticletitle{A Real-World WebAgent with Planning, Long Context Understanding, and Program Synthesis}.
\newblock \bibinfo{journal}{\emph{ICLR}} (\bibinfo{year}{2024}).
\newblock


\bibitem[Hannun et~al\mbox{.}(2023)]%
        {mlx2023}
\bibfield{author}{\bibinfo{person}{Awni Hannun}, \bibinfo{person}{Jagrit Digani}, \bibinfo{person}{Angelos Katharopoulos}, {and} \bibinfo{person}{Ronan Collobert}.} \bibinfo{year}{2023}\natexlab{}.
\newblock \bibinfo{title}{MLX: Efficient and flexible machine learning on Apple silicon}.
\newblock \bibinfo{howpublished}{\url{https://github.com/ml-explore}}.
\newblock
\newblock
\shownote{Accessed on June 8, 2024}.


\bibitem[He et~al\mbox{.}(2024)]%
        {he2024largelanguagemodelsinference}
\bibfield{author}{\bibinfo{person}{Ying He}, \bibinfo{person}{Jingcheng Fang}, \bibinfo{person}{F~Richard Yu}, {and} \bibinfo{person}{Victor~C Leung}.} \bibinfo{year}{2024}\natexlab{}.
\newblock \showarticletitle{Large Language Models (LLMs) Inference Offloading and Resource Allocation in Cloud-Edge Computing: An Active Inference Approach}.
\newblock \bibinfo{journal}{\emph{IEEE Transactions on Mobile Computing}} (\bibinfo{year}{2024}).
\newblock


\bibitem[Hedderich et~al\mbox{.}(2021)]%
        {hedderichSurveyRecentApproaches2021}
\bibfield{author}{\bibinfo{person}{Michael~A. Hedderich}, \bibinfo{person}{Lukas Lange}, \bibinfo{person}{Heike Adel}, \bibinfo{person}{Jannik Str{\"o}tgen}, {and} \bibinfo{person}{Dietrich Klakow}.} \bibinfo{year}{2021}\natexlab{}.
\newblock \showarticletitle{A Survey on Recent Approaches for Natural Language Processing in Low-Resource Scenarios}.
\newblock \bibinfo{journal}{\emph{NAACL}} (\bibinfo{year}{2021}).
\newblock


\bibitem[Ho et~al\mbox{.}(2023)]%
        {hoLargeLanguageModels2023}
\bibfield{author}{\bibinfo{person}{Namgyu Ho}, \bibinfo{person}{Laura Schmid}, {and} \bibinfo{person}{Se-Young Yun}.} \bibinfo{year}{2023}\natexlab{}.
\newblock \showarticletitle{Large Language Models Are Reasoning Teachers}.
\newblock \bibinfo{journal}{\emph{ACL}} (\bibinfo{year}{2023}).
\newblock


\bibitem[Hong et~al\mbox{.}(2023)]%
        {hong20233dllm}
\bibfield{author}{\bibinfo{person}{Yining Hong}, \bibinfo{person}{Haoyu Zhen}, \bibinfo{person}{Peihao Chen}, \bibinfo{person}{Shuhong Zheng}, \bibinfo{person}{Yilun Du}, \bibinfo{person}{Zhenfang Chen}, {and} \bibinfo{person}{Chuang Gan}.} \bibinfo{year}{2023}\natexlab{}.
\newblock \showarticletitle{3d-llm: Injecting the 3d world into large language models}.
\newblock \bibinfo{journal}{\emph{NeurIPS}} (\bibinfo{year}{2023}).
\newblock


\bibitem[Hossain et~al\mbox{.}(2024)]%
        {hossain2024deepToggle}
\bibfield{author}{\bibinfo{person}{Soneya~Binta Hossain}, \bibinfo{person}{Nan Jiang}, \bibinfo{person}{Qiang Zhou}, \bibinfo{person}{Xiaopeng Li}, \bibinfo{person}{Wen-Hao Chiang}, \bibinfo{person}{Yingjun Lyu}, \bibinfo{person}{Hoan Nguyen}, {and} \bibinfo{person}{Omer Tripp}.} \bibinfo{year}{2024}\natexlab{}.
\newblock \showarticletitle{A deep dive into large language models for automated bug localization and repair}.
\newblock \bibinfo{journal}{\emph{Proceedings of the ACM on Software Engineering}} (\bibinfo{year}{2024}).
\newblock


\bibitem[Hsieh et~al\mbox{.}(2023)]%
        {hsiehDistillingStepbyStepOutperforming2023}
\bibfield{author}{\bibinfo{person}{Cheng-Yu Hsieh}, \bibinfo{person}{Chun-Liang Li}, \bibinfo{person}{Chih-kuan Yeh}, \bibinfo{person}{Hootan Nakhost}, \bibinfo{person}{Yasuhisa Fujii}, \bibinfo{person}{Alex Ratner}, \bibinfo{person}{Ranjay Krishna}, \bibinfo{person}{Chen-Yu Lee}, {and} \bibinfo{person}{Tomas Pfister}.} \bibinfo{year}{2023}\natexlab{}.
\newblock \showarticletitle{Distilling Step-by-Step! Outperforming Larger Language Models with Less Training Data and Smaller Model Sizes}.
\newblock \bibinfo{journal}{\emph{Findings of ACL}} (\bibinfo{year}{2023}).
\newblock


\bibitem[Hsu et~al\mbox{.}(2022)]%
        {hsuLanguageModelCompression2021}
\bibfield{author}{\bibinfo{person}{Yen-Chang Hsu}, \bibinfo{person}{Ting Hua}, \bibinfo{person}{Sungen Chang}, \bibinfo{person}{Qian Lou}, \bibinfo{person}{Yilin Shen}, {and} \bibinfo{person}{Hongxia Jin}.} \bibinfo{year}{2022}\natexlab{}.
\newblock \showarticletitle{Language Model Compression with Weighted Low-Rank Factorization}.
\newblock \bibinfo{journal}{\emph{ICLR}} (\bibinfo{year}{2022}).
\newblock


\bibitem[Hu and Li(2024)]%
        {hu2024edge}
\bibfield{author}{\bibinfo{person}{Chenghao Hu} {and} \bibinfo{person}{Baochun Li}.} \bibinfo{year}{2024}\natexlab{}.
\newblock \showarticletitle{When the Edge Meets Transformers: Distributed Inference with Transformer Models}.
\newblock \bibinfo{journal}{\emph{IEEE ICDCS}} (\bibinfo{year}{2024}).
\newblock


\bibitem[Hu et~al\mbox{.}(2021)]%
        {huLoRALowRankAdaptation2021}
\bibfield{author}{\bibinfo{person}{Edward~J. Hu}, \bibinfo{person}{Yelong Shen}, \bibinfo{person}{Phillip Wallis}, \bibinfo{person}{Zeyuan {Allen-Zhu}}, \bibinfo{person}{Yuanzhi Li}, \bibinfo{person}{Shean Wang}, \bibinfo{person}{Lu Wang}, {and} \bibinfo{person}{Weizhu Chen}.} \bibinfo{year}{2021}\natexlab{}.
\newblock \showarticletitle{LoRA: Low-Rank Adaptation of Large Language Models}.
\newblock \bibinfo{journal}{\emph{ICLR}} (\bibinfo{year}{2021}).
\newblock


\bibitem[Huang et~al\mbox{.}(2023)]%
        {huang2023innermonologue}
\bibfield{author}{\bibinfo{person}{Wenlong Huang}, \bibinfo{person}{Fei Xia}, \bibinfo{person}{Ted Xiao}, \bibinfo{person}{Harris Chan}, \bibinfo{person}{Jacky Liang}, \bibinfo{person}{Pete Florence}, \bibinfo{person}{Andy Zeng}, \bibinfo{person}{Jonathan Tompson}, \bibinfo{person}{Igor Mordatch}, \bibinfo{person}{Yevgen Chebotar}, {et~al\mbox{.}}} \bibinfo{year}{2023}\natexlab{}.
\newblock \showarticletitle{Inner Monologue: Embodied Reasoning through Planning with Language Models}.
\newblock \bibinfo{journal}{\emph{CoRL}} (\bibinfo{year}{2023}).
\newblock


\bibitem[Intel(2020)]%
        {intelneuralcompressor}
\bibfield{author}{\bibinfo{person}{Intel}.} \bibinfo{year}{2020}\natexlab{}.
\newblock \bibinfo{title}{Intel Neural Compressor}.
\newblock \bibinfo{howpublished}{\url{https://github.com/intel/neural-compressor}}.
\newblock
\newblock
\shownote{Accessed on December 27, 2024}.


\bibitem[Intel(2022)]%
        {inteli913900news}
\bibfield{author}{\bibinfo{person}{Intel}.} \bibinfo{year}{2022}\natexlab{}.
\newblock \bibinfo{title}{Intel Launches 13th Gen Intel Core Processor Family Alongside New Intel Unison Solution}.
\newblock \bibinfo{howpublished}{\url{https://www.intel.com/content/www/us/en/newsroom/news/13th-gen-core-launch.html}}.
\newblock
\newblock
\shownote{Accessed on December 11, 2024}.


\bibitem[Jain et~al\mbox{.}(2020)]%
        {jain2020spatula}
\bibfield{author}{\bibinfo{person}{Samvit Jain}, \bibinfo{person}{Xun Zhang}, \bibinfo{person}{Yuhao Zhou}, \bibinfo{person}{Ganesh Ananthanarayanan}, \bibinfo{person}{Junchen Jiang}, \bibinfo{person}{Yuanchao Shu}, \bibinfo{person}{Paramvir Bahl}, {and} \bibinfo{person}{Joseph Gonzalez}.} \bibinfo{year}{2020}\natexlab{}.
\newblock \showarticletitle{Spatula: Efficient cross-camera video analytics on large camera networks}.
\newblock \bibinfo{journal}{\emph{IEEE/ACM SEC}} (\bibinfo{year}{2020}).
\newblock


\bibitem[Jeong et~al\mbox{.}(2023)]%
        {jeong2023winclip}
\bibfield{author}{\bibinfo{person}{Jongheon Jeong}, \bibinfo{person}{Yang Zou}, \bibinfo{person}{Taewan Kim}, \bibinfo{person}{Dongqing Zhang}, \bibinfo{person}{Avinash Ravichandran}, {and} \bibinfo{person}{Onkar Dabeer}.} \bibinfo{year}{2023}\natexlab{}.
\newblock \showarticletitle{Winclip: Zero-/few-shot anomaly classification and segmentation}.
\newblock \bibinfo{journal}{\emph{IEEE/CVF CVPR}} (\bibinfo{year}{2023}).
\newblock


\bibitem[Ji et~al\mbox{.}(2021)]%
        {ji2021distribution}
\bibfield{author}{\bibinfo{person}{Tianchu Ji}, \bibinfo{person}{Shraddhan Jain}, \bibinfo{person}{Michael Ferdman}, \bibinfo{person}{Peter Milder}, \bibinfo{person}{H~Andrew Schwartz}, {and} \bibinfo{person}{Niranjan Balasubramanian}.} \bibinfo{year}{2021}\natexlab{}.
\newblock \showarticletitle{On the Distribution, Sparsity, and Inference-time Quantization of Attention Values in Transformers}.
\newblock \bibinfo{journal}{\emph{ACL-IJCNLP}} (\bibinfo{year}{2021}).
\newblock


\bibitem[Ji et~al\mbox{.}(2024)]%
        {jiFeaturebasedLowRankCompression2024}
\bibfield{author}{\bibinfo{person}{Yixin Ji}, \bibinfo{person}{Yang Xiang}, \bibinfo{person}{Juntao Li}, \bibinfo{person}{Wei Chen}, \bibinfo{person}{Zhongyi Liu}, \bibinfo{person}{Kehai Chen}, {and} \bibinfo{person}{Min Zhang}.} \bibinfo{year}{2024}\natexlab{}.
\newblock \showarticletitle{Feature-Based Low-Rank Compression of Large Language Models via Bayesian Optimization}.
\newblock \bibinfo{journal}{\emph{Findings of EMNLP}} (\bibinfo{year}{2024}).
\newblock


\bibitem[Jiang et~al\mbox{.}(2023b)]%
        {jiang2023llmlingua}
\bibfield{author}{\bibinfo{person}{Huiqiang Jiang}, \bibinfo{person}{Qianhui Wu}, \bibinfo{person}{Chin-Yew Lin}, \bibinfo{person}{Yuqing Yang}, {and} \bibinfo{person}{Lili Qiu}.} \bibinfo{year}{2023}\natexlab{b}.
\newblock \showarticletitle{LLMLingua: Compressing Prompts for Accelerated Inference of Large Language Models}.
\newblock \bibinfo{journal}{\emph{EMNLP}} (\bibinfo{year}{2023}).
\newblock


\bibitem[Jiang et~al\mbox{.}(2019)]%
        {wang19hotedgevideo}
\bibfield{author}{\bibinfo{person}{Junchen Jiang}, \bibinfo{person}{Yuhao Zhou}, \bibinfo{person}{Ganesh Ananthanarayanan}, \bibinfo{person}{Yuanchao Shu}, {and} \bibinfo{person}{Andrew~A. Chien}.} \bibinfo{year}{2019}\natexlab{}.
\newblock \showarticletitle{{Networked Cameras Are the New Big Data Clusters}}. In \bibinfo{booktitle}{\emph{ACM Workshop on Hot Topics in Video Analytics and Intelligent Edges}}.
\newblock


\bibitem[Jiang et~al\mbox{.}(2021)]%
        {jiang2021flexible}
\bibfield{author}{\bibinfo{person}{Shiqi Jiang}, \bibinfo{person}{Zhiqi Lin}, \bibinfo{person}{Yuanchun Li}, \bibinfo{person}{Yuanchao Shu}, {and} \bibinfo{person}{Yunxin Liu}.} \bibinfo{year}{2021}\natexlab{}.
\newblock \showarticletitle{Flexible high-resolution object detection on edge devices with tunable latency}.
\newblock \bibinfo{journal}{\emph{ACM MobiCom}} (\bibinfo{year}{2021}).
\newblock


\bibitem[Jiang et~al\mbox{.}(2023a)]%
        {jiangLionAdversarialDistillation2023}
\bibfield{author}{\bibinfo{person}{Yuxin Jiang}, \bibinfo{person}{Chunkit Chan}, \bibinfo{person}{Mingyang Chen}, {and} \bibinfo{person}{Wei Wang}.} \bibinfo{year}{2023}\natexlab{a}.
\newblock \showarticletitle{Lion: Adversarial Distillation of Proprietary Large Language Models}.
\newblock \bibinfo{journal}{\emph{EMNLP}} (\bibinfo{year}{2023}).
\newblock


\bibitem[Jiao et~al\mbox{.}(2020)]%
        {jiaoTinyBERTDistillingBERT2020}
\bibfield{author}{\bibinfo{person}{Xiaoqi Jiao}, \bibinfo{person}{Yichun Yin}, \bibinfo{person}{Lifeng Shang}, \bibinfo{person}{Xin Jiang}, \bibinfo{person}{Xiao Chen}, \bibinfo{person}{Linlin Li}, \bibinfo{person}{Fang Wang}, {and} \bibinfo{person}{Qun Liu}.} \bibinfo{year}{2020}\natexlab{}.
\newblock \showarticletitle{TinyBERT: Distilling BERT for Natural Language Understanding}.
\newblock \bibinfo{journal}{\emph{EMNLP}} (\bibinfo{year}{2020}).
\newblock


\bibitem[Kang et~al\mbox{.}(2024)]%
        {kang2024quantitativeAutoFL}
\bibfield{author}{\bibinfo{person}{Sungmin Kang}, \bibinfo{person}{Gabin An}, {and} \bibinfo{person}{Shin Yoo}.} \bibinfo{year}{2024}\natexlab{}.
\newblock \showarticletitle{A quantitative and qualitative evaluation of LLM-based explainable fault localization}.
\newblock \bibinfo{journal}{\emph{Proceedings of the ACM on Software Engineering}} (\bibinfo{year}{2024}).
\newblock


\bibitem[Kaplan et~al\mbox{.}(2020)]%
        {kaplan2020scalingLaws}
\bibfield{author}{\bibinfo{person}{Jared Kaplan}, \bibinfo{person}{Sam McCandlish}, \bibinfo{person}{Tom Henighan}, \bibinfo{person}{Tom~B Brown}, \bibinfo{person}{Benjamin Chess}, \bibinfo{person}{Rewon Child}, \bibinfo{person}{Scott Gray}, \bibinfo{person}{Alec Radford}, \bibinfo{person}{Jeffrey Wu}, {and} \bibinfo{person}{Dario Amodei}.} \bibinfo{year}{2020}\natexlab{}.
\newblock \showarticletitle{Scaling laws for neural language models}.
\newblock \bibinfo{journal}{\emph{arXiv preprint arXiv:2001.08361}} (\bibinfo{year}{2020}).
\newblock


\bibitem[Khani et~al\mbox{.}(2023)]%
        {khani2023recl}
\bibfield{author}{\bibinfo{person}{Mehrdad Khani}, \bibinfo{person}{Ganesh Ananthanarayanan}, \bibinfo{person}{Kevin Hsieh}, \bibinfo{person}{Junchen Jiang}, \bibinfo{person}{Ravi Netravali}, \bibinfo{person}{Yuanchao Shu}, \bibinfo{person}{Mohammad Alizadeh}, {and} \bibinfo{person}{Victor Bahl}.} \bibinfo{year}{2023}\natexlab{}.
\newblock \showarticletitle{RECL: Responsive Resource-Efficient continuous learning for video analytics}.
\newblock \bibinfo{journal}{\emph{USENIX NSDI}} (\bibinfo{year}{2023}).
\newblock


\bibitem[Kim et~al\mbox{.}(2024a)]%
        {kim2024languageRCI}
\bibfield{author}{\bibinfo{person}{Geunwoo Kim}, \bibinfo{person}{Pierre Baldi}, {and} \bibinfo{person}{Stephen McAleer}.} \bibinfo{year}{2024}\natexlab{a}.
\newblock \showarticletitle{Language models can solve computer tasks}.
\newblock \bibinfo{journal}{\emph{NeurIPS}} (\bibinfo{year}{2024}).
\newblock


\bibitem[Kim and Cho(2021)]%
        {kim2021lengthAdaptiveTransformer}
\bibfield{author}{\bibinfo{person}{Gyuwan Kim} {and} \bibinfo{person}{Kyunghyun Cho}.} \bibinfo{year}{2021}\natexlab{}.
\newblock \showarticletitle{Length-Adaptive Transformer: Train Once with Length Drop, Use Anytime with Search}.
\newblock \bibinfo{journal}{\emph{ACL-IJCNLP}} (\bibinfo{year}{2021}).
\newblock


\bibitem[Kim et~al\mbox{.}(2023)]%
        {kimTokenScaledLogitDistillation}
\bibfield{author}{\bibinfo{person}{Minsoo Kim}, \bibinfo{person}{Sihwa Lee}, \bibinfo{person}{Janghwan Lee}, \bibinfo{person}{Sukjin Hong}, \bibinfo{person}{Du-Seong Chang}, \bibinfo{person}{Wonyong Sung}, {and} \bibinfo{person}{Jungwook Choi}.} \bibinfo{year}{2023}\natexlab{}.
\newblock \showarticletitle{Token-scaled logit distillation for ternary weight generative language models}.
\newblock \bibinfo{journal}{\emph{NeurIPS}} (\bibinfo{year}{2023}).
\newblock


\bibitem[Kim et~al\mbox{.}(2024b)]%
        {kim20CTransformer6182024}
\bibfield{author}{\bibinfo{person}{Sangyeob Kim}, \bibinfo{person}{Sangjin Kim}, \bibinfo{person}{Wooyoung Jo}, \bibinfo{person}{Soyeon Kim}, \bibinfo{person}{Seongyon Hong}, {and} \bibinfo{person}{Hoi-Jun Yoo}.} \bibinfo{year}{2024}\natexlab{b}.
\newblock \showarticletitle{20.5 C-Transformer: A 2.6-18.1$\mu$J/Token Homogeneous DNN-Transformer/Spiking-Transformer Processor with Big-Little Network and Implicit Weight Generation for Large Language Models}.
\newblock \bibinfo{journal}{\emph{IEEE ISSCC}} (\bibinfo{year}{2024}).
\newblock


\bibitem[Kim et~al\mbox{.}(2022)]%
        {kim2022learnedTokenPruning}
\bibfield{author}{\bibinfo{person}{Sehoon Kim}, \bibinfo{person}{Sheng Shen}, \bibinfo{person}{David Thorsley}, \bibinfo{person}{Amir Gholami}, \bibinfo{person}{Woosuk Kwon}, \bibinfo{person}{Joseph Hassoun}, {and} \bibinfo{person}{Kurt Keutzer}.} \bibinfo{year}{2022}\natexlab{}.
\newblock \showarticletitle{Learned token pruning for transformers}.
\newblock \bibinfo{journal}{\emph{ACM SIGKDD}} (\bibinfo{year}{2022}).
\newblock


\bibitem[Koh et~al\mbox{.}(2022)]%
        {koh2022empiricalDocumentSummarization}
\bibfield{author}{\bibinfo{person}{Huan~Yee Koh}, \bibinfo{person}{Jiaxin Ju}, \bibinfo{person}{Ming Liu}, {and} \bibinfo{person}{Shirui Pan}.} \bibinfo{year}{2022}\natexlab{}.
\newblock \showarticletitle{An empirical survey on long document summarization: Datasets, models, and metrics}.
\newblock \bibinfo{journal}{\emph{Comput. Surveys}} (\bibinfo{year}{2022}).
\newblock


\bibitem[Kong et~al\mbox{.}(2022)]%
        {kongAcceleratingInferencePretrained2022}
\bibfield{author}{\bibinfo{person}{Jun Kong}, \bibinfo{person}{Jin Wang}, \bibinfo{person}{Liang-Chih Yu}, {and} \bibinfo{person}{Xuejie Zhang}.} \bibinfo{year}{2022}\natexlab{}.
\newblock \showarticletitle{Accelerating Inference for Pretrained Language Models by Unified Multi-Perspective Early Exiting}.
\newblock \bibinfo{journal}{\emph{COLING}} (\bibinfo{year}{2022}).
\newblock


\bibitem[Kurtic et~al\mbox{.}(2022)]%
        {kurticOptimalBERTSurgeon2022}
\bibfield{author}{\bibinfo{person}{Eldar Kurtic}, \bibinfo{person}{Daniel Campos}, \bibinfo{person}{Tuan Nguyen}, \bibinfo{person}{Elias Frantar}, \bibinfo{person}{Mark Kurtz}, \bibinfo{person}{Benjamin Fineran}, \bibinfo{person}{Michael Goin}, {and} \bibinfo{person}{Dan Alistarh}.} \bibinfo{year}{2022}\natexlab{}.
\newblock \showarticletitle{The Optimal BERT Surgeon: Scalable and Accurate Second-Order Pruning for Large Language Models}.
\newblock \bibinfo{journal}{\emph{EMNLP}} (\bibinfo{year}{2022}).
\newblock


\bibitem[Kwon et~al\mbox{.}(2023)]%
        {kwonEfficientMemoryManagement2023}
\bibfield{author}{\bibinfo{person}{Woosuk Kwon}, \bibinfo{person}{Zhuohan Li}, \bibinfo{person}{Siyuan Zhuang}, \bibinfo{person}{Ying Sheng}, \bibinfo{person}{Lianmin Zheng}, \bibinfo{person}{Cody~Hao Yu}, \bibinfo{person}{Joseph Gonzalez}, \bibinfo{person}{Hao Zhang}, {and} \bibinfo{person}{Ion Stoica}.} \bibinfo{year}{2023}\natexlab{}.
\newblock \showarticletitle{Efficient Memory Management for Large Language Model Serving with PagedAttention}.
\newblock \bibinfo{journal}{\emph{ACM SOSP}} (\bibinfo{year}{2023}).
\newblock


\bibitem[Labrak et~al\mbox{.}(2024)]%
        {labrak2024biomistral}
\bibfield{author}{\bibinfo{person}{Yanis Labrak}, \bibinfo{person}{Adrien Bazoge}, \bibinfo{person}{Emmanuel Morin}, \bibinfo{person}{Pierre-Antoine Gourraud}, \bibinfo{person}{Mickael Rouvier}, {and} \bibinfo{person}{Richard Dufour}.} \bibinfo{year}{2024}\natexlab{}.
\newblock \showarticletitle{Biomistral: A collection of open-source pretrained large language models for medical domains}.
\newblock \bibinfo{journal}{\emph{Findings of ACL}} (\bibinfo{year}{2024}).
\newblock


\bibitem[Lai and Nissim(2024)]%
        {lai2024surveyfigurative}
\bibfield{author}{\bibinfo{person}{Huiyuan Lai} {and} \bibinfo{person}{Malvina Nissim}.} \bibinfo{year}{2024}\natexlab{}.
\newblock \showarticletitle{A survey on automatic generation of figurative language: From rule-based systems to large language models}.
\newblock \bibinfo{journal}{\emph{Comput. Surveys}} (\bibinfo{year}{2024}).
\newblock


\bibitem[Lalapura et~al\mbox{.}(2022)]%
        {lalapuraRecurrentNeuralNetworks2022b}
\bibfield{author}{\bibinfo{person}{Varsha~S. Lalapura}, \bibinfo{person}{J. Amudha}, {and} \bibinfo{person}{Hariramn~Selvamuruga Satheesh}.} \bibinfo{year}{2022}\natexlab{}.
\newblock \showarticletitle{Recurrent Neural Networks for Edge Intelligence: A Survey}.
\newblock \bibinfo{journal}{\emph{Comput. Surveys}} (\bibinfo{year}{2022}).
\newblock


\bibitem[Lan et~al\mbox{.}(2019)]%
        {lanALBERTLiteBERT2019}
\bibfield{author}{\bibinfo{person}{Zhenzhong Lan}, \bibinfo{person}{Mingda Chen}, \bibinfo{person}{Sebastian Goodman}, \bibinfo{person}{Kevin Gimpel}, \bibinfo{person}{Piyush Sharma}, {and} \bibinfo{person}{Radu Soricut}.} \bibinfo{year}{2019}\natexlab{}.
\newblock \showarticletitle{ALBERT: A Lite BERT for Self-supervised Learning of Language Representations}.
\newblock \bibinfo{journal}{\emph{ICLR}} (\bibinfo{year}{2019}).
\newblock


\bibitem[Laskaridis et~al\mbox{.}(2024)]%
        {laskaridisMELTingPointMobile2024}
\bibfield{author}{\bibinfo{person}{Stefanos Laskaridis}, \bibinfo{person}{Kleomenis Kateveas}, \bibinfo{person}{Lorenzo Minto}, {and} \bibinfo{person}{Hamed Haddadi}.} \bibinfo{year}{2024}\natexlab{}.
\newblock \showarticletitle{MELTing Point: Mobile Evaluation of Language Transformers}.
\newblock \bibinfo{journal}{\emph{ACM MobiCom}} (\bibinfo{year}{2024}).
\newblock


\bibitem[Lee et~al\mbox{.}(2024b)]%
        {leeOWQOutlierAwareWeight2024}
\bibfield{author}{\bibinfo{person}{Changhun Lee}, \bibinfo{person}{Jungyu Jin}, \bibinfo{person}{Taesu Kim}, \bibinfo{person}{Hyungjun Kim}, {and} \bibinfo{person}{Eunhyeok Park}.} \bibinfo{year}{2024}\natexlab{b}.
\newblock \showarticletitle{OWQ: Outlier-Aware Weight Quantization for Efficient Fine-Tuning and Inference of Large Language Models}.
\newblock \bibinfo{journal}{\emph{AAAI}} (\bibinfo{year}{2024}).
\newblock


\bibitem[Lee et~al\mbox{.}(2024a)]%
        {lee2024autonomous}
\bibfield{author}{\bibinfo{person}{Juhyeon Lee}, \bibinfo{person}{Insung Bahk}, \bibinfo{person}{Hoseung Kim}, \bibinfo{person}{Sinjin Jeong}, \bibinfo{person}{Suyeon Lee}, {and} \bibinfo{person}{Donghyun Min}.} \bibinfo{year}{2024}\natexlab{a}.
\newblock \showarticletitle{An Autonomous Parallelization of Transformer Model Inference on Heterogeneous Edge Devices}.
\newblock \bibinfo{journal}{\emph{ACM ICS}} (\bibinfo{year}{2024}).
\newblock


\bibitem[Leviathan et~al\mbox{.}(2023)]%
        {leviathan2023fastinferencespeculativedecoding}
\bibfield{author}{\bibinfo{person}{Yaniv Leviathan}, \bibinfo{person}{Matan Kalman}, {and} \bibinfo{person}{Yossi Matias}.} \bibinfo{year}{2023}\natexlab{}.
\newblock \showarticletitle{Fast inference from transformers via speculative decoding}.
\newblock \bibinfo{journal}{\emph{ICML}} (\bibinfo{year}{2023}).
\newblock


\bibitem[Li et~al\mbox{.}(2020)]%
        {li2020efficientTransformerPruning}
\bibfield{author}{\bibinfo{person}{Bingbing Li}, \bibinfo{person}{Zhenglun Kong}, \bibinfo{person}{Tianyun Zhang}, \bibinfo{person}{Ji Li}, \bibinfo{person}{Zhengang Li}, \bibinfo{person}{Hang Liu}, {and} \bibinfo{person}{Caiwen Ding}.} \bibinfo{year}{2020}\natexlab{}.
\newblock \showarticletitle{Efficient Transformer-based Large Scale Language Representations using Hardware-friendly Block Structured Pruning}.
\newblock \bibinfo{journal}{\emph{EMNLP}} (\bibinfo{year}{2020}).
\newblock


\bibitem[Li et~al\mbox{.}(2024a)]%
        {li2024sheetcopilot}
\bibfield{author}{\bibinfo{person}{Hongxin Li}, \bibinfo{person}{Jingran Su}, \bibinfo{person}{Yuntao Chen}, \bibinfo{person}{Qing Li}, {and} \bibinfo{person}{ZHAO-XIANG ZHANG}.} \bibinfo{year}{2024}\natexlab{a}.
\newblock \showarticletitle{SheetCopilot: Bringing software productivity to the next level through large language models}.
\newblock \bibinfo{journal}{\emph{NeurIPS}} (\bibinfo{year}{2024}).
\newblock


\bibitem[Li et~al\mbox{.}(2023b)]%
        {liSymbolicChainofThoughtDistillation2023}
\bibfield{author}{\bibinfo{person}{Liunian~Harold Li}, \bibinfo{person}{Jack Hessel}, \bibinfo{person}{Youngjae Yu}, \bibinfo{person}{Xiang Ren}, \bibinfo{person}{Kai-Wei Chang}, {and} \bibinfo{person}{Yejin Choi}.} \bibinfo{year}{2023}\natexlab{b}.
\newblock \showarticletitle{Symbolic Chain-of-Thought Distillation: Small Models Can Also ``Think'' Step-by-Step}.
\newblock \bibinfo{journal}{\emph{ACL}} (\bibinfo{year}{2023}).
\newblock


\bibitem[Li et~al\mbox{.}(2023a)]%
        {li2023textbooks}
\bibfield{author}{\bibinfo{person}{Yuanzhi Li}, \bibinfo{person}{S{\'e}bastien Bubeck}, \bibinfo{person}{Ronen Eldan}, \bibinfo{person}{Allie Del~Giorno}, \bibinfo{person}{Suriya Gunasekar}, {and} \bibinfo{person}{Yin~Tat Lee}.} \bibinfo{year}{2023}\natexlab{a}.
\newblock \showarticletitle{Textbooks are all you need ii: phi-1.5 technical report}.
\newblock \bibinfo{journal}{\emph{arXiv preprint arXiv:2309.05463}} (\bibinfo{year}{2023}).
\newblock


\bibitem[Li et~al\mbox{.}(2024b)]%
        {li2024personal}
\bibfield{author}{\bibinfo{person}{Yuanchun Li}, \bibinfo{person}{Hao Wen}, \bibinfo{person}{Weijun Wang}, \bibinfo{person}{Xiangyu Li}, \bibinfo{person}{Yizhen Yuan}, \bibinfo{person}{Guohong Liu}, \bibinfo{person}{Jiacheng Liu}, \bibinfo{person}{Wenxing Xu}, \bibinfo{person}{Xiang Wang}, \bibinfo{person}{Yi Sun}, {et~al\mbox{.}}} \bibinfo{year}{2024}\natexlab{b}.
\newblock \showarticletitle{Personal llm agents: Insights and survey about the capability, efficiency and security}.
\newblock \bibinfo{journal}{\emph{arXiv preprint arXiv:2401.05459}} (\bibinfo{year}{2024}).
\newblock


\bibitem[Li et~al\mbox{.}(2023c)]%
        {liLoSparseStructuredCompression2023}
\bibfield{author}{\bibinfo{person}{Yixiao Li}, \bibinfo{person}{Yifan Yu}, \bibinfo{person}{Qingru Zhang}, \bibinfo{person}{Chen Liang}, \bibinfo{person}{Pengcheng He}, \bibinfo{person}{Weizhu Chen}, {and} \bibinfo{person}{Tuo Zhao}.} \bibinfo{year}{2023}\natexlab{c}.
\newblock \showarticletitle{LoSparse: Structured Compression of Large Language Models Based on Low-Rank and Sparse Approximation}.
\newblock \bibinfo{journal}{\emph{ICML}} (\bibinfo{year}{2023}).
\newblock


\bibitem[Liang et~al\mbox{.}(2023)]%
        {liangLessMoreTaskaware2023}
\bibfield{author}{\bibinfo{person}{Chen Liang}, \bibinfo{person}{Simiao Zuo}, \bibinfo{person}{Qingru Zhang}, \bibinfo{person}{Pengcheng He}, \bibinfo{person}{Weizhu Chen}, {and} \bibinfo{person}{Tuo Zhao}.} \bibinfo{year}{2023}\natexlab{}.
\newblock \showarticletitle{Less Is More: Task-aware Layer-wise Distillation for Language Model Compression}.
\newblock \bibinfo{journal}{\emph{ICML}} (\bibinfo{year}{2023}).
\newblock


\bibitem[Lieber et~al\mbox{.}(2024)]%
        {lieber2024jamba}
\bibfield{author}{\bibinfo{person}{Opher Lieber}, \bibinfo{person}{Barak Lenz}, \bibinfo{person}{Hofit Bata}, \bibinfo{person}{Gal Cohen}, \bibinfo{person}{Jhonathan Osin}, \bibinfo{person}{Itay Dalmedigos}, \bibinfo{person}{Erez Safahi}, \bibinfo{person}{Shaked Meirom}, \bibinfo{person}{Yonatan Belinkov}, \bibinfo{person}{Shai Shalev-Shwartz}, {et~al\mbox{.}}} \bibinfo{year}{2024}\natexlab{}.
\newblock \showarticletitle{Jamba: A hybrid transformer-mamba language model}.
\newblock \bibinfo{journal}{\emph{arXiv preprint arXiv:2403.19887}} (\bibinfo{year}{2024}).
\newblock


\bibitem[Lin et~al\mbox{.}(2024)]%
        {linAWQActivationawareWeight2024}
\bibfield{author}{\bibinfo{person}{Ji Lin}, \bibinfo{person}{Jiaming Tang}, \bibinfo{person}{Haotian Tang}, \bibinfo{person}{Shang Yang}, \bibinfo{person}{Wei-Ming Chen}, \bibinfo{person}{Wei-Chen Wang}, \bibinfo{person}{Guangxuan Xiao}, \bibinfo{person}{Xingyu Dang}, \bibinfo{person}{Chuang Gan}, {and} \bibinfo{person}{Song Han}.} \bibinfo{year}{2024}\natexlab{}.
\newblock \showarticletitle{AWQ: Activation-aware Weight Quantization for On-Device LLM Compression and Acceleration}.
\newblock \bibinfo{journal}{\emph{MLSys}} (\bibinfo{year}{2024}).
\newblock


\bibitem[Liu et~al\mbox{.}(2024a)]%
        {liu2024deepseekv3}
\bibfield{author}{\bibinfo{person}{Aixin Liu}, \bibinfo{person}{Bei Feng}, \bibinfo{person}{Bing Xue}, \bibinfo{person}{Bingxuan Wang}, \bibinfo{person}{Bochao Wu}, \bibinfo{person}{Chengda Lu}, \bibinfo{person}{Chenggang Zhao}, \bibinfo{person}{Chengqi Deng}, \bibinfo{person}{Chenyu Zhang}, \bibinfo{person}{Chong Ruan}, {et~al\mbox{.}}} \bibinfo{year}{2024}\natexlab{a}.
\newblock \showarticletitle{Deepseek-v3 technical report}.
\newblock \bibinfo{journal}{\emph{arXiv preprint arXiv:2412.19437}} (\bibinfo{year}{2024}).
\newblock


\bibitem[Liu et~al\mbox{.}(2023a)]%
        {liuQLLMAccurateEfficient2023}
\bibfield{author}{\bibinfo{person}{Jing Liu}, \bibinfo{person}{Ruihao Gong}, \bibinfo{person}{Xiuying Wei}, \bibinfo{person}{Zhiwei Dong}, \bibinfo{person}{Jianfei Cai}, {and} \bibinfo{person}{Bohan Zhuang}.} \bibinfo{year}{2023}\natexlab{a}.
\newblock \showarticletitle{QLLM: Accurate and Efficient Low-Bitwidth Quantization for Large Language Models}.
\newblock \bibinfo{journal}{\emph{ICLR}} (\bibinfo{year}{2023}).
\newblock


\bibitem[Liu et~al\mbox{.}(2023b)]%
        {liu2023preNaturalLanguageProcessing}
\bibfield{author}{\bibinfo{person}{Pengfei Liu}, \bibinfo{person}{Weizhe Yuan}, \bibinfo{person}{Jinlan Fu}, \bibinfo{person}{Zhengbao Jiang}, \bibinfo{person}{Hiroaki Hayashi}, {and} \bibinfo{person}{Graham Neubig}.} \bibinfo{year}{2023}\natexlab{b}.
\newblock \showarticletitle{Pre-train, prompt, and predict: A systematic survey of prompting methods in natural language processing}.
\newblock \bibinfo{journal}{\emph{Comput. Surveys}} (\bibinfo{year}{2023}).
\newblock


\bibitem[Liu et~al\mbox{.}(2020)]%
        {liuFastBERTSelfdistillingBERT2020}
\bibfield{author}{\bibinfo{person}{Weijie Liu}, \bibinfo{person}{Peng Zhou}, \bibinfo{person}{Zhiruo Wang}, \bibinfo{person}{Zhe Zhao}, \bibinfo{person}{Haotang Deng}, {and} \bibinfo{person}{Qi Ju}.} \bibinfo{year}{2020}\natexlab{}.
\newblock \showarticletitle{FastBERT: A Self-distilling BERT with Adaptive Inference Time}.
\newblock \bibinfo{journal}{\emph{ACL}} (\bibinfo{year}{2020}).
\newblock


\bibitem[Liu et~al\mbox{.}(2024b)]%
        {liu2024mobilellm}
\bibfield{author}{\bibinfo{person}{Zechun Liu}, \bibinfo{person}{Changsheng Zhao}, \bibinfo{person}{Forrest Iandola}, \bibinfo{person}{Chen Lai}, \bibinfo{person}{Yuandong Tian}, \bibinfo{person}{Igor Fedorov}, \bibinfo{person}{Yunyang Xiong}, \bibinfo{person}{Ernie Chang}, \bibinfo{person}{Yangyang Shi}, \bibinfo{person}{Raghuraman Krishnamoorthi}, {et~al\mbox{.}}} \bibinfo{year}{2024}\natexlab{b}.
\newblock \showarticletitle{Mobilellm: Optimizing sub-billion parameter language models for on-device use cases}.
\newblock \bibinfo{journal}{\emph{ICML}} (\bibinfo{year}{2024}).
\newblock


\bibitem[Lu et~al\mbox{.}(2021)]%
        {lu2021sanger}
\bibfield{author}{\bibinfo{person}{Liqiang Lu}, \bibinfo{person}{Yicheng Jin}, \bibinfo{person}{Hangrui Bi}, \bibinfo{person}{Zizhang Luo}, \bibinfo{person}{Peng Li}, \bibinfo{person}{Tao Wang}, {and} \bibinfo{person}{Yun Liang}.} \bibinfo{year}{2021}\natexlab{}.
\newblock \showarticletitle{Sanger: A co-design framework for enabling sparse attention using reconfigurable architecture}.
\newblock \bibinfo{journal}{\emph{IEEE/ACM MICRO}} (\bibinfo{year}{2021}).
\newblock


\bibitem[Lu et~al\mbox{.}(2024)]%
        {lu2024multimodal}
\bibfield{author}{\bibinfo{person}{Ming~Y Lu}, \bibinfo{person}{Bowen Chen}, \bibinfo{person}{Drew~FK Williamson}, \bibinfo{person}{Richard~J Chen}, \bibinfo{person}{Melissa Zhao}, \bibinfo{person}{Aaron~K Chow}, \bibinfo{person}{Kenji Ikemura}, \bibinfo{person}{Ahrong Kim}, \bibinfo{person}{Dimitra Pouli}, \bibinfo{person}{Ankush Patel}, {et~al\mbox{.}}} \bibinfo{year}{2024}\natexlab{}.
\newblock \showarticletitle{A multimodal generative AI copilot for human pathology}.
\newblock \bibinfo{journal}{\emph{Nature}} (\bibinfo{year}{2024}).
\newblock


\bibitem[Lu et~al\mbox{.}(2022)]%
        {lu2022turbo}
\bibfield{author}{\bibinfo{person}{Yan Lu}, \bibinfo{person}{Shiqi Jiang}, \bibinfo{person}{Ting Cao}, {and} \bibinfo{person}{Yuanchao Shu}.} \bibinfo{year}{2022}\natexlab{}.
\newblock \showarticletitle{Turbo: Opportunistic enhancement for edge video analytics}.
\newblock \bibinfo{journal}{\emph{ACM SenSys}} (\bibinfo{year}{2022}).
\newblock


\bibitem[Lu et~al\mbox{.}(2019)]%
        {lu19sec}
\bibfield{author}{\bibinfo{person}{Yan Lu}, \bibinfo{person}{Yuanchao Shu}, \bibinfo{person}{Xu Tan}, \bibinfo{person}{Yunxin Liu}, \bibinfo{person}{Mengyu Zhou}, \bibinfo{person}{Qi Chen}, {and} \bibinfo{person}{Dan Pei}.} \bibinfo{year}{2019}\natexlab{}.
\newblock \showarticletitle{{Collaborative Learning between Cloud and End Devices: An Empirical Study on Location Prediction}}. In \bibinfo{booktitle}{\emph{ACM/IEEE SEC}}.
\newblock


\bibitem[Lu et~al\mbox{.}(2023)]%
        {lu2023multiview}
\bibfield{author}{\bibinfo{person}{Yan Lu}, \bibinfo{person}{Zhun Zhong}, {and} \bibinfo{person}{Yuanchao Shu}.} \bibinfo{year}{2023}\natexlab{}.
\newblock \showarticletitle{Multi-view domain adaptive object detection on camera networks}.
\newblock \bibinfo{journal}{\emph{AAAI}} (\bibinfo{year}{2023}).
\newblock


\bibitem[Ma et~al\mbox{.}(2024a)]%
        {ma2024hpipeparallelismheterogeneous}
\bibfield{author}{\bibinfo{person}{Ruilong Ma}, \bibinfo{person}{Xiang Yang}, \bibinfo{person}{Jingyu Wang}, \bibinfo{person}{Qi Qi}, \bibinfo{person}{Haifeng Sun}, \bibinfo{person}{Jing Wang}, \bibinfo{person}{Zirui Zhuang}, {and} \bibinfo{person}{Jianxin Liao}.} \bibinfo{year}{2024}\natexlab{a}.
\newblock \showarticletitle{HPipe: Large Language Model Pipeline Parallelism for Long Context on Heterogeneous Cost-effective Devices}.
\newblock \bibinfo{journal}{\emph{NAACL-HLT}} (\bibinfo{year}{2024}).
\newblock


\bibitem[Ma et~al\mbox{.}(2023)]%
        {maLLMPrunerStructuralPruning2023}
\bibfield{author}{\bibinfo{person}{Xinyin Ma}, \bibinfo{person}{Gongfan Fang}, {and} \bibinfo{person}{Xinchao Wang}.} \bibinfo{year}{2023}\natexlab{}.
\newblock \showarticletitle{LLM-Pruner: On the Structural Pruning of Large Language Models}.
\newblock \bibinfo{journal}{\emph{NeurIPS}} (\bibinfo{year}{2023}).
\newblock


\bibitem[Ma et~al\mbox{.}(2024b)]%
        {ma2024comprehensive}
\bibfield{author}{\bibinfo{person}{Xinbei Ma}, \bibinfo{person}{Zhuosheng Zhang}, {and} \bibinfo{person}{Hai Zhao}.} \bibinfo{year}{2024}\natexlab{b}.
\newblock \showarticletitle{Coco-agent: A comprehensive cognitive mllm agent for smartphone gui automation}.
\newblock \bibinfo{journal}{\emph{Findings of ACL}} (\bibinfo{year}{2024}).
\newblock


\bibitem[Mehta et~al\mbox{.}(2021)]%
        {mehta2020delight}
\bibfield{author}{\bibinfo{person}{Sachin Mehta}, \bibinfo{person}{Marjan Ghazvininejad}, \bibinfo{person}{Srinivasan Iyer}, \bibinfo{person}{Luke Zettlemoyer}, {and} \bibinfo{person}{Hannaneh Hajishirzi}.} \bibinfo{year}{2021}\natexlab{}.
\newblock \showarticletitle{Delight: Deep and light-weight transformer}.
\newblock \bibinfo{journal}{\emph{ICLR}} (\bibinfo{year}{2021}).
\newblock


\bibitem[Mehta et~al\mbox{.}(2024)]%
        {mehta2024openelm}
\bibfield{author}{\bibinfo{person}{Sachin Mehta}, \bibinfo{person}{Mohammad~Hossein Sekhavat}, \bibinfo{person}{Qingqing Cao}, \bibinfo{person}{Maxwell Horton}, \bibinfo{person}{Yanzi Jin}, \bibinfo{person}{Chenfan Sun}, \bibinfo{person}{Iman Mirzadeh}, \bibinfo{person}{Mahyar Najibi}, \bibinfo{person}{Dmitry Belenko}, \bibinfo{person}{Peter Zatloukal}, {et~al\mbox{.}}} \bibinfo{year}{2024}\natexlab{}.
\newblock \showarticletitle{OpenELM: An Efficient Language Model Family with Open-source Training and Inference Framework}.
\newblock \bibinfo{journal}{\emph{arXiv preprint arXiv:2404.14619}} (\bibinfo{year}{2024}).
\newblock


\bibitem[Microsoft(2018)]%
        {onnxruntime}
\bibfield{author}{\bibinfo{person}{Microsoft}.} \bibinfo{year}{2018}\natexlab{}.
\newblock \bibinfo{title}{ONNX Runtime is a cross-platform inference and training machine-learning accelerator.}
\newblock \bibinfo{howpublished}{\url{https://github.com/microsoft/onnxruntime}}.
\newblock
\newblock
\shownote{Accessed on December 29, 2024}.


\bibitem[Microsoft(2023)]%
        {copilot}
\bibfield{author}{\bibinfo{person}{Microsoft}.} \bibinfo{year}{2023}\natexlab{}.
\newblock \bibinfo{title}{Microsoft Copilot.}
\newblock \bibinfo{howpublished}{\url{https://copilot.microsoft.com/}}.
\newblock


\bibitem[Mojan~Javaheripi(2023)]%
        {hughesPhi2SurprisingPower2023}
\bibfield{author}{\bibinfo{person}{Sébastien~Bubeck Mojan~Javaheripi}.} \bibinfo{year}{2023}\natexlab{}.
\newblock \bibinfo{title}{Phi-2: The Surprising Power of Small Language Models}.
\newblock \bibinfo{howpublished}{\url{https://www.microsoft.com/en-us/research/blog/phi-2-the-surprising-power-of-small-language-models/}}.
\newblock
\newblock
\shownote{Accessed on June 1, 2024}.


\bibitem[Niu et~al\mbox{.}(2021)]%
        {niuDNNFusionAcceleratingDeep2021}
\bibfield{author}{\bibinfo{person}{Wei Niu}, \bibinfo{person}{Jiexiong Guan}, \bibinfo{person}{Yanzhi Wang}, \bibinfo{person}{Gagan Agrawal}, {and} \bibinfo{person}{Bin Ren}.} \bibinfo{year}{2021}\natexlab{}.
\newblock \showarticletitle{DNNFusion: Accelerating Deep Neural Networks Execution with Advanced Operator Fusion}.
\newblock \bibinfo{journal}{\emph{ACM PLDI}} (\bibinfo{year}{2021}).
\newblock


\bibitem[Niu et~al\mbox{.}(2024)]%
        {niuSmartMemLayoutTransformation2024}
\bibfield{author}{\bibinfo{person}{Wei Niu}, \bibinfo{person}{Md~Musfiqur~Rahman Sanim}, \bibinfo{person}{Zhihao Shu}, \bibinfo{person}{Jiexiong Guan}, \bibinfo{person}{Xipeng Shen}, \bibinfo{person}{Miao Yin}, \bibinfo{person}{Gagan Agrawal}, {and} \bibinfo{person}{Bin Ren}.} \bibinfo{year}{2024}\natexlab{}.
\newblock \showarticletitle{SmartMem: Layout Transformation Elimination and Adaptation for Efficient DNN Execution on Mobile}.
\newblock \bibinfo{journal}{\emph{ACM ASPLOS}} (\bibinfo{year}{2024}).
\newblock


\bibitem[NVIDIA(2023)]%
        {tensorrtllm}
\bibfield{author}{\bibinfo{person}{NVIDIA}.} \bibinfo{year}{2023}\natexlab{}.
\newblock \bibinfo{title}{TensorRT-LLM}.
\newblock \bibinfo{howpublished}{\url{https://github.com/NVIDIA/TensorRT-LLM}}.
\newblock
\newblock
\shownote{Accessed on June 9, 2024}.


\bibitem[NVIDIA(2024)]%
        {nvidia2024NemotronMini4b}
\bibfield{author}{\bibinfo{person}{NVIDIA}.} \bibinfo{year}{2024}\natexlab{}.
\newblock \bibinfo{title}{Minitron-4B-Base}.
\newblock \bibinfo{howpublished}{\url{https://huggingface.co/nvidia/Minitron-4B-Base}}.
\newblock
\newblock
\shownote{Accessed on December 19, 2024}.


\bibitem[OpenAI(2022)]%
        {IntroducingChatGPTOpenAI}
\bibfield{author}{\bibinfo{person}{OpenAI}.} \bibinfo{year}{2022}\natexlab{}.
\newblock \bibinfo{title}{Introducing ChatGPT}.
\newblock \bibinfo{howpublished}{\url{https://openai.com/index/chatgpt/}}.
\newblock
\newblock
\shownote{Accessed on July 5, 2024}.


\bibitem[Ouyang et~al\mbox{.}(2022)]%
        {ouyang2022trainingFollowingInstructions}
\bibfield{author}{\bibinfo{person}{Long Ouyang}, \bibinfo{person}{Jeff Wu}, \bibinfo{person}{Xu Jiang}, \bibinfo{person}{Diogo Almeida}, \bibinfo{person}{Carroll~L Wainwright}, \bibinfo{person}{Pamela Mishkin}, \bibinfo{person}{Chong Zhang}, \bibinfo{person}{Sandhini Agarwal}, \bibinfo{person}{Katarina Slama}, \bibinfo{person}{Alex Ray}, {et~al\mbox{.}}} \bibinfo{year}{2022}\natexlab{}.
\newblock \showarticletitle{Training language models to follow instructions with human feedback}.
\newblock \bibinfo{journal}{\emph{NeurIPS}} (\bibinfo{year}{2022}).
\newblock


\bibitem[Padmanabhan et~al\mbox{.}(2023)]%
        {padmanabhan2023gemel}
\bibfield{author}{\bibinfo{person}{Arthi Padmanabhan}, \bibinfo{person}{Neil Agarwal}, \bibinfo{person}{Anand Iyer}, \bibinfo{person}{Ganesh Ananthanarayanan}, \bibinfo{person}{Yuanchao Shu}, \bibinfo{person}{Nikolaos Karianakis}, \bibinfo{person}{Guoqing~Harry Xu}, {and} \bibinfo{person}{Ravi Netravali}.} \bibinfo{year}{2023}\natexlab{}.
\newblock \showarticletitle{Gemel: Model Merging for Memory-Efficient, Real-Time Video Analytics at the Edge}.
\newblock \bibinfo{journal}{\emph{USENIX NSDI}} (\bibinfo{year}{2023}).
\newblock


\bibitem[Padmanabhan et~al\mbox{.}(2024)]%
        {padmanabhanPropagatingKnowledgeUpdates}
\bibfield{author}{\bibinfo{person}{Shankar Padmanabhan}, \bibinfo{person}{Yasumasa Onoe}, \bibinfo{person}{Michael Zhang}, \bibinfo{person}{Greg Durrett}, {and} \bibinfo{person}{Eunsol Choi}.} \bibinfo{year}{2024}\natexlab{}.
\newblock \showarticletitle{Propagating knowledge updates to lms through distillation}.
\newblock \bibinfo{journal}{\emph{NeurIPS}} (\bibinfo{year}{2024}).
\newblock


\bibitem[Pan et~al\mbox{.}(2024)]%
        {pan2024vlp}
\bibfield{author}{\bibinfo{person}{Chenbin Pan}, \bibinfo{person}{Burhaneddin Yaman}, \bibinfo{person}{Tommaso Nesti}, \bibinfo{person}{Abhirup Mallik}, \bibinfo{person}{Alessandro~G Allievi}, \bibinfo{person}{Senem Velipasalar}, {and} \bibinfo{person}{Liu Ren}.} \bibinfo{year}{2024}\natexlab{}.
\newblock \showarticletitle{VLP: Vision Language Planning for Autonomous Driving}.
\newblock \bibinfo{journal}{\emph{IEEE/CVF CVPR}} (\bibinfo{year}{2024}).
\newblock


\bibitem[Parmar et~al\mbox{.}(2024)]%
        {parmar2024nemotron4-15b}
\bibfield{author}{\bibinfo{person}{Jupinder Parmar}, \bibinfo{person}{Shrimai Prabhumoye}, \bibinfo{person}{Joseph Jennings}, \bibinfo{person}{Mostofa Patwary}, \bibinfo{person}{Sandeep Subramanian}, \bibinfo{person}{Dan Su}, \bibinfo{person}{Chen Zhu}, \bibinfo{person}{Deepak Narayanan}, \bibinfo{person}{Aastha Jhunjhunwala}, \bibinfo{person}{Ayush Dattagupta}, {et~al\mbox{.}}} \bibinfo{year}{2024}\natexlab{}.
\newblock \showarticletitle{Nemotron-4 15B Technical Report}.
\newblock \bibinfo{journal}{\emph{arXiv preprint arXiv:2402.16819}} (\bibinfo{year}{2024}).
\newblock


\bibitem[Pi(2019)]%
        {raspberrypi4b}
\bibfield{author}{\bibinfo{person}{Raspberry Pi}.} \bibinfo{year}{2019}\natexlab{}.
\newblock \bibinfo{title}{Raspberry Pi 4 on sale now.}
\newblock \bibinfo{howpublished}{\url{https://www.raspberrypi.com/news/raspberry-pi-4-on-sale-now-from-35/}}.
\newblock
\newblock
\shownote{Accessed on December 29, 2024}.


\bibitem[Pytorch(2023)]%
        {executorch}
\bibfield{author}{\bibinfo{person}{Pytorch}.} \bibinfo{year}{2023}\natexlab{}.
\newblock \bibinfo{title}{ExecuTorch}.
\newblock \bibinfo{howpublished}{\url{https://github.com/pytorch/executorch}}.
\newblock
\newblock
\shownote{Accessed on December 12, 2024}.


\bibitem[Qi et~al\mbox{.}(2024)]%
        {qi2024interactive}
\bibfield{author}{\bibinfo{person}{Biqing Qi}, \bibinfo{person}{Xinquan Chen}, \bibinfo{person}{Junqi Gao}, \bibinfo{person}{Dong Li}, \bibinfo{person}{Jianxing Liu}, \bibinfo{person}{Ligang Wu}, {and} \bibinfo{person}{Bowen Zhou}.} \bibinfo{year}{2024}\natexlab{}.
\newblock \showarticletitle{Interactive continual learning: Fast and slow thinking}.
\newblock \bibinfo{journal}{\emph{IEEE/CVF CVPR}} (\bibinfo{year}{2024}).
\newblock


\bibitem[Qu et~al\mbox{.}(2024)]%
        {qu2024mobileedgeintelligence}
\bibfield{author}{\bibinfo{person}{Guanqiao Qu}, \bibinfo{person}{Qiyuan Chen}, \bibinfo{person}{Wei Wei}, \bibinfo{person}{Zheng Lin}, \bibinfo{person}{Xianhao Chen}, {and} \bibinfo{person}{Kaibin Huang}.} \bibinfo{year}{2024}\natexlab{}.
\newblock \showarticletitle{Mobile Edge Intelligence for Large Language Models: A Contemporary Survey}.
\newblock \bibinfo{journal}{\emph{arXiv preprint arXiv:2407.18921}} (\bibinfo{year}{2024}).
\newblock


\bibitem[Qualcomm(2024)]%
        {SnapdragonSeriesMobile}
\bibfield{author}{\bibinfo{person}{Qualcomm}.} \bibinfo{year}{2024}\natexlab{}.
\newblock \bibinfo{title}{Snapdragon 8 Series Mobile Platforms {\textbar} Qualcomm}.
\newblock \bibinfo{howpublished}{\url{https://www.qualcomm.com/products/mobile/snapdragon/smartphones/snapdragon-8-series-mobile-platforms}}.
\newblock
\newblock
\shownote{Accessed on July 17, 2024}.


\bibitem[Rakowski(2023)]%
        {GeminiNanoWeb}
\bibfield{author}{\bibinfo{person}{Brian Rakowski}.} \bibinfo{year}{2023}\natexlab{}.
\newblock \bibinfo{title}{Pixel 8 Pro — the first smartphone with AI built in — is now running Gemini Nano, plus more AI updates coming to the Pixel portfolio}.
\newblock \bibinfo{howpublished}{\url{https://blog.google/products/pixel/pixel-feature-drop-december-2023/}}.
\newblock
\newblock
\shownote{Accessed on December 9, 2024}.


\bibitem[Rawassizadeh and Rong(2023)]%
        {rawassizadehODSearchFastResource2022}
\bibfield{author}{\bibinfo{person}{Reza Rawassizadeh} {and} \bibinfo{person}{Yi Rong}.} \bibinfo{year}{2023}\natexlab{}.
\newblock \showarticletitle{ODSearch: Fast and Resource Efficient On-device Natural Language Search for Fitness Trackers' Data}.
\newblock \bibinfo{journal}{\emph{Proceedings of the ACM on Interactive, Mobile, Wearable and Ubiquitous Technologies}} (\bibinfo{year}{2023}).
\newblock


\bibitem[Ruan et~al\mbox{.}(2024)]%
        {ruan2024observationalScalingLaws}
\bibfield{author}{\bibinfo{person}{Yangjun Ruan}, \bibinfo{person}{Chris~J Maddison}, {and} \bibinfo{person}{Tatsunori Hashimoto}.} \bibinfo{year}{2024}\natexlab{}.
\newblock \showarticletitle{Observational Scaling Laws and the Predictability of Language Model Performance}.
\newblock \bibinfo{journal}{\emph{arXiv preprint arXiv:2405.10938}} (\bibinfo{year}{2024}).
\newblock


\bibitem[Saha et~al\mbox{.}(2023)]%
        {sahaMatrixCompressionRandomized2023}
\bibfield{author}{\bibinfo{person}{Rajarshi Saha}, \bibinfo{person}{Varun Srivastava}, {and} \bibinfo{person}{Mert Pilanci}.} \bibinfo{year}{2023}\natexlab{}.
\newblock \showarticletitle{Matrix Compression via Randomized Low Rank and Low Precision Factorization}.
\newblock \bibinfo{journal}{\emph{NeurIPS}} (\bibinfo{year}{2023}).
\newblock


\bibitem[Sanh et~al\mbox{.}(2020)]%
        {sanhMovementPruningAdaptive2020}
\bibfield{author}{\bibinfo{person}{Victor Sanh}, \bibinfo{person}{Thomas Wolf}, {and} \bibinfo{person}{Alexander Rush}.} \bibinfo{year}{2020}\natexlab{}.
\newblock \showarticletitle{Movement Pruning: Adaptive Sparsity by Fine-Tuning}.
\newblock \bibinfo{journal}{\emph{NeurIPS}} (\bibinfo{year}{2020}).
\newblock


\bibitem[Shao et~al\mbox{.}(2023)]%
        {shaoOmniQuantOmnidirectionallyCalibrated2023}
\bibfield{author}{\bibinfo{person}{Wenqi Shao}, \bibinfo{person}{Mengzhao Chen}, \bibinfo{person}{Zhaoyang Zhang}, \bibinfo{person}{Peng Xu}, \bibinfo{person}{Lirui Zhao}, \bibinfo{person}{Zhiqian Li}, \bibinfo{person}{Kaipeng Zhang}, \bibinfo{person}{Peng Gao}, \bibinfo{person}{Yu Qiao}, {and} \bibinfo{person}{Ping Luo}.} \bibinfo{year}{2023}\natexlab{}.
\newblock \showarticletitle{OmniQuant: Omnidirectionally Calibrated Quantization for Large Language Models}.
\newblock \bibinfo{journal}{\emph{ICLR}} (\bibinfo{year}{2023}).
\newblock


\bibitem[Shen et~al\mbox{.}(2020)]%
        {shenQBERTHessianBased2020a}
\bibfield{author}{\bibinfo{person}{Sheng Shen}, \bibinfo{person}{Zhen Dong}, \bibinfo{person}{Jiayu Ye}, \bibinfo{person}{Linjian Ma}, \bibinfo{person}{Zhewei Yao}, \bibinfo{person}{Amir Gholami}, \bibinfo{person}{Michael~W. Mahoney}, {and} \bibinfo{person}{Kurt Keutzer}.} \bibinfo{year}{2020}\natexlab{}.
\newblock \showarticletitle{Q-BERT: Hessian Based Ultra Low Precision Quantization of BERT}.
\newblock \bibinfo{journal}{\emph{AAAI}} (\bibinfo{year}{2020}).
\newblock


\bibitem[Shen et~al\mbox{.}(2024)]%
        {shenAgileQuantActivationGuidedQuantization2024}
\bibfield{author}{\bibinfo{person}{Xuan Shen}, \bibinfo{person}{Peiyan Dong}, \bibinfo{person}{Lei Lu}, \bibinfo{person}{Zhenglun Kong}, \bibinfo{person}{Zhengang Li}, \bibinfo{person}{Ming Lin}, \bibinfo{person}{Chao Wu}, {and} \bibinfo{person}{Yanzhi Wang}.} \bibinfo{year}{2024}\natexlab{}.
\newblock \showarticletitle{Agile-Quant: Activation-Guided Quantization for Faster Inference of LLMs on the Edge}.
\newblock \bibinfo{journal}{\emph{AAAI}} (\bibinfo{year}{2024}).
\newblock


\bibitem[Sheng et~al\mbox{.}(2023)]%
        {shengFlexGenHighThroughputGenerative2023}
\bibfield{author}{\bibinfo{person}{Ying Sheng}, \bibinfo{person}{Lianmin Zheng}, \bibinfo{person}{Binhang Yuan}, \bibinfo{person}{Zhuohan Li}, \bibinfo{person}{Max Ryabinin}, \bibinfo{person}{Beidi Chen}, \bibinfo{person}{Percy Liang}, \bibinfo{person}{Christopher Re}, \bibinfo{person}{Ion Stoica}, {and} \bibinfo{person}{Ce Zhang}.} \bibinfo{year}{2023}\natexlab{}.
\newblock \showarticletitle{FlexGen: High-Throughput Generative Inference of Large Language Models with a Single GPU}.
\newblock \bibinfo{journal}{\emph{ICML}} (\bibinfo{year}{2023}).
\newblock


\bibitem[Shridhar et~al\mbox{.}(2023)]%
        {shridharDistillingReasoningCapabilities2023}
\bibfield{author}{\bibinfo{person}{Kumar Shridhar}, \bibinfo{person}{Alessandro Stolfo}, {and} \bibinfo{person}{Mrinmaya Sachan}.} \bibinfo{year}{2023}\natexlab{}.
\newblock \showarticletitle{Distilling Reasoning Capabilities into Smaller Language Models}.
\newblock \bibinfo{journal}{\emph{ACL}} (\bibinfo{year}{2023}).
\newblock


\bibitem[Shuvo et~al\mbox{.}(2022)]%
        {shuvo2022efficientaccelerationOnEdgeDevice}
\bibfield{author}{\bibinfo{person}{Md~Maruf~Hossain Shuvo}, \bibinfo{person}{Syed~Kamrul Islam}, \bibinfo{person}{Jianlin Cheng}, {and} \bibinfo{person}{Bashir~I Morshed}.} \bibinfo{year}{2022}\natexlab{}.
\newblock \showarticletitle{Efficient acceleration of deep learning inference on resource-constrained edge devices: A review}.
\newblock \bibinfo{journal}{\emph{Proc. IEEE}} (\bibinfo{year}{2022}).
\newblock


\bibitem[Song et~al\mbox{.}(2024)]%
        {song2024powerinfer}
\bibfield{author}{\bibinfo{person}{Yixin Song}, \bibinfo{person}{Zeyu Mi}, \bibinfo{person}{Haotong Xie}, {and} \bibinfo{person}{Haibo Chen}.} \bibinfo{year}{2024}\natexlab{}.
\newblock \showarticletitle{Powerinfer: Fast large language model serving with a consumer-grade gpu}.
\newblock \bibinfo{journal}{\emph{ACM SOSP}} (\bibinfo{year}{2024}).
\newblock


\bibitem[Sridharan et~al\mbox{.}(2023)]%
        {sridharan2023xformer}
\bibfield{author}{\bibinfo{person}{Shrihari Sridharan}, \bibinfo{person}{Jacob~R Stevens}, \bibinfo{person}{Kaushik Roy}, {and} \bibinfo{person}{Anand Raghunathan}.} \bibinfo{year}{2023}\natexlab{}.
\newblock \showarticletitle{X-former: In-memory acceleration of transformers}.
\newblock \bibinfo{journal}{\emph{IEEE Transactions on Very Large Scale Integration (VLSI) Systems}} (\bibinfo{year}{2023}).
\newblock


\bibitem[Su et~al\mbox{.}(2024)]%
        {su2024roformerrotarypositionembedding}
\bibfield{author}{\bibinfo{person}{Jianlin Su}, \bibinfo{person}{Yu Lu}, \bibinfo{person}{Shengfeng Pan}, \bibinfo{person}{Ahmed Murtadha}, \bibinfo{person}{Bo Wen}, {and} \bibinfo{person}{Yunfeng Liu}.} \bibinfo{year}{2024}\natexlab{}.
\newblock \showarticletitle{RoFormer: Enhanced Transformer with Rotary Position Embedding}.
\newblock \bibinfo{journal}{\emph{Elsevier Neurocomputing}} (\bibinfo{year}{2024}).
\newblock


\bibitem[Sun et~al\mbox{.}(2024a)]%
        {sun2024simpleeffectivepruningapproach}
\bibfield{author}{\bibinfo{person}{Mingjie Sun}, \bibinfo{person}{Zhuang Liu}, \bibinfo{person}{Anna Bair}, {and} \bibinfo{person}{J.~Zico Kolter}.} \bibinfo{year}{2024}\natexlab{a}.
\newblock \showarticletitle{A Simple and Effective Pruning Approach for Large Language Models}.
\newblock \bibinfo{journal}{\emph{ICLR}} (\bibinfo{year}{2024}).
\newblock


\bibitem[Sun et~al\mbox{.}(2024b)]%
        {sun2024spectr}
\bibfield{author}{\bibinfo{person}{Ziteng Sun}, \bibinfo{person}{Ananda~Theertha Suresh}, \bibinfo{person}{Jae~Hun Ro}, \bibinfo{person}{Ahmad Beirami}, \bibinfo{person}{Himanshu Jain}, {and} \bibinfo{person}{Felix Yu}.} \bibinfo{year}{2024}\natexlab{b}.
\newblock \showarticletitle{Spectr: Fast speculative decoding via optimal transport}.
\newblock \bibinfo{journal}{\emph{NeurIPS}} (\bibinfo{year}{2024}).
\newblock


\bibitem[Sun et~al\mbox{.}(2020)]%
        {sunMobileBERTCompactTaskAgnostic2020}
\bibfield{author}{\bibinfo{person}{Zhiqing Sun}, \bibinfo{person}{Hongkun Yu}, \bibinfo{person}{Xiaodan Song}, \bibinfo{person}{Renjie Liu}, \bibinfo{person}{Yiming Yang}, {and} \bibinfo{person}{Denny Zhou}.} \bibinfo{year}{2020}\natexlab{}.
\newblock \showarticletitle{MobileBERT: A Compact Task-Agnostic BERT for Resource-Limited Devices}.
\newblock \bibinfo{journal}{\emph{ACL}} (\bibinfo{year}{2020}).
\newblock


\bibitem[Tambe et~al\mbox{.}(2021)]%
        {tambeEdgeBERTSentenceLevelEnergy2021}
\bibfield{author}{\bibinfo{person}{Thierry Tambe}, \bibinfo{person}{Coleman Hooper}, \bibinfo{person}{Lillian Pentecost}, \bibinfo{person}{Tianyu Jia}, \bibinfo{person}{En-Yu Yang}, \bibinfo{person}{Marco Donato}, \bibinfo{person}{Victor Sanh}, \bibinfo{person}{Paul Whatmough}, \bibinfo{person}{Alexander~M. Rush}, \bibinfo{person}{David Brooks}, {and} \bibinfo{person}{Gu-Yeon Wei}.} \bibinfo{year}{2021}\natexlab{}.
\newblock \showarticletitle{EdgeBERT: Sentence-Level Energy Optimizations for Latency-Aware Multi-Task NLP Inference}.
\newblock \bibinfo{journal}{\emph{IEEE/ACM MICRO}} (\bibinfo{year}{2021}).
\newblock


\bibitem[Tambe et~al\mbox{.}(2020)]%
        {tambe2020AdaptivFloat}
\bibfield{author}{\bibinfo{person}{Thierry Tambe}, \bibinfo{person}{En-Yu Yang}, \bibinfo{person}{Zishen Wan}, \bibinfo{person}{Yuntian Deng}, \bibinfo{person}{Vijay~Janapa Reddi}, \bibinfo{person}{Alexander Rush}, \bibinfo{person}{David Brooks}, {and} \bibinfo{person}{Gu-Yeon Wei}.} \bibinfo{year}{2020}\natexlab{}.
\newblock \showarticletitle{Algorithm-hardware co-design of adaptive floating-point encodings for resilient deep learning inference}.
\newblock \bibinfo{journal}{\emph{ACM/IEEE DAC}} (\bibinfo{year}{2020}).
\newblock


\bibitem[Tambe et~al\mbox{.}(2023a)]%
        {tambe202322STP}
\bibfield{author}{\bibinfo{person}{Thierry Tambe}, \bibinfo{person}{Jeff Zhang}, \bibinfo{person}{Coleman Hooper}, \bibinfo{person}{Tianyu Jia}, \bibinfo{person}{Paul~N Whatmough}, \bibinfo{person}{Joseph Zuckerman}, \bibinfo{person}{Maico~Cassel Dos~Santos}, \bibinfo{person}{Erik~Jens Loscalzo}, \bibinfo{person}{Davide Giri}, \bibinfo{person}{Kenneth Shepard}, {et~al\mbox{.}}} \bibinfo{year}{2023}\natexlab{a}.
\newblock \showarticletitle{22.9 A 12nm 18.1 TFLOPs/W sparse transformer processor with entropy-based early exit, mixed-precision predication and fine-grained power management}.
\newblock \bibinfo{journal}{\emph{IEEE ISSCC}} (\bibinfo{year}{2023}).
\newblock


\bibitem[Tambe et~al\mbox{.}(2023b)]%
        {tambe2212nm182023}
\bibfield{author}{\bibinfo{person}{Thierry Tambe}, \bibinfo{person}{Jeff Zhang}, \bibinfo{person}{Coleman Hooper}, \bibinfo{person}{Tianyu Jia}, \bibinfo{person}{Paul~N. Whatmough}, \bibinfo{person}{Joseph Zuckerman}, \bibinfo{person}{Maico Cassel~Dos Santos}, \bibinfo{person}{Erik~Jens Loscalzo}, \bibinfo{person}{Davide Giri}, \bibinfo{person}{Kenneth Shepard}, \bibinfo{person}{Luca Carloni}, \bibinfo{person}{Alexander Rush}, \bibinfo{person}{David Brooks}, {and} \bibinfo{person}{Gu-Yeon Wei}.} \bibinfo{year}{2023}\natexlab{b}.
\newblock \showarticletitle{22.9 A 12nm 18.1TFLOPs/W Sparse Transformer Processor with Entropy-Based Early Exit, Mixed-Precision Predication and Fine-Grained Power Management}.
\newblock \bibinfo{journal}{\emph{IEEE ISSCC}} (\bibinfo{year}{2023}).
\newblock


\bibitem[Tang et~al\mbox{.}(2024)]%
        {tang2024graphgpt}
\bibfield{author}{\bibinfo{person}{Jiabin Tang}, \bibinfo{person}{Yuhao Yang}, \bibinfo{person}{Wei Wei}, \bibinfo{person}{Lei Shi}, \bibinfo{person}{Lixin Su}, \bibinfo{person}{Suqi Cheng}, \bibinfo{person}{Dawei Yin}, {and} \bibinfo{person}{Chao Huang}.} \bibinfo{year}{2024}\natexlab{}.
\newblock \showarticletitle{Graphgpt: Graph instruction tuning for large language models}.
\newblock \bibinfo{journal}{\emph{ACM SIGIR}} (\bibinfo{year}{2024}), \bibinfo{pages}{491--500}.
\newblock


\bibitem[Team et~al\mbox{.}(2023)]%
        {team2023gemini}
\bibfield{author}{\bibinfo{person}{Gemini Team}, \bibinfo{person}{Rohan Anil}, \bibinfo{person}{Sebastian Borgeaud}, \bibinfo{person}{Jean-Baptiste Alayrac}, \bibinfo{person}{Jiahui Yu}, \bibinfo{person}{Radu Soricut}, \bibinfo{person}{Johan Schalkwyk}, \bibinfo{person}{Andrew~M Dai}, \bibinfo{person}{Anja Hauth}, \bibinfo{person}{Katie Millican}, {et~al\mbox{.}}} \bibinfo{year}{2023}\natexlab{}.
\newblock \showarticletitle{Gemini: a family of highly capable multimodal models}.
\newblock \bibinfo{journal}{\emph{arXiv preprint arXiv:2312.11805}} (\bibinfo{year}{2023}).
\newblock


\bibitem[Team et~al\mbox{.}(2024a)]%
        {team2024gemma}
\bibfield{author}{\bibinfo{person}{Gemma Team}, \bibinfo{person}{Thomas Mesnard}, \bibinfo{person}{Cassidy Hardin}, \bibinfo{person}{Robert Dadashi}, \bibinfo{person}{Surya Bhupatiraju}, \bibinfo{person}{Shreya Pathak}, \bibinfo{person}{Laurent Sifre}, \bibinfo{person}{Morgane Rivi{\`e}re}, \bibinfo{person}{Mihir~Sanjay Kale}, \bibinfo{person}{Juliette Love}, {et~al\mbox{.}}} \bibinfo{year}{2024}\natexlab{a}.
\newblock \showarticletitle{Gemma: Open models based on gemini research and technology}.
\newblock \bibinfo{journal}{\emph{arXiv preprint arXiv:2403.08295}} (\bibinfo{year}{2024}).
\newblock


\bibitem[Team et~al\mbox{.}(2024b)]%
        {GoogleGemma2}
\bibfield{author}{\bibinfo{person}{Gemma Team}, \bibinfo{person}{Morgane Riviere}, \bibinfo{person}{Shreya Pathak}, \bibinfo{person}{Pier~Giuseppe Sessa}, \bibinfo{person}{Cassidy Hardin}, \bibinfo{person}{Surya Bhupatiraju}, \bibinfo{person}{L{\'e}onard Hussenot}, \bibinfo{person}{Thomas Mesnard}, \bibinfo{person}{Bobak Shahriari}, \bibinfo{person}{Alexandre Ram{\'e}}, {et~al\mbox{.}}} \bibinfo{year}{2024}\natexlab{b}.
\newblock \showarticletitle{Gemma 2: Improving open language models at a practical size}.
\newblock \bibinfo{journal}{\emph{arXiv preprint arXiv:2408.00118}} (\bibinfo{year}{2024}).
\newblock


\bibitem[Team(2023)]%
        {mlc-llm}
\bibfield{author}{\bibinfo{person}{MLC Team}.} \bibinfo{year}{2023}\natexlab{}.
\newblock \bibinfo{title}{MLC LLM}.
\newblock \bibinfo{howpublished}{\url{https://github.com/mlc-ai/mlc-llm}}.
\newblock
\newblock
\shownote{Accessed on June 6, 2024}.


\bibitem[Team(2024)]%
        {QualcommSnapdragon8Gen3MobilePlatform}
\bibfield{author}{\bibinfo{person}{Qualcomm Team}.} \bibinfo{year}{2024}\natexlab{}.
\newblock \bibinfo{title}{Snapdragon 8 Gen 3 Mobile Platform}.
\newblock \bibinfo{howpublished}{\url{https://www.qualcomm.com/products/mobile/snapdragon/smartphones/snapdragon-8-series-mobile-platforms/snapdragon-8-gen-3-mobile-platform}}.
\newblock
\newblock
\shownote{Accessed on December 3, 2024}.


\bibitem[Tian et~al\mbox{.}(2024a)]%
        {tian2024large}
\bibfield{author}{\bibinfo{person}{Wenbin Tian}, \bibinfo{person}{Chaojie Gu}, \bibinfo{person}{Miao Guo}, \bibinfo{person}{Shibo He}, \bibinfo{person}{Jiawen Kang}, \bibinfo{person}{Dusit Niyato}, {and} \bibinfo{person}{Jiming Chen}.} \bibinfo{year}{2024}\natexlab{a}.
\newblock \showarticletitle{Large-Scale Deterministic Networks: Architecture, Enabling Technologies, Case Study and Future Directions}.
\newblock \bibinfo{journal}{\emph{IEEE Network}} (\bibinfo{year}{2024}).
\newblock


\bibitem[Tian et~al\mbox{.}(2024b)]%
        {tian2024drivevlm}
\bibfield{author}{\bibinfo{person}{Xiaoyu Tian}, \bibinfo{person}{Junru Gu}, \bibinfo{person}{Bailin Li}, \bibinfo{person}{Yicheng Liu}, \bibinfo{person}{Yang Wang}, \bibinfo{person}{Zhiyong Zhao}, \bibinfo{person}{Kun Zhan}, \bibinfo{person}{Peng Jia}, \bibinfo{person}{Xianpeng Lang}, {and} \bibinfo{person}{Hang Zhao}.} \bibinfo{year}{2024}\natexlab{b}.
\newblock \showarticletitle{Drivevlm: The convergence of autonomous driving and large vision-language models}.
\newblock \bibinfo{journal}{\emph{arXiv preprint arXiv:2402.12289}} (\bibinfo{year}{2024}).
\newblock


\bibitem[Tian et~al\mbox{.}(2024c)]%
        {tian2024dialoguesummarization}
\bibfield{author}{\bibinfo{person}{Yuanhe Tian}, \bibinfo{person}{Fei Xia}, {and} \bibinfo{person}{Yan Song}.} \bibinfo{year}{2024}\natexlab{c}.
\newblock \showarticletitle{Dialogue summarization with mixture of experts based on large language models}.
\newblock \bibinfo{journal}{\emph{ACL}} (\bibinfo{year}{2024}).
\newblock


\bibitem[Touvron et~al\mbox{.}(2023a)]%
        {touvron2023llama}
\bibfield{author}{\bibinfo{person}{Hugo Touvron}, \bibinfo{person}{Thibaut Lavril}, \bibinfo{person}{Gautier Izacard}, \bibinfo{person}{Xavier Martinet}, \bibinfo{person}{Marie-Anne Lachaux}, \bibinfo{person}{Timoth{\'e}e Lacroix}, \bibinfo{person}{Baptiste Rozi{\`e}re}, \bibinfo{person}{Naman Goyal}, \bibinfo{person}{Eric Hambro}, \bibinfo{person}{Faisal Azhar}, {et~al\mbox{.}}} \bibinfo{year}{2023}\natexlab{a}.
\newblock \showarticletitle{Llama: Open and efficient foundation language models}.
\newblock \bibinfo{journal}{\emph{arXiv preprint arXiv:2302.13971}} (\bibinfo{year}{2023}).
\newblock


\bibitem[Touvron et~al\mbox{.}(2023b)]%
        {touvron2023llama2}
\bibfield{author}{\bibinfo{person}{Hugo Touvron}, \bibinfo{person}{Louis Martin}, \bibinfo{person}{Kevin Stone}, \bibinfo{person}{Peter Albert}, \bibinfo{person}{Amjad Almahairi}, \bibinfo{person}{Yasmine Babaei}, \bibinfo{person}{Nikolay Bashlykov}, \bibinfo{person}{Soumya Batra}, \bibinfo{person}{Prajjwal Bhargava}, \bibinfo{person}{Shruti Bhosale}, {et~al\mbox{.}}} \bibinfo{year}{2023}\natexlab{b}.
\newblock \showarticletitle{Llama 2: Open foundation and fine-tuned chat models}.
\newblock \bibinfo{journal}{\emph{arXiv preprint arXiv:2307.09288}} (\bibinfo{year}{2023}).
\newblock


\bibitem[Treviso et~al\mbox{.}(2023)]%
        {trevisoEfficientMethodsNatural2023}
\bibfield{author}{\bibinfo{person}{Marcos Treviso}, \bibinfo{person}{Ji-Ung Lee}, \bibinfo{person}{Tianchu Ji}, \bibinfo{person}{Betty {van Aken}}, \bibinfo{person}{Qingqing Cao}, \bibinfo{person}{Manuel~R. Ciosici}, \bibinfo{person}{Michael Hassid}, \bibinfo{person}{Kenneth Heafield}, \bibinfo{person}{Sara Hooker}, \bibinfo{person}{Colin Raffel}, \bibinfo{person}{Pedro~H. Martins}, \bibinfo{person}{Andr{\'e} F.~T. Martins}, \bibinfo{person}{Jessica~Zosa Forde}, \bibinfo{person}{Peter Milder}, \bibinfo{person}{Edwin Simpson}, \bibinfo{person}{Noam Slonim}, \bibinfo{person}{Jesse Dodge}, \bibinfo{person}{Emma Strubell}, \bibinfo{person}{Niranjan Balasubramanian}, \bibinfo{person}{Leon Derczynski}, \bibinfo{person}{Iryna Gurevych}, {and} \bibinfo{person}{Roy Schwartz}.} \bibinfo{year}{2023}\natexlab{}.
\newblock \showarticletitle{Efficient Methods for Natural Language Processing: A Survey}.
\newblock \bibinfo{journal}{\emph{Transactions of the Association for Computational Linguistics}} (\bibinfo{year}{2023}).
\newblock


\bibitem[Tseng et~al\mbox{.}(2024)]%
        {tseng2024quipbetter}
\bibfield{author}{\bibinfo{person}{Albert Tseng}, \bibinfo{person}{Jerry Chee}, \bibinfo{person}{Qingyao Sun}, \bibinfo{person}{Volodymyr Kuleshov}, {and} \bibinfo{person}{Christopher De~Sa}.} \bibinfo{year}{2024}\natexlab{}.
\newblock \showarticletitle{Quip\#: Even better LLM quantization with hadamard incoherence and lattice codebooks}.
\newblock \bibinfo{journal}{\emph{ICML}} (\bibinfo{year}{2024}).
\newblock


\bibitem[Tu et~al\mbox{.}(2022)]%
        {tu202228nmTranCIM}
\bibfield{author}{\bibinfo{person}{Fengbin Tu}, \bibinfo{person}{Zihan Wu}, \bibinfo{person}{Yiqi Wang}, \bibinfo{person}{Ling Liang}, \bibinfo{person}{Liu Liu}, \bibinfo{person}{Yufei Ding}, \bibinfo{person}{Leibo Liu}, \bibinfo{person}{Shaojun Wei}, \bibinfo{person}{Yuan Xie}, {and} \bibinfo{person}{Shouyi Yin}.} \bibinfo{year}{2022}\natexlab{}.
\newblock \showarticletitle{A 28nm 15.59 $\mu$J/token full-digital bitline-transpose CIM-based sparse transformer accelerator with pipeline/parallel reconfigurable modes}.
\newblock \bibinfo{journal}{\emph{IEEE ISSCC}} (\bibinfo{year}{2022}).
\newblock


\bibitem[Tu et~al\mbox{.}(2023)]%
        {tu202316MulTCIM}
\bibfield{author}{\bibinfo{person}{Fengbin Tu}, \bibinfo{person}{Zihan Wu}, \bibinfo{person}{Yiqi Wang}, \bibinfo{person}{Weiwei Wu}, \bibinfo{person}{Leibo Liu}, \bibinfo{person}{Yang Hu}, \bibinfo{person}{Shaojun Wei}, {and} \bibinfo{person}{Shouyi Yin}.} \bibinfo{year}{2023}\natexlab{}.
\newblock \showarticletitle{16.1 MuITCIM: A 28nm $2.24 \mu\mathrm{J}$/Token Attention-Token-Bit Hybrid Sparse Digital CIM-Based Accelerator for Multimodal Transformers}.
\newblock \bibinfo{journal}{\emph{IEEE ISSCC}} (\bibinfo{year}{2023}).
\newblock


\bibitem[Tuli and Jha(2023)]%
        {tuliAccelTranSparsityAwareAccelerator2023}
\bibfield{author}{\bibinfo{person}{Shikhar Tuli} {and} \bibinfo{person}{Niraj~K. Jha}.} \bibinfo{year}{2023}\natexlab{}.
\newblock \showarticletitle{AccelTran: A Sparsity-Aware Accelerator for Dynamic Inference With Transformers}.
\newblock \bibinfo{journal}{\emph{IEEE Transactions on Computer-Aided Design of Integrated Circuits and Systems}} (\bibinfo{year}{2023}).
\newblock


\bibitem[Vaswani et~al\mbox{.}(2017)]%
        {vaswani2017attention}
\bibfield{author}{\bibinfo{person}{Ashish Vaswani}, \bibinfo{person}{Noam Shazeer}, \bibinfo{person}{Niki Parmar}, \bibinfo{person}{Jakob Uszkoreit}, \bibinfo{person}{Llion Jones}, \bibinfo{person}{Aidan~N Gomez}, \bibinfo{person}{\L~ukasz Kaiser}, {and} \bibinfo{person}{Illia Polosukhin}.} \bibinfo{year}{2017}\natexlab{}.
\newblock \showarticletitle{Attention is all you need}.
\newblock \bibinfo{journal}{\emph{NeurIPS}} (\bibinfo{year}{2017}).
\newblock


\bibitem[Wan et~al\mbox{.}(2024a)]%
        {wan2024knowledgefusion}
\bibfield{author}{\bibinfo{person}{Fanqi Wan}, \bibinfo{person}{Xinting Huang}, \bibinfo{person}{Deng Cai}, \bibinfo{person}{Xiaojun Quan}, \bibinfo{person}{Wei Bi}, {and} \bibinfo{person}{Shuming Shi}.} \bibinfo{year}{2024}\natexlab{a}.
\newblock \showarticletitle{Knowledge fusion of large language models}.
\newblock \bibinfo{journal}{\emph{ICLR}} (\bibinfo{year}{2024}).
\newblock


\bibitem[Wan et~al\mbox{.}(2024b)]%
        {wanEfficientLargeLanguage2024}
\bibfield{author}{\bibinfo{person}{Zhongwei Wan}, \bibinfo{person}{Xin Wang}, \bibinfo{person}{Che Liu}, \bibinfo{person}{Samiul Alam}, \bibinfo{person}{Yu Zheng}, \bibinfo{person}{Jiachen Liu}, \bibinfo{person}{Zhongnan Qu}, \bibinfo{person}{Shen Yan}, \bibinfo{person}{Yi Zhu}, \bibinfo{person}{Quanlu Zhang}, \bibinfo{person}{Mosharaf Chowdhury}, {and} \bibinfo{person}{Mi Zhang}.} \bibinfo{year}{2024}\natexlab{b}.
\newblock \showarticletitle{Efficient Large Language Models: A Survey}.
\newblock \bibinfo{journal}{\emph{Transactions on Machine Learning Research}} (\bibinfo{year}{2024}).
\newblock


\bibitem[Wang et~al\mbox{.}(2021b)]%
        {wangSpAttenEfficientSparse2021}
\bibfield{author}{\bibinfo{person}{Hanrui Wang}, \bibinfo{person}{Zhekai Zhang}, {and} \bibinfo{person}{Song Han}.} \bibinfo{year}{2021}\natexlab{b}.
\newblock \showarticletitle{SpAtten: Efficient Sparse Attention Architecture with Cascade Token and Head Pruning}.
\newblock \bibinfo{journal}{\emph{IEEE HPCA}} (\bibinfo{year}{2021}).
\newblock


\bibitem[Wang et~al\mbox{.}(2025a)]%
        {10734312}
\bibfield{author}{\bibinfo{person}{Rongkai Wang}, \bibinfo{person}{Yiyang Jing}, \bibinfo{person}{Chaojie Gu}, \bibinfo{person}{Shibo He}, {and} \bibinfo{person}{Jiming Chen}.} \bibinfo{year}{2025}\natexlab{a}.
\newblock \showarticletitle{{End-to-End Multitarget Flexible Job Shop Scheduling With Deep Reinforcement Learning}}.
\newblock \bibinfo{journal}{\emph{IEEE Internet of Things Journal}} \bibinfo{volume}{12}, \bibinfo{number}{4} (\bibinfo{year}{2025}), \bibinfo{pages}{4420--4434}.
\newblock


\bibitem[Wang et~al\mbox{.}(2021a)]%
        {wangMiniLMv2MultiHeadSelfAttention2021}
\bibfield{author}{\bibinfo{person}{Wenhui Wang}, \bibinfo{person}{Hangbo Bao}, \bibinfo{person}{Shaohan Huang}, \bibinfo{person}{Li Dong}, {and} \bibinfo{person}{Furu Wei}.} \bibinfo{year}{2021}\natexlab{a}.
\newblock \showarticletitle{MiniLMv2: Multi-Head Self-Attention Relation Distillation for Compressing Pretrained Transformers}.
\newblock \bibinfo{journal}{\emph{ACL-IJCNLP}} (\bibinfo{year}{2021}).
\newblock


\bibitem[Wang et~al\mbox{.}(2020b)]%
        {wangMiniLMDeepSelfAttention2020}
\bibfield{author}{\bibinfo{person}{Wenhui Wang}, \bibinfo{person}{Furu Wei}, \bibinfo{person}{Li Dong}, \bibinfo{person}{Hangbo Bao}, \bibinfo{person}{Nan Yang}, {and} \bibinfo{person}{Ming Zhou}.} \bibinfo{year}{2020}\natexlab{b}.
\newblock \showarticletitle{MiniLM: Deep Self-Attention Distillation for Task-Agnostic Compression of Pre-Trained Transformers}.
\newblock \bibinfo{journal}{\emph{NeurIPS}} (\bibinfo{year}{2020}).
\newblock


\bibitem[Wang et~al\mbox{.}(2020a)]%
        {wangConvergenceEdgeComputing2020a}
\bibfield{author}{\bibinfo{person}{Xiaofei Wang}, \bibinfo{person}{Yiwen Han}, \bibinfo{person}{Victor C.~M. Leung}, \bibinfo{person}{Dusit Niyato}, \bibinfo{person}{Xueqiang Yan}, {and} \bibinfo{person}{Xu Chen}.} \bibinfo{year}{2020}\natexlab{a}.
\newblock \showarticletitle{Convergence of Edge Computing and Deep Learning: A Comprehensive Survey}.
\newblock \bibinfo{journal}{\emph{IEEE Communications Surveys \& Tutorials}} (\bibinfo{year}{2020}).
\newblock


\bibitem[Wang et~al\mbox{.}(2023)]%
        {wang2023tabi}
\bibfield{author}{\bibinfo{person}{Yiding Wang}, \bibinfo{person}{Kai Chen}, \bibinfo{person}{Haisheng Tan}, {and} \bibinfo{person}{Kun Guo}.} \bibinfo{year}{2023}\natexlab{}.
\newblock \showarticletitle{Tabi: An efficient multi-level inference system for large language models}.
\newblock \bibinfo{journal}{\emph{ACM EuroSys}} (\bibinfo{year}{2023}).
\newblock


\bibitem[Wang et~al\mbox{.}(2024)]%
        {wang2024end}
\bibfield{author}{\bibinfo{person}{Yingchao Wang}, \bibinfo{person}{Chen Yang}, \bibinfo{person}{Shulin Lan}, \bibinfo{person}{Liehuang Zhu}, {and} \bibinfo{person}{Yan Zhang}.} \bibinfo{year}{2024}\natexlab{}.
\newblock \showarticletitle{End-edge-cloud collaborative computing for deep learning: A comprehensive survey}.
\newblock \bibinfo{journal}{\emph{IEEE Communications Surveys \& Tutorials}} (\bibinfo{year}{2024}).
\newblock


\bibitem[Wang et~al\mbox{.}(2025b)]%
        {wang2025feddfa}
\bibfield{author}{\bibinfo{person}{Zichen Wang}, \bibinfo{person}{Feng Yan}, \bibinfo{person}{Tianyi Wang}, \bibinfo{person}{Cong Wang}, \bibinfo{person}{Yuanchao Shu}, \bibinfo{person}{Peng Cheng}, {and} \bibinfo{person}{Jiming Chen}.} \bibinfo{year}{2025}\natexlab{b}.
\newblock \showarticletitle{Fed-DFA: Federated Distillation for Heterogeneous Model Fusion through the Adversarial Lens}.
\newblock \bibinfo{journal}{\emph{AAAI}} (\bibinfo{year}{2025}).
\newblock


\bibitem[Wen et~al\mbox{.}(2024)]%
        {wenAutoDroidLLMpoweredTask2024}
\bibfield{author}{\bibinfo{person}{Hao Wen}, \bibinfo{person}{Yuanchun Li}, \bibinfo{person}{Guohong Liu}, \bibinfo{person}{Shanhui Zhao}, \bibinfo{person}{Tao Yu}, \bibinfo{person}{Toby Jia-Jun Li}, \bibinfo{person}{Shiqi Jiang}, \bibinfo{person}{Yunhao Liu}, \bibinfo{person}{Yaqin Zhang}, {and} \bibinfo{person}{Yunxin Liu}.} \bibinfo{year}{2024}\natexlab{}.
\newblock \showarticletitle{AutoDroid: LLM-powered Task Automation in Android}.
\newblock \bibinfo{journal}{\emph{ACM MobiCom}} (\bibinfo{year}{2024}).
\newblock


\bibitem[Wu et~al\mbox{.}(2024)]%
        {wuLaMiniLMDiverseHerd2024}
\bibfield{author}{\bibinfo{person}{Minghao Wu}, \bibinfo{person}{Abdul Waheed}, \bibinfo{person}{Chiyu Zhang}, \bibinfo{person}{Muhammad Abdul-Mageed}, {and} \bibinfo{person}{Alham Aji}.} \bibinfo{year}{2024}\natexlab{}.
\newblock \showarticletitle{LaMini-LM: A Diverse Herd of Distilled Models from Large-Scale Instructions}.
\newblock \bibinfo{journal}{\emph{EACL}} (\bibinfo{year}{2024}).
\newblock


\bibitem[Xia et~al\mbox{.}(2024)]%
        {xiaShearedLLaMAAccelerating2023}
\bibfield{author}{\bibinfo{person}{Mengzhou Xia}, \bibinfo{person}{Tianyu Gao}, \bibinfo{person}{Zhiyuan Zeng}, {and} \bibinfo{person}{Danqi Chen}.} \bibinfo{year}{2024}\natexlab{}.
\newblock \showarticletitle{Sheared LLaMA: Accelerating Language Model Pre-training via Structured Pruning}.
\newblock \bibinfo{journal}{\emph{ICML}} (\bibinfo{year}{2024}).
\newblock


\bibitem[Xia et~al\mbox{.}(2022)]%
        {xiaStructuredPruningLearns2022}
\bibfield{author}{\bibinfo{person}{Mengzhou Xia}, \bibinfo{person}{Zexuan Zhong}, {and} \bibinfo{person}{Danqi Chen}.} \bibinfo{year}{2022}\natexlab{}.
\newblock \showarticletitle{Structured Pruning Learns Compact and Accurate Models}.
\newblock \bibinfo{journal}{\emph{ACL}} (\bibinfo{year}{2022}).
\newblock


\bibitem[Xiang et~al\mbox{.}(2024)]%
        {xiang2024languageEmbodied}
\bibfield{author}{\bibinfo{person}{Jiannan Xiang}, \bibinfo{person}{Tianhua Tao}, \bibinfo{person}{Yi Gu}, \bibinfo{person}{Tianmin Shu}, \bibinfo{person}{Zirui Wang}, \bibinfo{person}{Zichao Yang}, {and} \bibinfo{person}{Zhiting Hu}.} \bibinfo{year}{2024}\natexlab{}.
\newblock \showarticletitle{Language models meet world models: Embodied experiences enhance language models}.
\newblock \bibinfo{journal}{\emph{NeurIPS}} (\bibinfo{year}{2024}).
\newblock


\bibitem[Xiao et~al\mbox{.}(2023)]%
        {xiaoSmoothQuantAccurateEfficient2023}
\bibfield{author}{\bibinfo{person}{Guangxuan Xiao}, \bibinfo{person}{Ji Lin}, \bibinfo{person}{Mickael Seznec}, \bibinfo{person}{Hao Wu}, \bibinfo{person}{Julien Demouth}, {and} \bibinfo{person}{Song Han}.} \bibinfo{year}{2023}\natexlab{}.
\newblock \showarticletitle{SmoothQuant: Accurate and Efficient Post-Training Quantization for Large Language Models}.
\newblock \bibinfo{journal}{\emph{ICML}} (\bibinfo{year}{2023}).
\newblock


\bibitem[Xu and McAuley(2023)]%
        {xuSurveyModelCompression2023}
\bibfield{author}{\bibinfo{person}{Canwen Xu} {and} \bibinfo{person}{Julian McAuley}.} \bibinfo{year}{2023}\natexlab{}.
\newblock \showarticletitle{A survey on model compression and acceleration for pretrained language models}.
\newblock \bibinfo{journal}{\emph{AAAI}} (\bibinfo{year}{2023}).
\newblock


\bibitem[Xu et~al\mbox{.}(2024d)]%
        {xu2024EdgeLLMSpeculative}
\bibfield{author}{\bibinfo{person}{Daliang Xu}, \bibinfo{person}{Wangsong Yin}, \bibinfo{person}{Hao Zhang}, \bibinfo{person}{Xin Jin}, \bibinfo{person}{Ying Zhang}, \bibinfo{person}{Shiyun Wei}, \bibinfo{person}{Mengwei Xu}, {and} \bibinfo{person}{Xuanzhe Liu}.} \bibinfo{year}{2024}\natexlab{d}.
\newblock \showarticletitle{EdgeLLM: Fast On-device LLM Inference with Speculative Decoding}.
\newblock \bibinfo{journal}{\emph{IEEE Transactions on Mobile Computing}} (\bibinfo{year}{2024}).
\newblock


\bibitem[Xu et~al\mbox{.}(2023)]%
        {xu2023mesen}
\bibfield{author}{\bibinfo{person}{Lilin Xu}, \bibinfo{person}{Chaojie Gu}, \bibinfo{person}{Rui Tan}, \bibinfo{person}{Shibo He}, {and} \bibinfo{person}{Jiming Chen}.} \bibinfo{year}{2023}\natexlab{}.
\newblock \showarticletitle{MESEN: Exploit Multimodal Data to Design Unimodal Human Activity Recognition with Few Labels}.
\newblock \bibinfo{journal}{\emph{ACM SenSys}} (\bibinfo{year}{2023}).
\newblock


\bibitem[Xu et~al\mbox{.}(2024b)]%
        {xu2024gestureprint}
\bibfield{author}{\bibinfo{person}{Lilin Xu}, \bibinfo{person}{Keyi Wang}, \bibinfo{person}{Chaojie Gu}, \bibinfo{person}{Xiuzhen Guo}, \bibinfo{person}{Shibo He}, {and} \bibinfo{person}{Jiming Chen}.} \bibinfo{year}{2024}\natexlab{b}.
\newblock \showarticletitle{GesturePrint: Enabling user identification for mmWave-based gesture recognition systems}.
\newblock \bibinfo{journal}{\emph{IEEE ICDCS}} (\bibinfo{year}{2024}).
\newblock


\bibitem[Xu et~al\mbox{.}(2024a)]%
        {xu2024besapruning}
\bibfield{author}{\bibinfo{person}{Peng Xu}, \bibinfo{person}{Wenqi Shao}, \bibinfo{person}{Mengzhao Chen}, \bibinfo{person}{Shitao Tang}, \bibinfo{person}{Kaipeng Zhang}, \bibinfo{person}{Peng Gao}, \bibinfo{person}{Fengwei An}, \bibinfo{person}{Yu Qiao}, {and} \bibinfo{person}{Ping Luo}.} \bibinfo{year}{2024}\natexlab{a}.
\newblock \showarticletitle{BESA: Pruning Large Language Models with Blockwise Parameter-Efficient Sparsity Allocation}.
\newblock \bibinfo{journal}{\emph{ICLR}} (\bibinfo{year}{2024}).
\newblock


\bibitem[Xu et~al\mbox{.}(2024c)]%
        {xu2024mentallm}
\bibfield{author}{\bibinfo{person}{Xuhai Xu}, \bibinfo{person}{Bingsheng Yao}, \bibinfo{person}{Yuanzhe Dong}, \bibinfo{person}{Saadia Gabriel}, \bibinfo{person}{Hong Yu}, \bibinfo{person}{James Hendler}, \bibinfo{person}{Marzyeh Ghassemi}, \bibinfo{person}{Anind~K Dey}, {and} \bibinfo{person}{Dakuo Wang}.} \bibinfo{year}{2024}\natexlab{c}.
\newblock \showarticletitle{Mental-llm: Leveraging large language models for mental health prediction via online text data}.
\newblock \bibinfo{journal}{\emph{ACM UbiComp}} (\bibinfo{year}{2024}).
\newblock


\bibitem[Yan et~al\mbox{.}(2024)]%
        {yan2024federa}
\bibfield{author}{\bibinfo{person}{Yuxuan Yan}, \bibinfo{person}{Qianqian Yang}, \bibinfo{person}{Shunpu Tang}, {and} \bibinfo{person}{Zhiguo Shi}.} \bibinfo{year}{2024}\natexlab{}.
\newblock \showarticletitle{Federa: Efficient fine-tuning of language models in federated learning leveraging weight decomposition}.
\newblock \bibinfo{journal}{\emph{arXiv preprint arXiv:2404.18848}} (\bibinfo{year}{2024}).
\newblock


\bibitem[Yang et~al\mbox{.}(2024c)]%
        {yang2024qwen2}
\bibfield{author}{\bibinfo{person}{An Yang}, \bibinfo{person}{Baosong Yang}, \bibinfo{person}{Binyuan Hui}, \bibinfo{person}{Bo Zheng}, \bibinfo{person}{Bowen Yu}, \bibinfo{person}{Chang Zhou}, \bibinfo{person}{Chengpeng Li}, \bibinfo{person}{Chengyuan Li}, \bibinfo{person}{Dayiheng Liu}, \bibinfo{person}{Fei Huang}, {et~al\mbox{.}}} \bibinfo{year}{2024}\natexlab{c}.
\newblock \showarticletitle{Qwen2 technical report}.
\newblock \bibinfo{journal}{\emph{arXiv preprint arXiv:2407.10671}} (\bibinfo{year}{2024}).
\newblock


\bibitem[Yang et~al\mbox{.}(2024b)]%
        {yang2024largeLLMAO}
\bibfield{author}{\bibinfo{person}{Aidan~ZH Yang}, \bibinfo{person}{Claire Le~Goues}, \bibinfo{person}{Ruben Martins}, {and} \bibinfo{person}{Vincent Hellendoorn}.} \bibinfo{year}{2024}\natexlab{b}.
\newblock \showarticletitle{Large language models for test-free fault localization}.
\newblock \bibinfo{journal}{\emph{ACM/IEEE ICSE}} (\bibinfo{year}{2024}).
\newblock


\bibitem[Yang et~al\mbox{.}(2024d)]%
        {yang2024mentallama}
\bibfield{author}{\bibinfo{person}{Kailai Yang}, \bibinfo{person}{Tianlin Zhang}, \bibinfo{person}{Ziyan Kuang}, \bibinfo{person}{Qianqian Xie}, \bibinfo{person}{Jimin Huang}, {and} \bibinfo{person}{Sophia Ananiadou}.} \bibinfo{year}{2024}\natexlab{d}.
\newblock \showarticletitle{MentaLLaMA: interpretable mental health analysis on social media with large language models}.
\newblock \bibinfo{journal}{\emph{ACM WWW}} (\bibinfo{year}{2024}).
\newblock


\bibitem[Yang et~al\mbox{.}(2024a)]%
        {yang2024maf}
\bibfield{author}{\bibinfo{person}{Yongjie Yang}, \bibinfo{person}{Tao Chen}, \bibinfo{person}{Yujing Huang}, \bibinfo{person}{Xiuzhen Guo}, {and} \bibinfo{person}{Longfei Shangguan}.} \bibinfo{year}{2024}\natexlab{a}.
\newblock \showarticletitle{MAF: Exploring Mobile Acoustic Field for Hand-to-Face Gesture Interactions}.
\newblock \bibinfo{journal}{\emph{ACM CHI}} (\bibinfo{year}{2024}).
\newblock


\bibitem[Yao et~al\mbox{.}(2022)]%
        {yaoZeroQuantEfficientAffordable2022}
\bibfield{author}{\bibinfo{person}{Zhewei Yao}, \bibinfo{person}{Reza Yazdani~Aminabadi}, \bibinfo{person}{Minjia Zhang}, \bibinfo{person}{Xiaoxia Wu}, \bibinfo{person}{Conglong Li}, {and} \bibinfo{person}{Yuxiong He}.} \bibinfo{year}{2022}\natexlab{}.
\newblock \showarticletitle{ZeroQuant: Efficient and Affordable Post-Training Quantization for Large-Scale Transformers}.
\newblock \bibinfo{journal}{\emph{NeurIPS}} (\bibinfo{year}{2022}).
\newblock


\bibitem[Yu et~al\mbox{.}(2023)]%
        {yuBoostTransformerbasedLanguage2023}
\bibfield{author}{\bibinfo{person}{Chong Yu}, \bibinfo{person}{Tao Chen}, {and} \bibinfo{person}{Zhongxue Gan}.} \bibinfo{year}{2023}\natexlab{}.
\newblock \showarticletitle{Boost Transformer-based Language Models with GPU-Friendly Sparsity and Quantization}.
\newblock \bibinfo{journal}{\emph{Findings of ACL}} (\bibinfo{year}{2023}).
\newblock


\bibitem[Yuan et~al\mbox{.}(2024)]%
        {yuan2024mobilefirmware}
\bibfield{author}{\bibinfo{person}{Jinliang Yuan}, \bibinfo{person}{Chen Yang}, \bibinfo{person}{Dongqi Cai}, \bibinfo{person}{Shihe Wang}, \bibinfo{person}{Xin Yuan}, \bibinfo{person}{Zeling Zhang}, \bibinfo{person}{Xiang Li}, \bibinfo{person}{Dingge Zhang}, \bibinfo{person}{Hanzi Mei}, \bibinfo{person}{Xianqing Jia}, {et~al\mbox{.}}} \bibinfo{year}{2024}\natexlab{}.
\newblock \showarticletitle{Mobile Foundation Model as Firmware}.
\newblock \bibinfo{journal}{\emph{ACM MobiCom}} (\bibinfo{year}{2024}).
\newblock


\bibitem[Zadeh et~al\mbox{.}(2020)]%
        {zadeh2020gobo}
\bibfield{author}{\bibinfo{person}{Ali~Hadi Zadeh}, \bibinfo{person}{Isak Edo}, \bibinfo{person}{Omar~Mohamed Awad}, {and} \bibinfo{person}{Andreas Moshovos}.} \bibinfo{year}{2020}\natexlab{}.
\newblock \showarticletitle{Gobo: Quantizing attention-based nlp models for low latency and energy efficient inference}.
\newblock \bibinfo{journal}{\emph{IEEE/ACM MICRO}} (\bibinfo{year}{2020}).
\newblock


\bibitem[Zadeh et~al\mbox{.}(2022)]%
        {zadeh2022mokey}
\bibfield{author}{\bibinfo{person}{Ali~Hadi Zadeh}, \bibinfo{person}{Mostafa Mahmoud}, \bibinfo{person}{Ameer Abdelhadi}, {and} \bibinfo{person}{Andreas Moshovos}.} \bibinfo{year}{2022}\natexlab{}.
\newblock \showarticletitle{Mokey: Enabling narrow fixed-point inference for out-of-the-box floating-point transformer models}.
\newblock \bibinfo{journal}{\emph{ACM/IEEE ISCA}} (\bibinfo{year}{2022}).
\newblock


\bibitem[Zeng et~al\mbox{.}(2024)]%
        {zeng2024consistentee}
\bibfield{author}{\bibinfo{person}{Ziqian Zeng}, \bibinfo{person}{Yihuai Hong}, \bibinfo{person}{Hongliang Dai}, \bibinfo{person}{Huiping Zhuang}, {and} \bibinfo{person}{Cen Chen}.} \bibinfo{year}{2024}\natexlab{}.
\newblock \showarticletitle{ConsistentEE: A Consistent and Hardness-Guided Early Exiting Method for Accelerating Language Models Inference}.
\newblock \bibinfo{journal}{\emph{AAAI}} (\bibinfo{year}{2024}).
\newblock


\bibitem[Zhang et~al\mbox{.}(2023)]%
        {zhangLiftingCurseCapacity2023}
\bibfield{author}{\bibinfo{person}{Chen Zhang}, \bibinfo{person}{Yang Yang}, \bibinfo{person}{Jiahao Liu}, \bibinfo{person}{Jingang Wang}, \bibinfo{person}{Yunsen Xian}, \bibinfo{person}{Benyou Wang}, {and} \bibinfo{person}{Dawei Song}.} \bibinfo{year}{2023}\natexlab{}.
\newblock \showarticletitle{Lifting the Curse of Capacity Gap in Distilling Language Models}.
\newblock \bibinfo{journal}{\emph{ACL}} (\bibinfo{year}{2023}).
\newblock


\bibitem[Zhang et~al\mbox{.}(2024e)]%
        {zhang2024llamatouch}
\bibfield{author}{\bibinfo{person}{Li Zhang}, \bibinfo{person}{Shihe Wang}, \bibinfo{person}{Xianqing Jia}, \bibinfo{person}{Zhihan Zheng}, \bibinfo{person}{Yunhe Yan}, \bibinfo{person}{Longxi Gao}, \bibinfo{person}{Yuanchun Li}, {and} \bibinfo{person}{Mengwei Xu}.} \bibinfo{year}{2024}\natexlab{e}.
\newblock \showarticletitle{LlamaTouch: A Faithful and Scalable Testbed for Mobile UI Automation Task Evaluation}.
\newblock \bibinfo{journal}{\emph{ACM UIST}} (\bibinfo{year}{2024}).
\newblock


\bibitem[Zhang et~al\mbox{.}(2024b)]%
        {zhang2024loraprunestructuredpruningmeets}
\bibfield{author}{\bibinfo{person}{Mingyang Zhang}, \bibinfo{person}{Hao Chen}, \bibinfo{person}{Chunhua Shen}, \bibinfo{person}{Zhen Yang}, \bibinfo{person}{Linlin Ou}, \bibinfo{person}{Xinyi Yu}, {and} \bibinfo{person}{Bohan Zhuang}.} \bibinfo{year}{2024}\natexlab{b}.
\newblock \showarticletitle{LoRAPrune: Structured Pruning Meets Low-Rank Parameter-Efficient Fine-Tuning}.
\newblock \bibinfo{journal}{\emph{Findlings of ACL}} (\bibinfo{year}{2024}).
\newblock


\bibitem[Zhang et~al\mbox{.}(2022)]%
        {zhang2022opt}
\bibfield{author}{\bibinfo{person}{Susan Zhang}, \bibinfo{person}{Stephen Roller}, \bibinfo{person}{Naman Goyal}, \bibinfo{person}{Mikel Artetxe}, \bibinfo{person}{Moya Chen}, \bibinfo{person}{Shuohui Chen}, \bibinfo{person}{Christopher Dewan}, \bibinfo{person}{Mona Diab}, \bibinfo{person}{Xian Li}, \bibinfo{person}{Xi~Victoria Lin}, {et~al\mbox{.}}} \bibinfo{year}{2022}\natexlab{}.
\newblock \showarticletitle{Opt: Open pre-trained transformer language models}.
\newblock \bibinfo{journal}{\emph{arXiv preprint arXiv:2205.01068}} (\bibinfo{year}{2022}).
\newblock


\bibitem[Zhang et~al\mbox{.}(2020)]%
        {zhang2020ternarybert}
\bibfield{author}{\bibinfo{person}{Wei Zhang}, \bibinfo{person}{Lu Hou}, \bibinfo{person}{Yichun Yin}, \bibinfo{person}{Lifeng Shang}, \bibinfo{person}{Xiao Chen}, \bibinfo{person}{Xin Jiang}, {and} \bibinfo{person}{Qun Liu}.} \bibinfo{year}{2020}\natexlab{}.
\newblock \showarticletitle{TernaryBERT: Distillation-aware Ultra-low Bit BERT}.
\newblock \bibinfo{journal}{\emph{EMNLP}} (\bibinfo{year}{2020}).
\newblock


\bibitem[Zhang and Debroy(2023)]%
        {zhangResourceManagementMobile2023a}
\bibfield{author}{\bibinfo{person}{Xiaojie Zhang} {and} \bibinfo{person}{Saptarshi Debroy}.} \bibinfo{year}{2023}\natexlab{}.
\newblock \showarticletitle{Resource Management in Mobile Edge Computing: A Comprehensive Survey}.
\newblock \bibinfo{journal}{\emph{Comput. Surveys}} (\bibinfo{year}{2023}).
\newblock


\bibitem[Zhang et~al\mbox{.}(2024d)]%
        {zhang2024beyondtheCloud}
\bibfield{author}{\bibinfo{person}{Xinyuan Zhang}, \bibinfo{person}{Jiangtian Nie}, \bibinfo{person}{Yudong Huang}, \bibinfo{person}{Gaochang Xie}, \bibinfo{person}{Zehui Xiong}, \bibinfo{person}{Jiang Liu}, \bibinfo{person}{Dusit Niyato}, {and} \bibinfo{person}{Xuemin~Sherman Shen}.} \bibinfo{year}{2024}\natexlab{d}.
\newblock \showarticletitle{Beyond the Cloud: Edge Inference for Generative Large Language Models in Wireless Networks}.
\newblock \bibinfo{journal}{\emph{IEEE Transactions on Wireless Communications}} (\bibinfo{year}{2024}).
\newblock


\bibitem[Zhang et~al\mbox{.}(2024a)]%
        {zhang2024plugandplay}
\bibfield{author}{\bibinfo{person}{Yingtao Zhang}, \bibinfo{person}{Haoli Bai}, \bibinfo{person}{Haokun Lin}, \bibinfo{person}{Jialin Zhao}, \bibinfo{person}{Lu Hou}, {and} \bibinfo{person}{Carlo~Vittorio Cannistraci}.} \bibinfo{year}{2024}\natexlab{a}.
\newblock \showarticletitle{Plug-and-play: An efficient post-training pruning method for large language models}.
\newblock \bibinfo{journal}{\emph{ICLR}} (\bibinfo{year}{2024}).
\newblock


\bibitem[Zhang et~al\mbox{.}(2024f)]%
        {zhang2024vulcan}
\bibfield{author}{\bibinfo{person}{Yiwen Zhang}, \bibinfo{person}{Xumiao Zhang}, \bibinfo{person}{Ganesh Ananthanarayanan}, \bibinfo{person}{Anand Iyer}, \bibinfo{person}{Yuanchao Shu}, \bibinfo{person}{Victor Bahl}, \bibinfo{person}{Z~Morley Mao}, {and} \bibinfo{person}{Mosharaf Chowdhury}.} \bibinfo{year}{2024}\natexlab{f}.
\newblock \showarticletitle{Vulcan: Automatic Query Planning for Live ML Analytics}.
\newblock \bibinfo{journal}{\emph{USENIX NSDI}} (\bibinfo{year}{2024}).
\newblock


\bibitem[Zhang et~al\mbox{.}(2024c)]%
        {zhang2024qhitter}
\bibfield{author}{\bibinfo{person}{Zhenyu Zhang}, \bibinfo{person}{Shiwei Liu}, \bibinfo{person}{Runjin Chen}, \bibinfo{person}{Bhavya Kailkhura}, \bibinfo{person}{Beidi Chen}, {and} \bibinfo{person}{Atlas Wang}.} \bibinfo{year}{2024}\natexlab{c}.
\newblock \showarticletitle{Q-Hitter: A Better Token Oracle for Efficient LLM Inference via Sparse-Quantized KV Cache}.
\newblock \bibinfo{journal}{\emph{MLSys}} (\bibinfo{year}{2024}).
\newblock


\bibitem[Zhao et~al\mbox{.}(2024)]%
        {zhao2024lingualinked}
\bibfield{author}{\bibinfo{person}{Junchen Zhao}, \bibinfo{person}{Yurun Song}, \bibinfo{person}{Ian Harris}, \bibinfo{person}{Sangeetha~Abdu Jyothi}, {et~al\mbox{.}}} \bibinfo{year}{2024}\natexlab{}.
\newblock \showarticletitle{LinguaLinked: Distributed Large Language Model Inference on Mobile Devices}.
\newblock \bibinfo{journal}{\emph{ACL}} (\bibinfo{year}{2024}).
\newblock


\bibitem[Zhong et~al\mbox{.}(2024)]%
        {zhong2024distserve}
\bibfield{author}{\bibinfo{person}{Yinmin Zhong}, \bibinfo{person}{Shengyu Liu}, \bibinfo{person}{Junda Chen}, \bibinfo{person}{Jianbo Hu}, \bibinfo{person}{Yibo Zhu}, \bibinfo{person}{Xuanzhe Liu}, \bibinfo{person}{Xin Jin}, {and} \bibinfo{person}{Hao Zhang}.} \bibinfo{year}{2024}\natexlab{}.
\newblock \showarticletitle{DistServe: Disaggregating Prefill and Decoding for Goodput-optimized Large Language Model Serving}.
\newblock \bibinfo{journal}{\emph{USENIX OSDI}} (\bibinfo{year}{2024}).
\newblock


\bibitem[Zhou et~al\mbox{.}(2022)]%
        {zhou2022transpim}
\bibfield{author}{\bibinfo{person}{Minxuan Zhou}, \bibinfo{person}{Weihong Xu}, \bibinfo{person}{Jaeyoung Kang}, {and} \bibinfo{person}{Tajana Rosing}.} \bibinfo{year}{2022}\natexlab{}.
\newblock \showarticletitle{Transpim: A memory-based acceleration via software-hardware co-design for transformer}.
\newblock \bibinfo{journal}{\emph{IEEE HPCA}} (\bibinfo{year}{2022}).
\newblock


\bibitem[Zhou et~al\mbox{.}(2024)]%
        {zhouanomalyclip}
\bibfield{author}{\bibinfo{person}{Qihang Zhou}, \bibinfo{person}{Guansong Pang}, \bibinfo{person}{Yu Tian}, \bibinfo{person}{Shibo He}, {and} \bibinfo{person}{Jiming Chen}.} \bibinfo{year}{2024}\natexlab{}.
\newblock \showarticletitle{AnomalyCLIP: Object-agnostic Prompt Learning for Zero-shot Anomaly Detection}.
\newblock \bibinfo{journal}{\emph{ICLR}} (\bibinfo{year}{2024}).
\newblock


\bibitem[Zhou et~al\mbox{.}(2020)]%
        {zhouBERTLosesPatience2020}
\bibfield{author}{\bibinfo{person}{Wangchunshu Zhou}, \bibinfo{person}{Canwen Xu}, \bibinfo{person}{Tao Ge}, \bibinfo{person}{Julian McAuley}, \bibinfo{person}{Ke Xu}, {and} \bibinfo{person}{Furu Wei}.} \bibinfo{year}{2020}\natexlab{}.
\newblock \showarticletitle{BERT Loses Patience: Fast and Robust Inference with Early Exit}.
\newblock \bibinfo{journal}{\emph{NeurIPS}} (\bibinfo{year}{2020}).
\newblock


\bibitem[Zhu et~al\mbox{.}(2024a)]%
        {zhu2024llmsALFA}
\bibfield{author}{\bibinfo{person}{Jiaqi Zhu}, \bibinfo{person}{Shaofeng Cai}, \bibinfo{person}{Fang Deng}, \bibinfo{person}{Beng~Chin Ooi}, {and} \bibinfo{person}{Junran Wu}.} \bibinfo{year}{2024}\natexlab{a}.
\newblock \showarticletitle{Do LLMs Understand Visual Anomalies? Uncovering LLM's Capabilities in Zero-shot Anomaly Detection}.
\newblock \bibinfo{journal}{\emph{ACM MM}} (\bibinfo{year}{2024}).
\newblock


\bibitem[Zhu(2021)]%
        {zhuLeeBERTLearnedEarly2021}
\bibfield{author}{\bibinfo{person}{Wei Zhu}.} \bibinfo{year}{2021}\natexlab{}.
\newblock \showarticletitle{LeeBERT: Learned Early Exit for BERT with Cross-Level Optimization}.
\newblock \bibinfo{journal}{\emph{ACL}} (\bibinfo{year}{2021}).
\newblock


\bibitem[Zhu et~al\mbox{.}(2024b)]%
        {zhu2024textrewriting}
\bibfield{author}{\bibinfo{person}{Yun Zhu}, \bibinfo{person}{Yinxiao Liu}, \bibinfo{person}{Felix Stahlberg}, \bibinfo{person}{Shankar Kumar}, \bibinfo{person}{Yu-Hui Chen}, \bibinfo{person}{Liangchen Luo}, \bibinfo{person}{Lei Shu}, \bibinfo{person}{Renjie Liu}, \bibinfo{person}{Jindong Chen}, {and} \bibinfo{person}{Lei Meng}.} \bibinfo{year}{2024}\natexlab{b}.
\newblock \showarticletitle{Towards an On-device Agent for Text Rewriting}.
\newblock \bibinfo{journal}{\emph{Findings of NAACL}} (\bibinfo{year}{2024}).
\newblock


\end{thebibliography}

\end{document}